\documentclass[11pt]{prop_axis} 
\usepackage[font=small,labelfont=bf]{caption}

\usepackage[super,comma]{natbib} 
\usepackage{apjfonts} 
\usepackage{graphicx} 
\usepackage{amssymb,amsmath} 
\usepackage{color} 
\usepackage{enumitem} 
\usepackage{wrapfig} 
\usepackage{tikz} 

\usepackage{multirow} 
\usepackage{tabularx} 
\usepackage{array} 
\usepackage{hyperref} 
\hypersetup{
    colorlinks,
    linkcolor={blue!80!black},
    citecolor={blue!80!black},
    urlcolor={blue!80!black}
}

\usepackage{sidecap}
\sidecaptionvpos{figure}{c} 

\newcolumntype{L}[1]{>{\raggedright\let\newline\\\arraybackslash\hspace{0pt}}p{#1}}
\newcolumntype{C}[1]{>{\centering\let\newline\\\arraybackslash\hspace{0pt}}p{#1}}
\newcolumntype{R}[1]{>{\raggedleft\let\newline\\\arraybackslash\hspace{0pt}}p{#1}}
\renewcommand\arraystretch{1.3}

\definecolor{headcolor}{rgb}{0.65,0.65,0.65}
\usepackage{fancyhdr}
\renewcommand{\topmargin}{-9mm}

\renewcommand{\textfloatsep}{4mm}

\newcommand{\lem}{\textit{LEM}}

\newcommand{\bsf}{\sffamily\bfseries}

\definecolor{callout}{rgb}{0.25,0.40,0.85}
\definecolor{synergies}{rgb}{0.20,0.45,0.99}
\definecolor{methods}{rgb}{0.20,0.70,0.45}

\definecolor{calllem}{rgb}{0.20,0.45,0.99}
\definecolor{tabledef}{rgb}{0.95,0.95,0.95}
\definecolor{tablealt}{rgb}{0.77,0.80,1.0}
\definecolor{tablelem}{rgb}{0.80,0.85,1.0}
\definecolor{whitelem}{rgb}{1.0,1.0,1.0}
\definecolor{greenlem}{rgb}{0.7,1.0,0.7}
\begin{document}

\baselineskip=13.2pt
\sloppy
\pagenumbering{roman}
\thispagestyle{empty}

\title{\textcolor{headcolor}{\LARGE\bsf 
Revolutionary Solar System Science Enabled by the Line Emission Mapper X-ray Probe}}
\maketitle
\vspace*{-10mm}

\begin{tikzpicture}[remember picture,overlay]
\node[anchor=north west,yshift=2pt,xshift=2pt]%
    at (current page.north west)
    {\includegraphics[height=20mm]{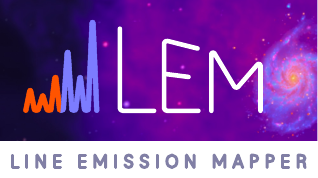}};
\end{tikzpicture}

\vspace*{-11mm}
\begin{center}
\begin{minipage}{17.5cm}
\hspace{-5mm}
\centering
W. R. Dunn$^{1,2}$, D. Koutroumpa$^{3}$, J. A. Carter$^{4}$, K. D. Kuntz$^{5,6}$, S. McEntee$^{7,8}$, T. Deskins$^{9}$, B. Parry$^{1,2}$, S. Wolk$^{10}$, C. Lisse$^{11}$, K. Dennerl$^{12}$, C.M. Jackman$^{7}$, D.M. Weigt$^{13}$, F. S. Porter$^{6}$, G. Branduardi-Raymont$^{14}$, D. Bodewits$^{9}$, F. Leppard$^{19}$, A Foster$^{10}$, G. R. Gladstone$^{15,16}$, V. Parmar$^{1,2}$, S. Brophy-Lee$^{7}$, C. Feldman$^{4}$, J-U. Ness$^{17}$,  R. Cumbee$^{6}$, M. Markevitch$^{6}$, R. Kraft$^{10}$, A. Bogdan$^{10}$, A. Bhardwaj$^{18}$, A. Wibisono$^{20}$, F. Mernier$^{6, 21}$, A. Ogorzalek $^{6, 21}$

\end{minipage}
\end{center}

\vfill



{\footnotesize
\noindent
$^{1}$ Department of Physics and Astronomy, University College London, London, UK \\
$^{2}$ Center for Planetary Science, University College London, UK \\
$^{3}$ LATMOS-OVSQ, CNRS, UVSQ Paris-Saclay, Sorbonne Université, 11 Boulevard d'Alembert, 78280, Guyancourt, France \\
$^{4}$ School of Physics and Astronomy, University of Leicester, Leicester LE1 7RH, UK\\
$^{5}$ Department of Physics and Astronomy, Johns Hopkins University, 3701 San Martin Drive, Baltimore, MD, 21218, USA \\
$^{6}$ NASA Goddard Space Flight Center, 8800 Greenbelt Road, Greenbelt, MD 20771, USA \\
$^{7}$ School of Cosmic Physics, DIAS Dunsink Observatory, Dublin Institute for Advanced Studies, Dublin, Ireland \\
$^{8}$ School of Physics, Trinity College Dublin, Dublin, Ireland \\
$^{9}$ Physics Department, Edmund C. Leach Science Center, Auburn University, AL 36832, USA \\
$^{10}$ Center for Astrophysics | Harvard \& Smithsonian, Cambridge, MA, 02138 US \\
$^{11}$ Space Department, Johns Hopkins University Applied Physics Laboratory, 11100 Johns Hopkins Rd, Laurel, MD 20723 \\
$^{12}$ Max-Planck-Institut für extraterrestrische Physik, Garching, Germany \\
$^{13}$ Department of Computer Science, Aalto University, 02150 Espoo, Finland \\
$^{14}$ Mullard Space Science Laboratory, University College London, Dorking, UK \\
$^{15}$ Southwest Research Institute, San Antonio, TX, USA \\
$^{16}$University of Texas at San Antonio, San Antonio, TX, USA \\
$^{17}$ European Space Astronomy Center, Madrid, Spain \\
$^{18}$ Physical Research Laboratory, Ahmedabad, India \\
$^{19}$ Department of Earth Sciences, The University of Hong Kong, Pokfulam, Hong Kong \\
$^{20}$ Royal Observatory Greenwich, London, UK \\
$^{21}$ Department of Astronomy, University of Maryland, College Park, MD 20742-2421, USA\\

\phantom{${^52}$}~\textcolor{blue}{\bsf \href{https://lem-observatory.org}{lem-observatory.org}}\\
\phantom{${^52}$}~\textcolor{blue}{\bsf X / twitter: \href{https://www.twitter.com/LEMXray}{LEMXray}}\\
\phantom{${^52}$}~\textcolor{blue}{\bsf facebook: \href{https://www.facebook.com/LEMXrayProbe}{LEMXrayProbe}}\\
}

\centerline{\em White Paper, October 2023}

\clearpage

\section*{Executive Summary}
\label{sec:summary}

The Line Emission Mapper’s (LEM’s) exquisite spectral resolution and effective area will open new research domains in Astrophysics, Planetary Science and Heliophysics. \textit{LEM} will provide step-change capabilities for the solar coronal X-ray scattering, fluorescence, solar wind charge exchange (SWCX) and auroral precipitation processes that dominate X-ray emissions in our Solar System and Heliosphere.
The observatory will enable novel X-ray measurements of historically inaccessible line species, thermal broadening, characteristic line ratios and Doppler shifts – a universally valuable new astrophysics diagnostic toolkit. These measurements will identify the underlying compositions, conditions and physical processes from km-scale ultra-cold comets to the MK solar wind in the heliopause at $\sim$120 au.  Here, we focus on the paradigm-shifts \textit{LEM} will provide for understanding the nature of the interaction between a star and its planets, especially the fundamental processes that govern the transfer of mass and energy within our Solar System, and the 
distribution of elements throughout the heliosphere.

In this White Paper we show how \textit{LEM} will enable a treasure trove of new scientific contributions that directly address key questions from the National Academies’ 2023-2032 Planetary Science (PS) and 2013-2022 Heliophysics (H) Decadal Strategies\cite{PSDec,HelioDec}. The topics we highlight include:

\begin{enumerate}

\item \textbf{The richest global trace element maps of the Lunar Surface} ever produced; insights that address Solar System and planetary formation, and provide invaluable context ahead of \textit{Artemis} and the \textit{Lunar Gateway} (PS - Q1; Q2; Q9.1). 

\item \textbf{Global maps of our Heliosphere} through Solar Wind Charge Exchange (SWCX) that trace the interstellar neutral distributions in interplanetary space and measure system-wide solar wind ion abundances and velocities; a key new understanding of our local astrosphere and a synergistic complement to \textit{NASA IMAP} observations of heliospheric interactions (H-Q3).

\item \textbf{Elemental and molecular composition measurements of the atmospheres of Mars, Venus, Jupiter, Saturn and Uranus}; noble gas and volatile abundances that distinguish planetary formation theories and provide invaluable new insights ahead of  \textit{NASA's Uranus Orbiter and Probe} Flagship (PS - Q1.1; Q2.1; Q7.1;Q7.2; Q12)

\item \textbf{Global elemental composition measurements of the surfaces of the Galilean Satellites}; essential insights for their formation, evolution and potential habitability. Synchronised with \textit{JUICE} and \textit{Europa Clipper} operations, these will provide important contemporaneous measurements for degeneracy-breaking of spacecraft IR molecular spectroscopic observations (PS - Q2.2; Q2.3; Q9; Q10)

\item \textbf{Direct global observations of atmospheric loss from Mars and Venus through SWCX} X-ray images of their planetary ‘halos’, providing global context to the localized, single-point in-situ measurements by \textit{MAVEN, enVision, Veritas} and \textit{DAVINCI+} (PS - Q6.5)

\item \textbf{Probing Solar System archeology and Heliospheric dynamics beyond the range of in-situ spacecraft} through observations of distant Comets and Kuiper Belt Objects, with the possibility of interstellar comet observations simultaneous with Comet Interceptor (H - Q1, Q3; PS - Q1; Q2; Q3). 

\item \textbf{Abundance data for solar wind species that are unavailable to in-situ spacecraft measurements and in locations that are yet to be visited by in-situ missions} through high resolution SWCX observations; such solar abundances are the baseline for chemical composition across the cosmos and are key for understanding space weather processes and Solar System formation (PS - Q1.1, Q3.6, H-Q1)

\item \textbf{The longest temporal baseline, high resolution 0.2-2 keV solar X-ray spectra}, as seen reflected from hydrogen in planetary atmospheres, a valuable ground truth for stellar populations across the sky (H-Q1)

\item \textbf{Identification of solar wind ion precipitation in the Jovian aurorae and the magnetospheric processes that produce the most powerful aurora in the Solar System.} Through CX, \textit{LEM} will remotely measure the influx of the solar wind to the Giant Planet atmospheres and measure the thermal and collisional velocities of precipitating ions in the aurorae. (PS-Q7.4)

\item \textbf{High Cadence (3-minute) imagery of small-scale structure ($\sim$0.22 $R_E$) on the terrestrial magnetosheath}. This provides new windows into the processes that govern magnetospheric physics and the impact of Space Weather on the Earth, and is ideally timed for partnerships with NASA's HelioSwarm mission to connect small to large scale magnetospheric processes (H-Q2, Q3, Q4)

\end{enumerate}

In these areas and more, \textit{LEM} will provide groundbreaking new National Academy of Sciences mandated Decadal science and serve as a new, essential planetary mission – all while remaining in an L1 halo orbit. It will do so by building on decades of X-ray studies from the \textit{Einstein Observatory, Chandra, ROSAT, XMM-Newton, NuSTAR, and NICER} and heliophysics and planetary missions like \textit{ACE, IMP-8, TIMED, Cassini, New Horizons, Messenger, Chandrayaan, Galileo, Juno and Voyager}. 
\textit{LEM} will open a new wavelength window on planetary science and heliophysics, addressing some of the most profound and fundamental questions for Solar System science. In turn, this new understanding will provide irreplaceable ground truths that span astrophysics.



\begin{figure}
	\centering
		\includegraphics[width=1\textwidth]{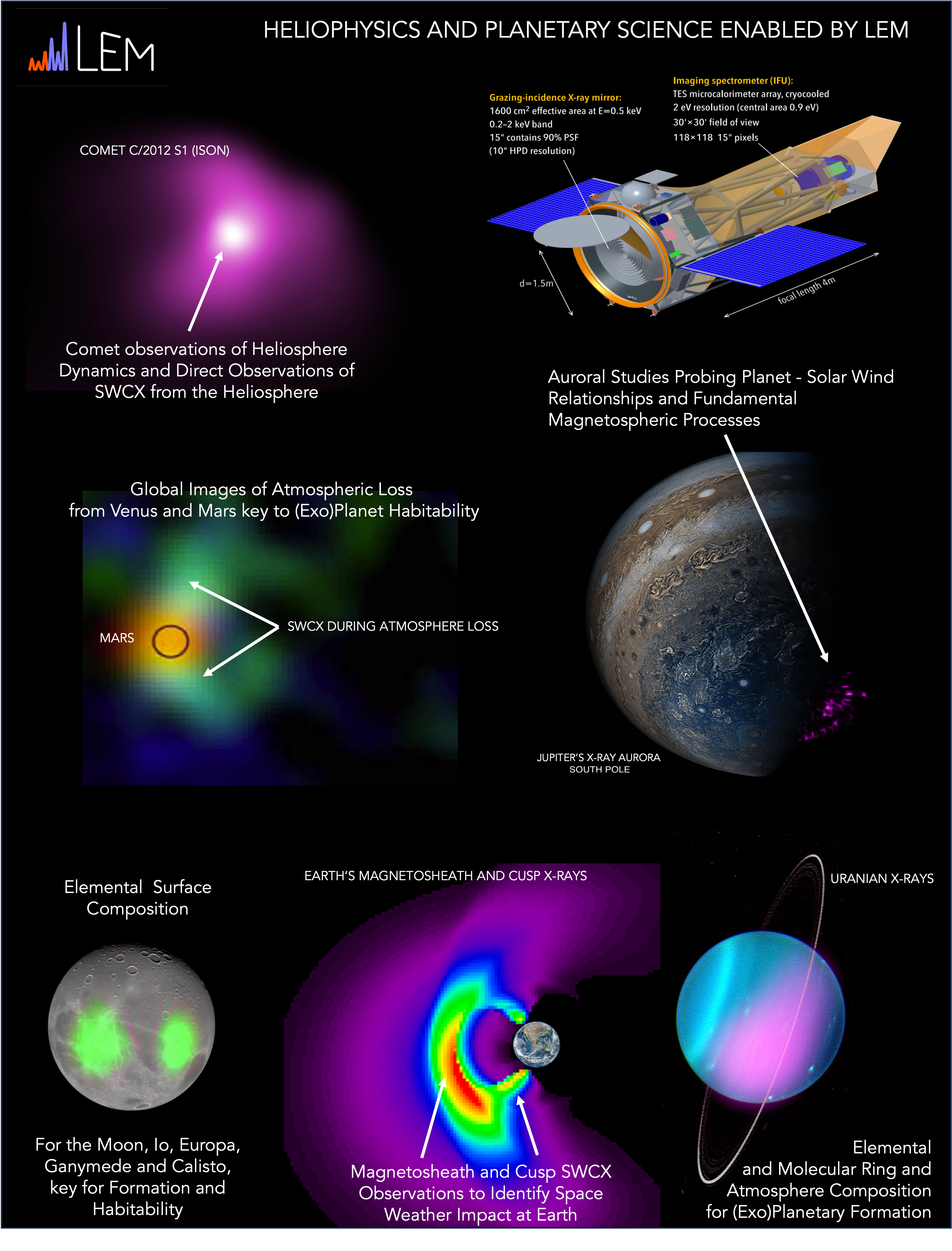}
			\caption{\textbf{Illustrative summary of Heliophysics and Planetary Science Enabled by  \textit{LEM} }. Image Credit:NASA/CXC/ESA/XMM-NEWTON/Kraft et al.\cite{Kraft22}/Snios/Dennerl/Dunn/Kuntz}
	\label{CoverPage}
\end{figure}

\twocolumn


\setcounter{page}{1}
\pagenumbering{arabic}

\section{Introduction}

\begin{figure*}[h]
	\centering
		\includegraphics[width=0.95\textwidth, trim={0cm 0cm 0cm 0cm},clip]{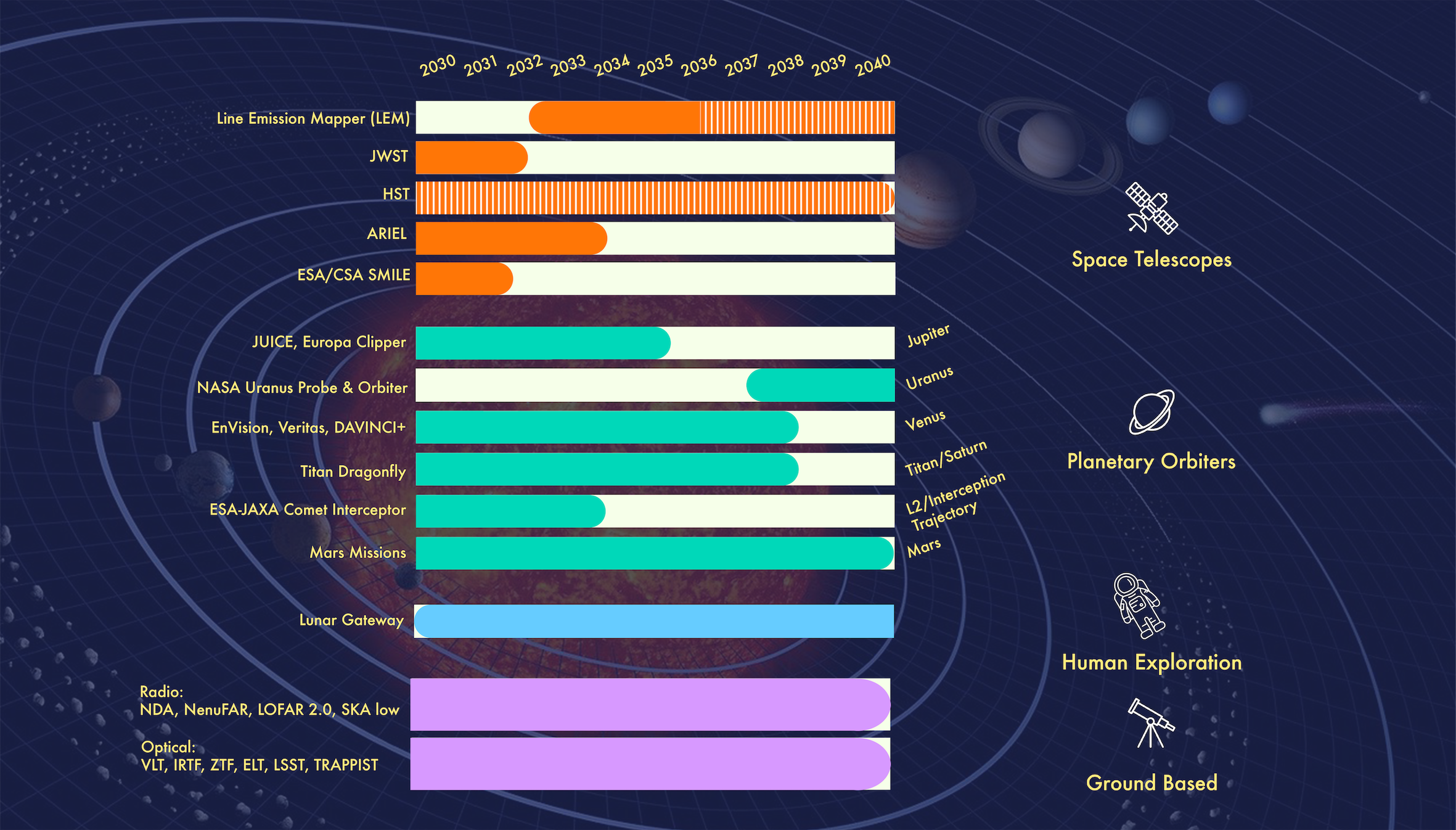}
			\caption{\textbf{Synergistic spacecraft and observatories active with the Line Emission Mapper.}}
	\label{OtherSpacecraft}
\end{figure*}

\subsection{LEM’s Breakthrough Capabilities for Planetary Science and Heliophysics}
\label{sec:intro}

Proposed as a mission concept for the NASA 2023 Astrophysics Probes call, the \textit{Line Emission Mapper (LEM)} consists of a micro-calorimeter array spanning a large Field of View (FoV) of 30 by 30 arcminutes\cite{Kraft22}. This array is capable of reaching an unprecedented energy resolution of $\sim$1 eV within a central 7×7 arcminute pixel configuration (within which all Solar System planets will fit) and 2 eV across the rest of the Field of View. \textit{LEM} covers the 0.2–2 keV energy band across which planetary bodies and solar wind charge exchange (SWCX) are brightest in X-rays, making it ideal for their study. The X-ray mirrors of  \textit{LEM}  will provide a collecting area ($\sim$1600 cm$^2$) and angular resolution ($\sim$15''), of the order of the \textit{EPIC-pn} instrument onboard \textit{XMM-Newton}, which has been highly successful for the study of planetary bodies \cite{Bhardwaj2007,GBR2022,Dennerl2010,Dunn2022Jupiter}. The combination of capabilities means that observations will have a much reduced background, enhancing signal to noise significantly and enabling detailed characterisation of planetary and heliophysics signals that have historically been beyond reach.  \textit{LEM} also possesses some sensitivity down to $\sim$0.05 keV, which, if preserved, will give a factor of several enhancement in signal from Solar System objects. This will provide access to never-before-measured lines offering key new compositional insights (see section 2 and 3), enabling studies of objects previously too dim to observe and probe variability on shorter timescales than ever before.
\\

Solar System bodies offer irreplaceable ground-truth in-situ spacecraft measurements (e.g. magnetic and electric fields, energetic particles and wave measurements) with which to explore conditions that are challenging to create in terrestrial laboratories and that identify source processes that drive X-ray emissions. Applying \textit{LEM’s} state-of-the-art spectroscopic capabilities to the heliosphere provides a powerful natural laboratory to identify fundamental atomic and plasma physics at astrophysical bodies across the Universe. For astrophysics in general, this anchors spectroscopic physical models to well-characterized bodies, enabling these models to be reliably and robustly extrapolated to objects beyond the reach of in-situ measurements.

\subsection{Multi-Spacecraft Synergies with LEM}
\label{sec:multispacecraft}

 \textit{LEM} will launch amidst the operation of a diverse fleet of spacecraft and other observatories, some of which are shown in Fig. \ref{OtherSpacecraft}. Throughout this white paper we highlight example multi-spacecraft synergies with  \textit{LEM}. Undoubtedly the discoveries by \textit{LEM} and these missions will also open new opportunities. There is a rich heritage proving that multi-instrument measurements enable and inspire scientific investigations beyond the sum of their parts\cite{Bunce2008,Clarke1998,Jia2018,YaoDunn2021} and \textit{LEM}’s partnerships will be no exception to this. 

In turn, this deeper understanding of the fundamental processes governing X-ray emissions in the Solar System will help to address problems and break degeneracies in astrophysics more broadly.

\subsection{Utilising \textit{LEM} to Identify What  Processes Govern Mass and Energy Transfer within the Heliosphere} \label{subsec:Processes}

We begin this white paper by exploring what  \textit{LEM} 's new capabilities enable at the fundamental level of spectral line identification of species and the underlying X-ray-production  processes. We then explore how identification of source species and processes can be leveraged to address some of the most important questions in planetary science and heliophysics (section 2 and 3). 

The majority of astrophysical emissions can be grouped into five X-ray processes: fluorescence, thermal/coronal (collisionally ionised and non-collisionally ionised emissions), scattering, charge exchange (CX) and particle acceleration. The Solar System possesses regions and objects that exhibit all of these processes \cite{Bhardwaj2007,GBR2022,Dennerl2010,Dunn2022Jupiter}. However, given current spectral resolution with \textit{Chandra ACIS} or \textit{XMM-Newton EPIC-pn} or sensitivity with \textit{XMM-Newton RGS} or \textit{Chandra LETG}, for planetary bodies and sources within the heliosphere, it is often difficult to distinguish these emission processes and therefore to unambiguously identify the underlying physics. \textit{LEM’s} spectral resolution and sensitivity will provide the needed instrumental step-change, revolutionising the use of remote X-ray observations to study planetary bodies and heliophysics.

\begin{figure}
	\centering
		\includegraphics[width=0.55\textwidth, trim={0cm 0cm 0cm 0cm},clip]{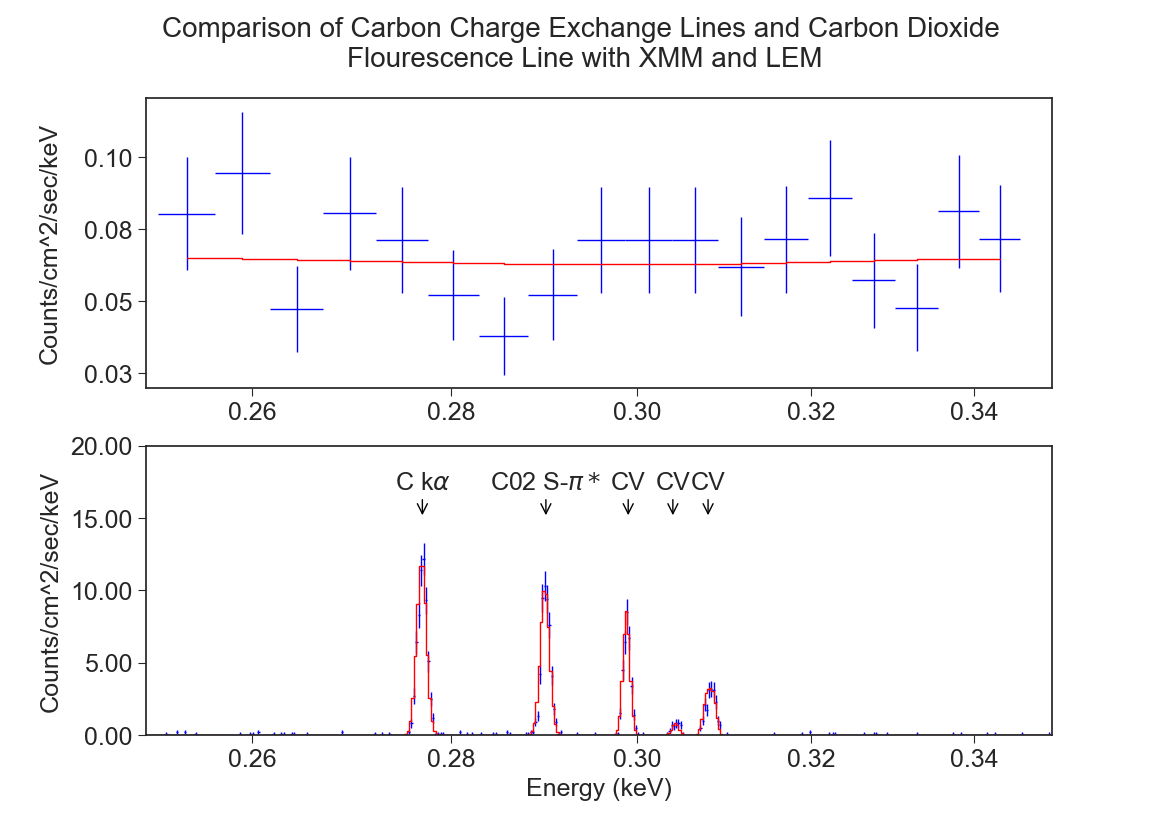}
			\caption{\textbf{Illustrative comparison of \textit{LEM} ’s capability to distinguish source species in comparison with current state-of-the-art}: this shows a spectrum from a neutral carbon k-$\alpha$ line, a neutral CO$_2$ $s\rightarrow\pi^{*}$ line and several CV charge exchange (CX) lines. The upper panel shows a simulated spectrum (blue crosses) as would be observed with \textit{XMM-Newton EPIC-pn}, while the lower panel shows the same spectrum as would be observed by \textit{LEM}. The red-line shows the theoretical model convolved with the instrument response. Figure from \cite{Parmar2023}.}
	\label{BrynFluorescenceCX}
\end{figure}

For example, Fig. \ref{BrynFluorescenceCX} shows spectral lines from fluorescence from neutral atomic carbon, neutral carbon dioxide molecules and from solar wind charge exchange (SWCX) of highly charged ($C^{5+}$) carbon ions with neutrals - emission processes and species that are particularly relevant for e.g. Venus\cite{Dennerl2002Venus,Dennerl2008}, Mars\cite{Dennerl2002Mars,Dennerl2006}, Titan \cite{Bertucci2008}, comets \cite{Snios2018} and Kuiper Belt objects \cite{Dunn2022Ice,Lisse2017}. The processes and conditions that lead to each emission are dramatically different and probe physics from the formation of planets and their habitability to the nature of their interaction with the solar wind. The top panel shows that with \textit{XMM-Newton EPIC-pn} (the current state of the art for combining sensitivity and spectral resolution) it is impossible to distinguish these lines, so that it has historically been challenging to distinguish the source ion/molecule/atom itself, let alone the underlying physical processes.  However, the lower panel shows that \textit{LEM} will enable the needed resolution of the lines. By doing this, it can unambiguously resolve the different emission sources and processes  (providing the line broadening velocities are less than 1000 km/s and dependent on the molecular species). \textbf{For planetary science,  \textit{LEM}  opens entirely new domains of science, enabling exploration of neutral atomic and molecular species as well as highly charged ions.} 

\begin{figure*}
	\centering
		\includegraphics[width=0.98\textwidth, trim={0cm 0cm 0cm 0cm},clip]{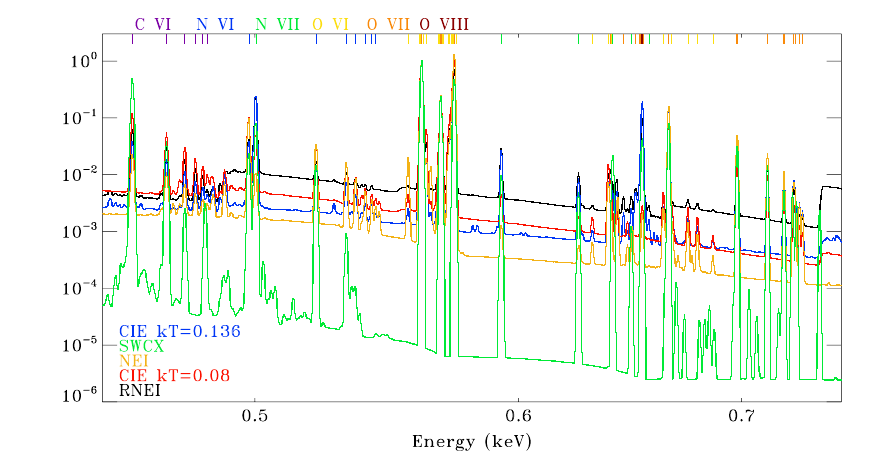}
			\caption{\textbf{\textit{LEM’s} capability to distinguish emission processes - spectra from 5 different X-ray emission processes.} The different example spectra presented are: collisionally ionised plasma with kT $\sim$0.136 (blue - representative of the halo); collisionally ionised plasma with kT $\sim$0.0808 (red - representative of the local hot bubble), overionised plasma recombining from kT of 0.136 to 0.808 (black) and under ionised plasma with a kT of 0.136 (orange), solar wind charge exchange (green).  Color codes for the tick marks on the upper axis highlight lines from different ion species: Purple C VI; Blue N VI; Green N VII; Yellow O VI; Orange O VII; Red OVIII. }
	\label{KuntzSpectralComparison}
\end{figure*}

Fig. \ref{BrynFluorescenceCX} shows how \textit{LEM} will distinguish neutral atomic and molecular lines from those of highly charged ions. However, identifying the processes that generate X-ray emissions often requires measurement of the line ratios between different lines from the same species \cite{Porquet2010}. \textit{LEM} also enables these studies. Fig. \ref{KuntzSpectralComparison} showcases that \textit{LEM} can distinguish the fundamental emission processes and plasma properties including charge exchange (CX), collisional ionisation, fluorescence and photoionisation to characterize the dominant transfers of energy in different astrophysical systems. 

\begin{figure}
	\centering
		\includegraphics[width=0.51\textwidth, trim={0cm 0cm 0cm 0cm},clip]{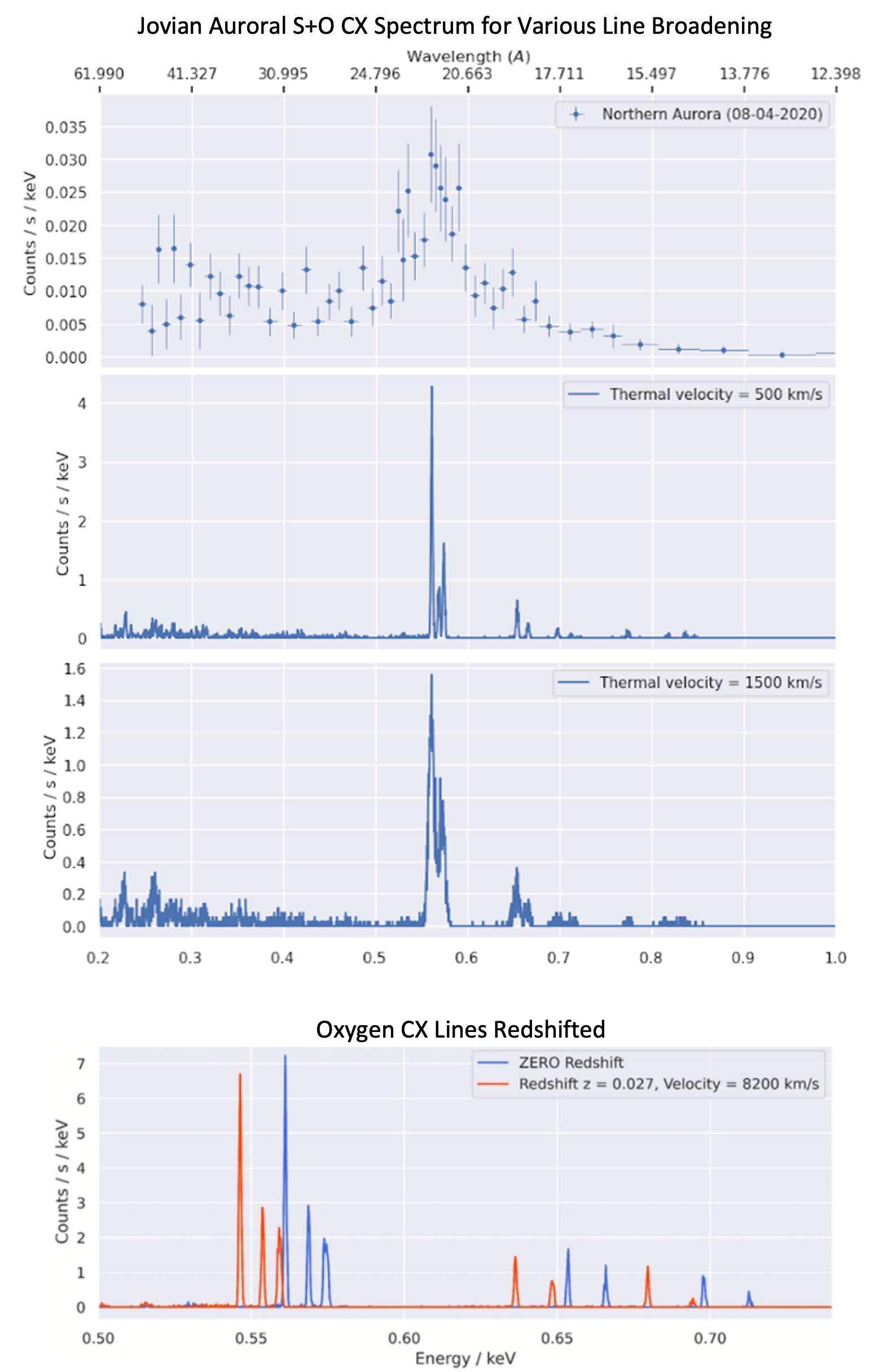}
			\caption{\textbf{Illustrative examples of \textit{LEM} ’s capability to measure line broadening and Doppler shifts for typical Solar System conditions:} The top panel shows an \textit{XMM-Newton EPIC-pn} observation of the X-ray spectrum from Jupiter's Northern aurora. This spectrum is best-fit with a model that represents precipitating magnetospheric sulphur and oxygen ions undergoing charge exchange with Jupiter's neutral atmosphere. However, the \textit{EPIC-pn} resolution is insufficient to identify broadening or doppler shifts at velocities that are typical for the system, so that the underlying plasma properties cannot be explored (see section 3 for details). The panels below show  \textit{LEM}  observations of different line broadening (middle two panels) or Doppler shifts (bottom panel) for this Jovian aurora CX spectrum, showcasing  \textit{LEM} 's capacity to remotely explore the underlying plasma properties.  We note that this is illustrative of the step-change in scientific exploration enabled by  \textit{LEM}, but that a real observation will convolve Doppler and thermal broadening so that care will need to be taken when interpreting the new observations made by \textit{LEM}. Figures from \cite{Parmar2023}.}
	\label{ParmarBroadeningAndShifts}
\end{figure}

Beyond identification of the photon production processes, \textit{LEM} will also probe the thermal and collisional properties of Solar System plasmas. Through broadening and shifts of spectral lines,  \textit{LEM}  is sensitive at the velocities relevant for Solar System bodies (i.e. a few hundred km/s). Fig. \ref{ParmarBroadeningAndShifts} shows  \textit{LEM}  measurements of different thermal and doppler-shifted lines. In turn, understanding the width and shift of the emission from different components enables the separation of different components along the line of sight, and a deep exploration of the underlying plasma properties that has historically been unavailable for Solar System objects.

Through these new measurement of line species, ratios, broadening and shifts, \textit{LEM} enables an unprecedented richness of Solar System study that will provide step changes towards the strategic goals of the planetary science and heliophysics decadals \cite{PSDec,HelioDec}.

\subsection{Symbiosis with Laboratory Atomic Physics Advancements}

By ushering-in this new era in astrophysics, heliophysics and planetary science, \textit{LEM} will guide terrestrial laboratory experiments to ensure atomic models are rich with the necessary cross sections to support astrophysical exploration, driving forward synergistic domains of astro- and atomic physics. At the same time, astrophysical environments, including locations within the Solar System, provide extreme conditions that in many cases are very difficult or impossible to recreate in terrestrial laboratories (low densities, high temperatures, extremely low and high magnetic field strengths), that expand the horizons and fundamental understanding of atomic and molecular physics. For example, an array of X-ray spectral lines that are labeled ‘forbidden’, are often the dominant emission lines for astrophysical bodies. This provides a universal context for Earth-based experiments.

Undoubtedly,  \textit{LEM}  will reveal and detect an array of spectral signatures that are uncatalogued or not yet characterised. Identifying these lines will require strong partnerships with laboratory atomic physics to ensure that signals can be correctly interpreted. Here, spatially resolved and better characterised Solar System objects can provide the necessary knowledge stepping stone to build the needed atomic physics progress.  \textbf{\textit{LEM}'s high resolution X-ray spectra necessitate and foster a strong symbiosis between astrophysics, solar system science and laboratory atomic physics}.

\section{Planetary Elemental Inventories: Identifying the Building Blocks of the Solar System}

The 2023 National Academies Planetary Science (PS) Decadal extensively discusses the vital importance of complete chemical inventories across the Solar System (e.g. PS Decadal questions: 1.1; 1.2; 2.1; 2.2; 2.4; 3.1; 3.3; 3.5; 7.1; 9.1; 9.2; 10.5, 12.2)\cite{PSDec}. These inventories are essential to inform models on planet formation and evolution and the prevalence of life beyond Earth. Models of prebiotic conditions and those that explore habitability require solar-system-wide elemental abundances: ``especially those deemed essential to terrestrial life (e.g., key building block elements like C, H, N, O, P, S)'' (PS Decadal\cite{PSDec}).  

Through fluorescence, X-rays observations directly enable remote characterisation of the elemental composition of an object. These finger-print spectral line signatures unambiguously distinguish elements and can also identify a range of molecules (e.g. Fig. \ref{BrynFluorescenceCX}). In this section, we explore how  \textit{LEM}  will provide a paradigm shift in our understanding of the elemental abundances across the Solar System from the solid surfaces of rocky planetary bodies, to the atmospheres of planets and finally to the rings and potentially life-harboring Icy Moons of the Gas and Ice Giants.

\subsection{What We Know}
Generally, chemical inventories in planetary science are collected through either single-point measurements or via global imaging-spectroscopy. In-situ particle measurements provide single-point, single-time measurements of plasma and sputtered surfaces, such as the paradigm-shifting measurements conducted by \textit{Cassini’s INMS}\cite{Waite2004} and \textit{CAPS}\cite{Linder1998} instruments that sampled the Enceladus plume material that populates the Saturnian magnetosphere or the revolutionary measurements made by the \textit{Huygens} probe’s descent through Titan’s atmosphere \cite{Lebreton2005}. More recently, sample return has begun to play an increasingly important role in the \textit{NASA, ESA, CAS} and \textit{JAXA} strategic objectives, and is showcased through the recent success of \textit{NASA's OSIRIS-REx}\cite{Bierhaus2018} and \textit{JAXA's Hayabusa}\cite{Tsuda2013} in returning meteor samples to Earth for exploration in terrestrial laboratories.

\begin{figure}
	\centering
		\includegraphics[width=0.48\textwidth, trim={0cm 0cm 0cm 0cm},clip]{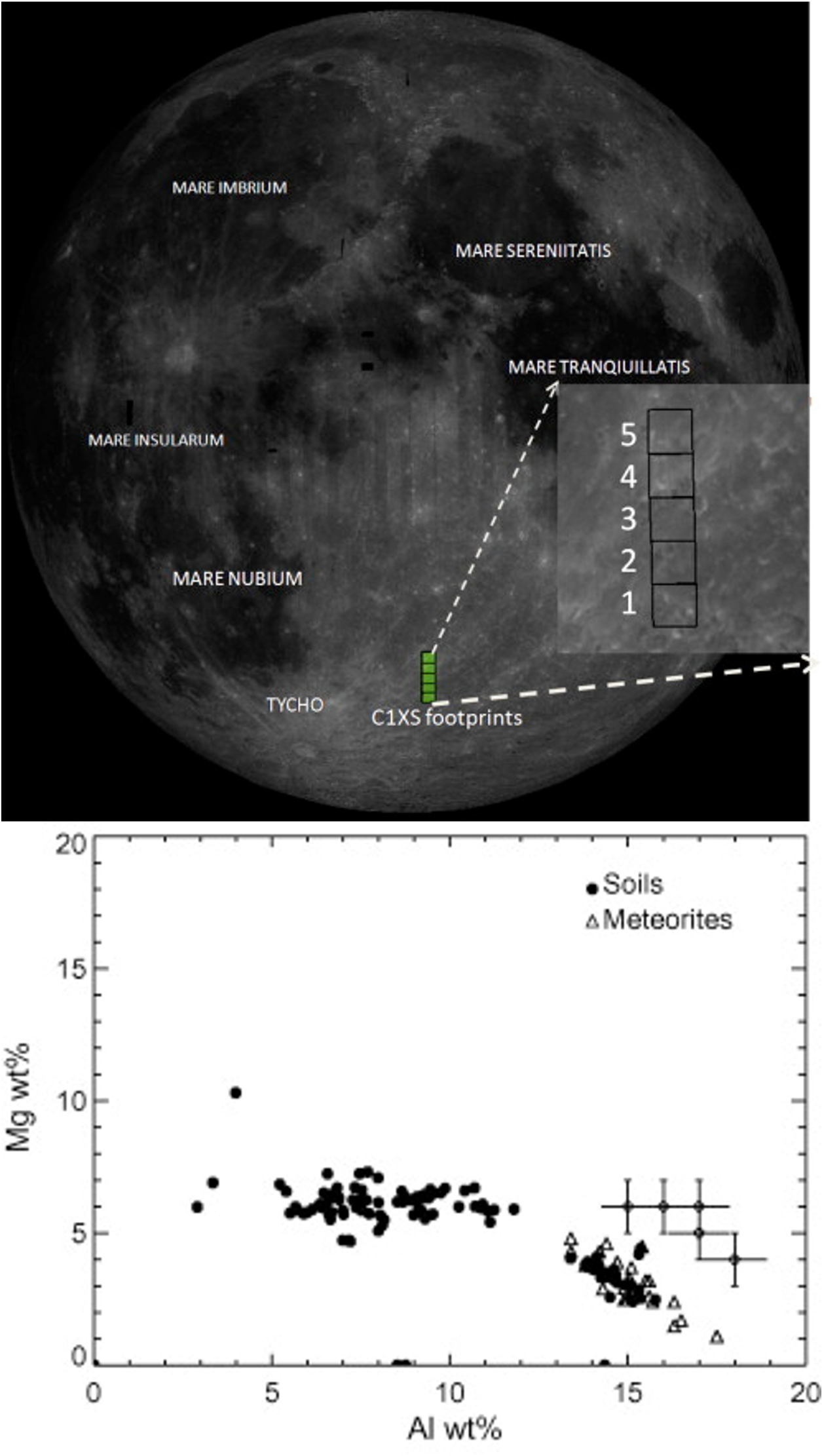}
			\caption{\textbf{\textit{Chandrayaan-1 C1XS} measurements of the lunar elemental abundances through observations of X-ray fluorescence.} Upper Panel: shows in green and in the inset (indicated by the white dashed arrows) the track of the spacecraft footprint within which X-ray measurements were taken. Lower Panel: Compositions derived from \textit{C1XS} plotted along with lunar soil averages \cite{Haskin1991,Morris1983} and feldspathic meteorite compositions \cite{Demidova2007}. Both figures are reproduced from Narendranath et al. (2011) \cite{Narendranath2011}.}
	\label{Chandrayaan}
\end{figure}

At a global scale, remote imaging is key and provides maps of the chemical inventories across planetary surfaces, atmospheres, and rings - essential context for sample return. UV, Infrared and sub-millimeter imaging spectroscopy enable identification of molecular constituents, while X-ray and Gamma-Ray observations provide direct measurements of the elemental composition (although we note that with new high spectral resolution, X-rays can also directly identify molecules - see Fig. \ref{BrynFluorescenceCX}). This mixture of wavebands is highly complementary with elemental measurements breaking degeneracies in molecular models and offering critical noble gas measurements. In turn, noble gas measurements are essential to constrain formation models and connections to the solar abundances, with significantly less chemical processing through the Solar System’s history than volatiles\cite{Atreya2022}.  

\subsubsection{Elemental Composition of the Surfaces of Rocky Bodies}

\begin{figure}
	\centering
		\includegraphics[width=0.45\textwidth, trim={0cm 0cm 0cm 0cm},clip]{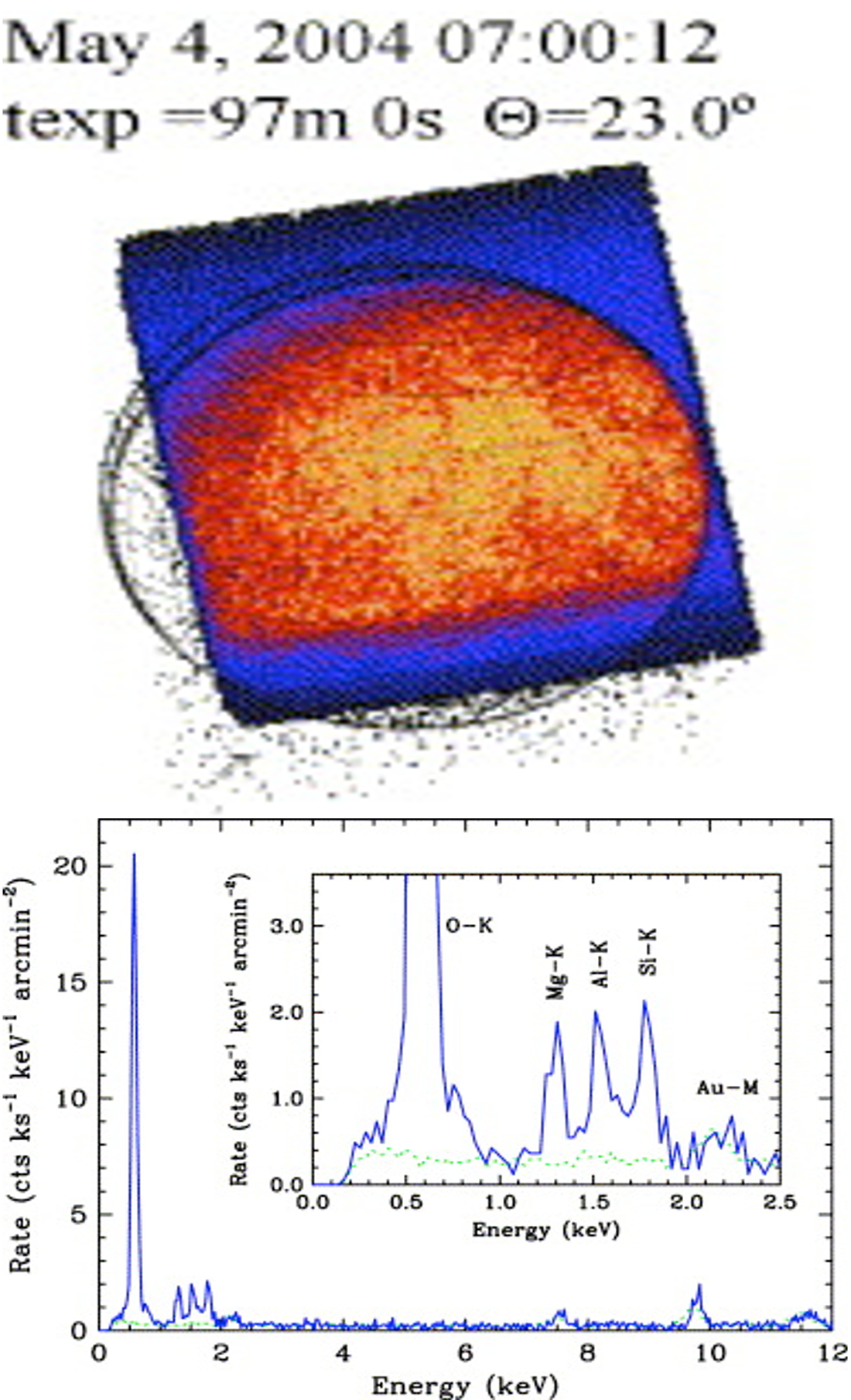}
			\caption{\textbf{Pre-\textit{LEM} X-ray Observations of the Moon:} Upper: \textit{Chandra HRC-I} observation of the Moon. The Moon slightly overfills the 30'×30' \textit{HRC-I} field of view. Lower: \textit{Chandra ACIS} spectrum of the bright side of the Moon. The green dotted curve is the detector background. K-shell fluorescence lines from O, Mg, Al, and Si are shifted up by 50 eV from their true values because of residual optical leak effects. Features at 2.2, 7.5, and 9.7 keV are intrinsic to the detector. Figures and caption content from Wargelin et al. (2004) \cite{Wargelin2004}.}
	\label{ChandraLunar}
\end{figure}

Throughout the PS decadal, the importance of understanding the surface and composition of the moon is discussed extensively, from formation and evolution of the Solar System (PS-Q1, Q2) to understanding the prebiotic context of Earth (PS-Q9.1). For example, Q2.4 states: \textit{``Presently there is no fully accepted, self-consistent model for the simultaneous formation of the giant planets during the nebular era. The most important data to obtain, in the next decade if possible, would be chronological constraints on the early bombardment of the Moon. Bombardment by outer Solar System objects should carry compositional (mainly volatile) and isotopic signals.''}\cite{PSDec}. Consequently, measurements of elemental abundances in a variety of lunar craters would provide valuable insights.

The use of X-ray fluorescence to study planetary bodies began in the \textit{Apollo} era when global elemental and mineralogical datasets were acquired by the Apollo X-ray Fluorescence and Gamma Ray Experiment and were critical in refining theories of lunar evolution \cite{Adler1973,Adler1977,Andre1977,Swinyard2009}. The resulting Al/Si ratios were consistent with a plagioclase-rich highland and basaltic maria in returned lunar samples, confirming that the lunar highland crust is differentiated and of feldspathic composition.  

Since then, semiconductor X-ray detectors have flown to the Moon on a range of multi-agency spacecraft exploring the origins of many other regions on the lunar surface. These spacecraft included \textit{SMART-1\cite{Racca2002}, Kaguya \cite{Kato2010}} and \textit{Chang’e-1\cite{Ouyang2010}} and \textit{2}\cite{Zou2014}. Most recently, India’s \textit{Chandrayaan-1}\cite{Goswami2009} and \textit{-2} spacecraft each carried X-ray spectrometers (\textit{C1XS}\cite{Grande2009} and \textit{CLASS}\cite{Pillai2021}), which place soil sample-return and meteorite measurements into the global context of elemental abundances (Fig. \ref{Chandrayaan}). These show clear variation in the abundances of elements across the surface. 
Lunar X-ray fluorescence studies have also been conducted by X-ray telescopes in Earth orbit\cite{Narendranath2011}. \textit{Chandra} (Fig. \ref{ChandraLunar}), resolved fluorescence lines from O, Mg, Al, and Si (Fig. \ref{ChandraLunar}) within a relatively short series of observations (of order ~10s of minute) and showed elemental abundance variations between the highlands and mare through the \textit{High Resolution Camera (HRC)} instrument\cite{Wargelin2004}. Understanding such elemental abundances is not only key to understanding lunar evolution, but also for the upcoming possibility of a sustained human presence on the surface, e.g. in the context of \textit{Artemis} and \textit{the Lunar Gateway}. 

Beyond the moon, \textit{``The analysis of asteroids, meteorites, comets, and other rocky bodies of the inner Solar System can provide key insights into how the chemical inventories (e.g., organics and volatiles) of early Earth may have evolved''} (PS Decadal\cite{PSDec}). X-ray fluorescence has already played an important role in cataloging the elemental inventories of asteroids \cite{Trombka2000}. For comets, such fluorescence signatures have often been hidden by the dominant spectral lines from SWCX and the low spectral resolution of current instruments. However, even with the current low resolution there have been hints of fluorescence emissions that could be used for compositional studies \cite{Snios2014,Snios2018}.  

Consequently, while X-ray fluorescence measurements of comets would be a revolutionary tool for probing Solar System formation, it has proven largely illusive for the current generation of instruments,  \textit{LEM}  will surpass such challenges.

\subsubsection{Elemental and Molecular Composition of Rocky Planet Atmospheres}


Fluorescence is also a valuable tool to probe the atmospheric composition of planets. The strategic Research for Q1.1 and Q1.2 in the PS Decadal is to measure: \textit{``the elemental and isotopic composition of the surface, and, where relevant, atmospheres of bodies formed from different nebular reservoirs (especially Uranus, Neptune and Mercury, but also of Venus, asteroids, Centaurs and Saturn)\cite{PSDec}.}

\begin{figure}
	\centering
		\includegraphics[width=0.5\textwidth, trim={0cm 0cm 0cm 0cm},clip]{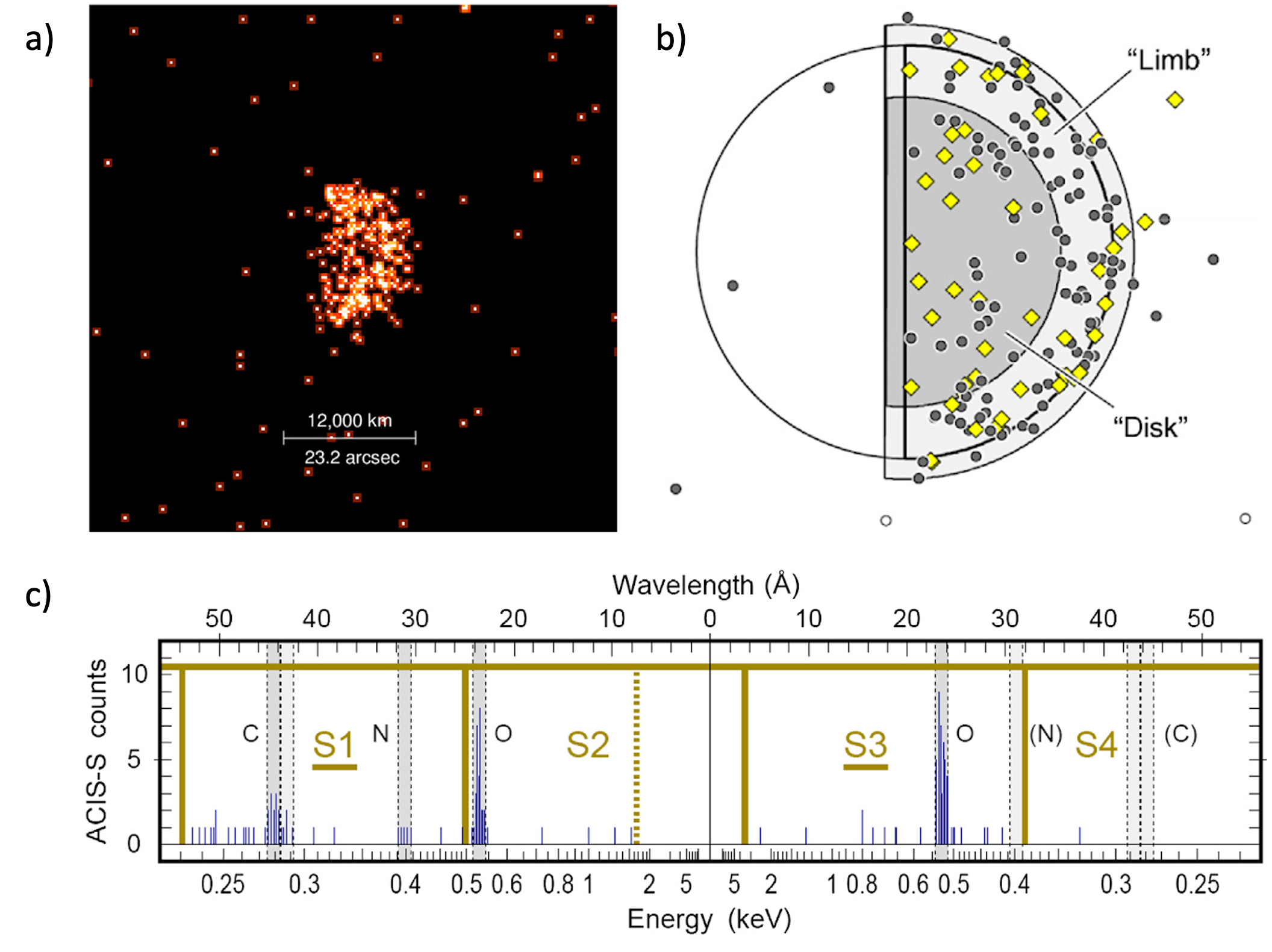}
			\caption{\textbf{Fluorescent X-ray emissions from Venus’s upper atmosphere:} a) \textit{Chandra ACIS-I} X-ray image of Venus on 13 January 2001, b) Location of individual photons relative to the disk and limb of the planet c) \textit{Chandra LETG} Spectral scan of Venus on the ACIS-S array. The observed C, N, and O fluorescent emission lines are enclosed by dashed lines; the width of these intervals matches the size of the Venus crescent during the observation (22.8''). Figures from Dennerl et al. (2002)\cite{Dennerl2002Venus}.}
	\label{VenusDennerl}
\end{figure}

\begin{figure}
	\centering
		\includegraphics[width=0.35\textwidth, trim={0cm 0cm 0cm 0cm},clip]{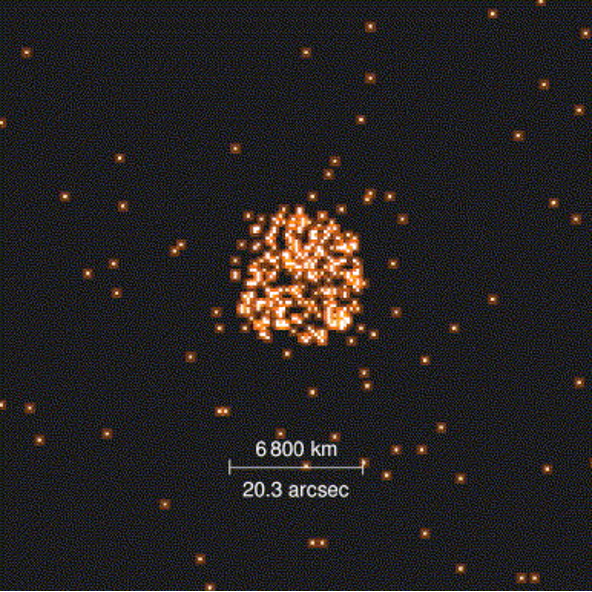}
			\caption{\textbf{Chandra ACIS-I X-ray image of Mars on 4 July 2001.} Figure from Dennerl, (2002)\cite{Dennerl2002Mars}.}
	\label{MarsDennerl}
\end{figure}

\textit{Chandra} and \textit{XMM-Newton}, have observed fluorescence signatures from the atmospheres of Venus and Mars (Fig. \ref{VenusDennerl} \& \ref{MarsDennerl}). For both planets, the observed emissions agreed well with models with a factor of $\sim$2 enhancement over predictions \cite{Dennerl2002Venus,Dennerl2002Mars,Cravens2001}. The Venus observations by the \textit{Chandra Low Energy Transmission Grating} showed carbon (0.28 keV) and oxygen (0.53 keV) k-$\alpha$ lines with traces of nitrogen k-$\alpha$ lines (Fig. \ref{VenusDennerl}). Interestingly, these high spectral resolution observations also hinted at the possibility of molecular identification as well through the 0.29 keV carbon s$\rightarrow\pi^*$ line in CO$_2$ \cite{Dennerl2002Venus}. Given sufficiently high sensitivity and spectral resolution, Fig. \ref{BrynFluorescenceCX} shows that it would be possible to distinguish molecular fluorescence lines from atomic lines, opening a new paradigm of X-ray molecular spectroscopy, and providing new molecular constraints for planetary objects.

The fluorescence emissions are clustered along the limb of the planets. For Venus they occur at an atmospheric height of 120-140 km above the surface, where the optical depth is minimized in comparison with the volume X-ray emissivity of the atmosphere (Fig. \ref{VenusDennerl})\cite{Dennerl2002Venus,Dennerl2008}. Consequently, X-ray fluorescence probes the composition of the upper atmospheres of planets, with the brightness of emissions depending sensitively on the density and chemical composition. With increasing depth into the atmosphere, the X-rays become increasingly  attenuated by the atmosphere.

X-ray measurements can also determine what is {\it not} in the upper atmosphere. Through comparison with the prevalence of dust storms at the time, Martian observations tested a connection between dust storms at the planetary surface and a resulting increase in the scattering of solar X-ray photons caused by increased dust in the upper atmosphere. No enhancement in scattered X-ray flux was detected during the global Martian dust storm\cite{Dennerl2002Mars}.

\begin{figure}
	\centering
		\includegraphics[width=0.5\textwidth, trim={0cm 0cm 0cm 0cm},clip]{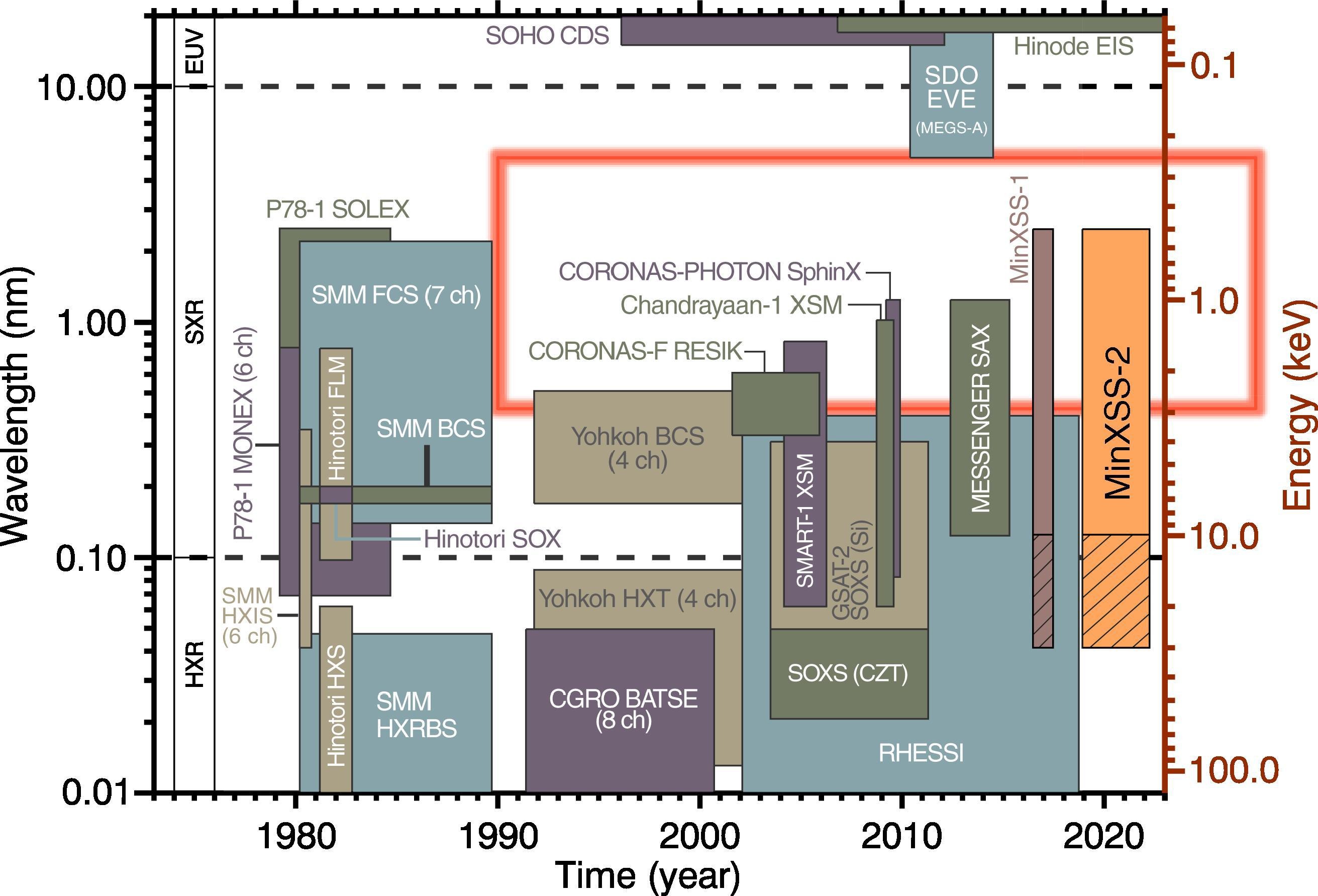}
			\caption{\textbf{Limited Coverage of the Solar Emission between 0.2-2 keV:} the energy/wavelength coverage and time of operation of spacecraft that have directly observed the Sun since 1980 (excluding broad bandpass instruments). Observations that have been acquired have typically been done as part of an instrument suite (e.g. Chandrayaan-1\cite{Goswami2009}) or as a dedicated instrument on a cubesat (e.g. MinXSS 1 and 2\cite{Mason2020}). Figure from Mason et al. (2020)\cite{Mason2020}.}
	\label{Mason}
\end{figure}

\subsubsection{Solar Composition Measurements}
Solar composition and variability is key to the heliophysics decadal (H-Q1).  Solar abundances are the benchmark against which many measurements of the Universe are compared and provide key insights into the gas cloud from which the Solar System formed : \textit{``The best representative of the molecular cloud composition is probably the Sun, which contains $>$99 \% of the mass of the present Solar System. Its composition is inferred from spectroscopic measurements, analyses of refractory elements in primitive meteorites, and direct sampling of the solar wind''} (PS Decadal). 

Many key spectral measurements of the Sun, that are used as reference points and ground truth measurements for other stars, occur between 0.2-2 keV. Fig. \ref{Mason} shows that relatively little coverage of the Sun has been obtained between 0.2-2 keV for the past three decades. The instruments that have taken dedicated solar measurements in this bandpass have offered relatively low spectral resolution and effective areas, at the cutting-edge of that which can be carried on cubesats or as part of an extended instrument suite. These solar emissions have never been explored at high resolution.

\begin{figure}
	\centering
		\includegraphics[width=0.48\textwidth, trim={0cm 0cm 0cm 0cm},clip]{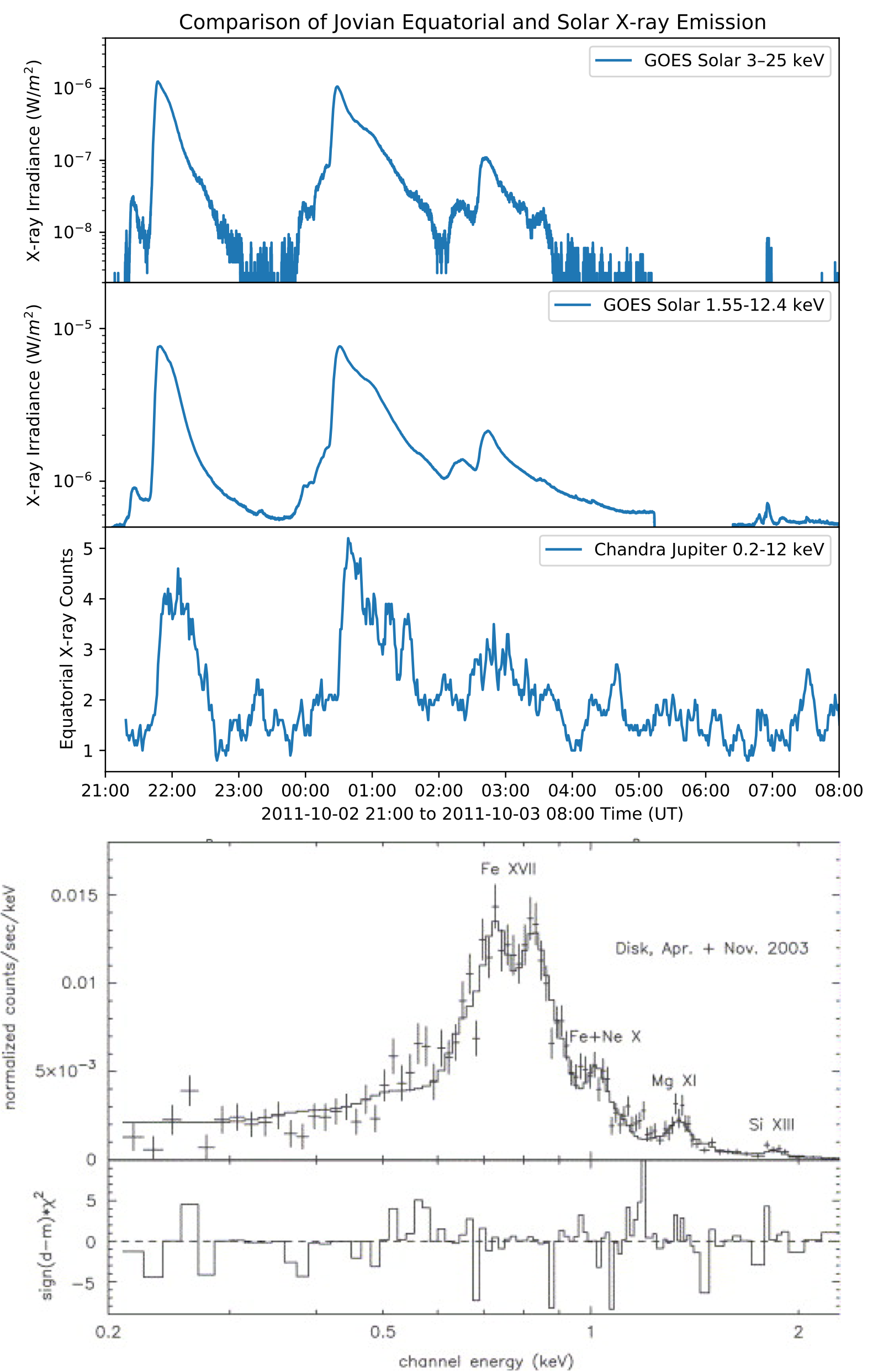}
			\caption{\textbf{Monitoring the Solar X-ray Spectrum and Variability through Scattering of Emission at the Gas Giant Equators:} Upper 3 panels compare \textit{GOES}  observations of the Solar X-ray irradiance with the 5-minute moving-mean X-ray photon counts from Jupiter’s equatorial region (third panel), as observed by  \textit{Chandra ACIS} in October 2011. The Jupiter observations have been shifted to account for the light travel time. The lower panels show combined \textit{XMM-Newton EPIC-pn} observations from April and November 2003 of Jupiter’s equatorial disk spectrum (crosses) overlaid with the best fit model (line). Labels on the spectrum indicate different spectral lines commonly observed from the Sun. The lower panel shows the model as the horizontal dashed line at zero, with contributions of the spectrum away from the model shown as the histograms above and below the line. Figures from Dunn, (2022a)\cite{Dunn2022Jupiter} and Branduardi-Raymont et al. (2007a) \cite{GBR2007Disk}.}
	\label{JupiterSolar}
\end{figure}

For the gas giants, Jupiter and Saturn, the atmospheres are dominated by hydrogen which elastically scatters solar photons so that the observed equatorial emissions are predominantly a solar-like spectrum, which varies with solar cycle and solar activity in a given observation \cite{Gladstone1998,Ness2004a,Ness2004b, Bhardwaj2005SolarJupiter, Bhardwaj2005SolarSaturn, GBR2007Disk, Dunn2016, Dunn2020a, Wibisono2023}. Fig. \ref{JupiterSolar} shows that the Jovian equatorial X-ray emission follows the solar X-ray emission in both time-series and spectrum, making the Jovian equator an excellent indirect monitor of solar activity. While X-ray observatories often cannot directly observe the Sun due to instrument safety, this offers indirect methods to utilise state-of-the-art X-ray observing capabilities to study our local star.

\subsubsection{Elemental and Molecular Composition of Gas and Ice Giant Atmospheres and Solar Composition}

Measurements of the elemental composition of the atmospheres of the Gas and Ice Giants are key drivers for the prevalence of atmospheric probes within the PS Decadal. Both the recommended flagship, \textit{Uranus Orbiter and Probe} mission, and the Decadal concept New Frontiers mission, the \textit{Saturn probe}, are motivated by the need to measure the composition of the atmospheres. However, to date, only the \textit{Galileo} probe \cite{Wong2004} has been used for the Gas Giants themselves and, at the time of writing, it is likely to be at least two decades until another such a probe measures in-situ the Gas/Ice Giant atmospheres. Until then, studies of the Gas/Ice Giant atmospheres, depend on remote observations. These remote observations provide invaluable local analogues for the rapidly-growing field of exoplanet atmosphere studies\cite{Tinetti2018}.

\begin{figure}
	\centering
		\includegraphics[width=0.4\textwidth, trim={0cm 0cm 0cm 0cm},clip]{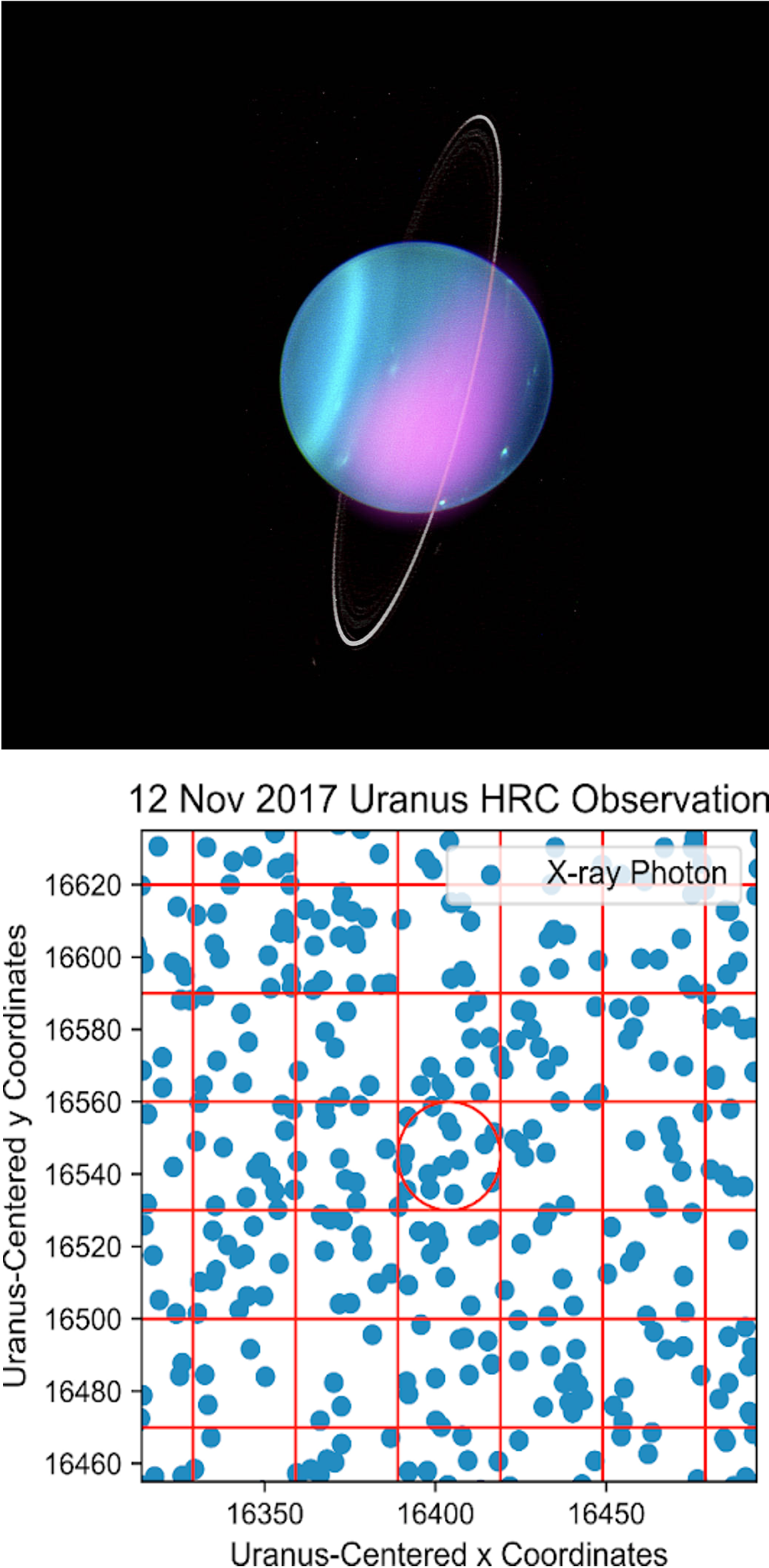}
			\caption{\textbf{The challenges with a high background and low sensitivity for current X-ray instrumentation observations of Uranus.} Upper: Aug 2002 \textit{Chandra ACIS} Observation (purple) overlaid on Keck IR image of Uranus. Here, the background could be filtered through the \textit{Chandra ACIS} energy resolution to enable scientific investigation of the 10-sigma signal. Lower: Uranus \textit{Chandra HRC} observation on 12 Nov 2017. The red circle indicates the region occupied by Uranus on the detector. Each detected X-ray photon is indicated by a blue dot. For \textit{Chandra HRC}, the background is too high vs the signal for a clear detection. \textit{LEM's} spectral resolution and sensitivity will be key for solving this problem. Image Credit: NASA/CXC/Keck/Dunn/Wolk and figure from Dunn et al. (2021)\cite{Dunn2021}.
}
	\label{Uranus}
\end{figure}

It is expected that $\sim$10\% of observed X-ray emission from the equators of Jupiter and Saturn is due to fluorescence\cite{Maurellis2000,Cravens2006}, but with current spectral resolution and effective area, it has been impossible to disentangle this from the scattered solar spectrum. For example, highly stripped OVII from solar emission produces spectral lines at 0.57 keV from He-like-$\alpha$ transitions, while neutral oxygen k-$\alpha$ fluorescence occurs at 0.53 keV. For spatially resolved observations with high sensitivity instruments like \textit{XMM-Newton’s EPIC-pn} and \textit{Chandra’s ACIS} instruments, these lines cannot be distinguished by the instrument spectral resolution. For Jupiter, there are also further complications from hints of spatially localised emissions that coincide with low surface magnetic field strength regions in which radiation belt particles would precipitate \cite{Waite1997,Bhardwaj2006,McEntee2022,Kollmann2021}.  Here, heightened spectral resolution would be invaluable for not only leveraging the X-ray emissions to explore the atmospheric constituents but to also explore the presence of direct radiation belt precipitation into the atmosphere and the processes that lead to this\cite{Kollmann2021,Dunn2023}.

\begin{figure*}
	\centering
		\includegraphics[width=0.9\textwidth, trim={0cm 0cm 0cm 0cm},clip]{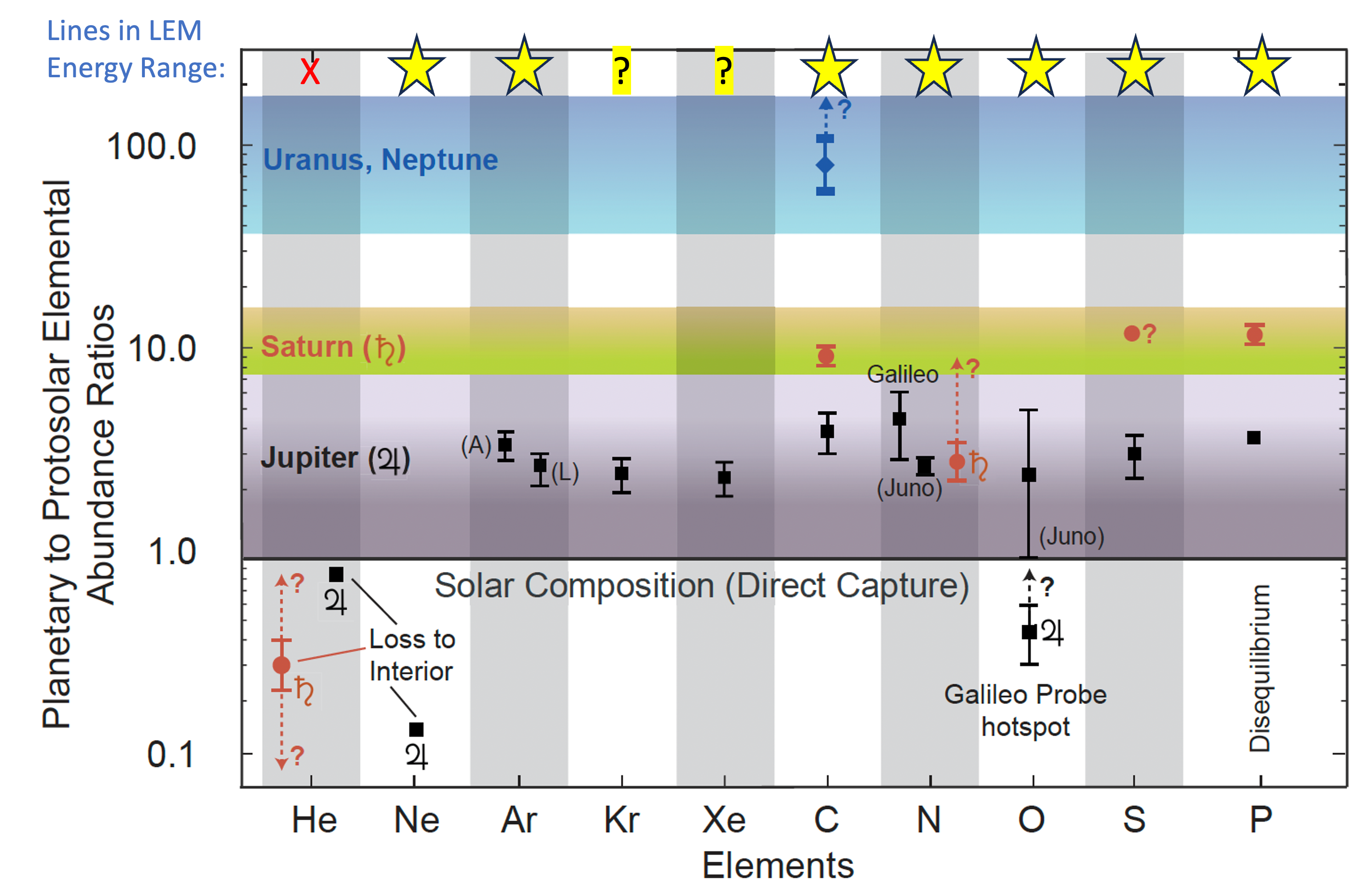}
			\caption{\textbf{Elemental abundance ratios in the outer planets.} The elemental abundances in the atmosphere of Saturn, with comparison to the Sun and the other giant planets. Values marked with a black `?' are either uncertain (N, S, He) or may not be representative of the actual ratio in the deep atmosphere. The original Figure and a detailed description of the measurement origins are in Atreya et al. (2022)\cite{Atreya2022} and reproduced in the PS Decadal\cite{PSDec}. To this figure we add an upper line to indicate which elements have X-ray fluorescence lines within the \textit{LEM} energy range. The red X indicates that there are not He lines. The yellow stars indicate that K or L-shell lines from these elements are within the \textit{LEM} energy range. The yellow '?' indicate that it is unlikely that the available lines are detected by LEM, because they are expected to be low probability transitions, but there are spectral lines within the LEM energy range.}
	\label{AtreyuAtmos}
\end{figure*}

Beyond Saturn, at the ‘Frozen Frontier’, exist the Ice Giants \cite{Fletcher2020}. The 2023 Planetary Decadal survey highlighted the Ice Giants, and in particular Uranus, as the top priority for the next NASA flagship mission\cite{PSDec}. The Ice Giants represent cornerstones in our understanding of planetary formation and evolution, and such planets appear to be the most common type of exoplanet in the Universe\cite{Wakeford2020}. The PS Decadal highlights the importance of compositional measurements: \textit{``compositional differences between the gas giants and ice giants may reflect their formation at different heliocentric distances with respect to condensation temperatures of ices containing oxygen, nitrogen, carbon, Noble gases, and possibly other elements such as sulfur.''}\cite{PSDec}. Fig. \ref{AtreyuAtmos} shows the lack of measurement of elemental abundances across Saturn, Uranus and Neptune and the value that \textit{LEM} can contribute to the field, with the measurements shown in Table 1. Here, X-ray measurements of fluorescence from the atmosphere will be invaluable. The reduced hydrogen abundance relative to the gas giants and the heightened presence of carbon, nitrogen and oxygen mean that the Ice Giants are likely to produce an atmospheric spectrum that is intermediate between Venus/Mars and Jupiter/Saturn.

Again, abundances of the Noble gases are key to distinguishing formation processes since they preserve the early Solar System record, free from modulation by 4.5 billion years of chemical processing\cite{Atreya2022}. This is highlighted in the PS decadal through question 7.1a which asks: \textit{``Are the Helium and Noble Gas Abundances Across the Giant Planets Consistent with Interior and Solar Evolution Models?''}\cite{PSDec}. Although we note that the abundances above the clouds (measured through X-ray fluorescence) may be very different to the lower layers of the atmosphere due to cloud condensation and chemistry. X-rays are increasingly attenuated with atmospheric depth so that it will not be possible to probe deeper into the atmosphere.

Prior to  \textit{LEM},  the challenges for Uranus observations have been a limited signal compared to the background. To date, there have been 3 Chandra X-ray observations of Uranus \cite{Dunn2021,Dunn2022Ice}. The first was taken in 2002, by \textit{Chandra-ACIS}, while the others (due to the decline in the \textit{ACIS} response\cite{Plucinsky2018}), utilised the \textit{Chandra High Resolution Camera (HRC)}, which provides no energy resolution. The relatively short ($\sim$few-hour) \textit{Chandra ACIS} observation from 2002 showed a 10-sigma detection of Uranus, and leveraged the instrument energy resolution to reduce the background for this detection. Unfortunately, the subsequent \textit{Chandra HRC-I} observations did not possess the energy resolution necessary to filter the background and consequently detections of planetary emission were within 2-sigma of the background. Figure \ref{Uranus} shows an example of the X-ray signal from a few-hour \textit{Chandra HRC} observation taken in November 2017. This observation showed intriguing hints of variability that might have been consistent with planetary aurorae\cite{Dunn2021} or ring fluorescence from energetic electron collisions from the Uranian radiation belts \cite{MaukFox2010}. Unfortunately, interpretation was inhibited by the lack of energy resolution and the high background. This problem diminishes for an instrument with heightened spectral resolution and a similar effective area, like \textit{LEM} . 

Since \textit{ROSAT}, no \textit{Chandra} or \textit{XMM-Newton} observations of Neptune\cite{Ness2000} have been conducted, but the time required for a detection of solar-driven atmospheric emission with current instrumentation is likely to be at least megaseconds \cite{Dunn2022Ice}, unless, like Uranus, Neptune shows enhancements in signal from e.g. auroral emissions.

Consequently, for Uranus, X-ray observations that provide higher sensitivity and can filter the background more effectively will provide a powerful new tool for Ice Giant exploration.

\subsubsection{Elemental and Molecular Composition of Icy Satellites and Rings}

\begin{figure}
	\centering
		\includegraphics[width=0.48\textwidth, trim={0cm 0cm 0cm 0cm},clip]{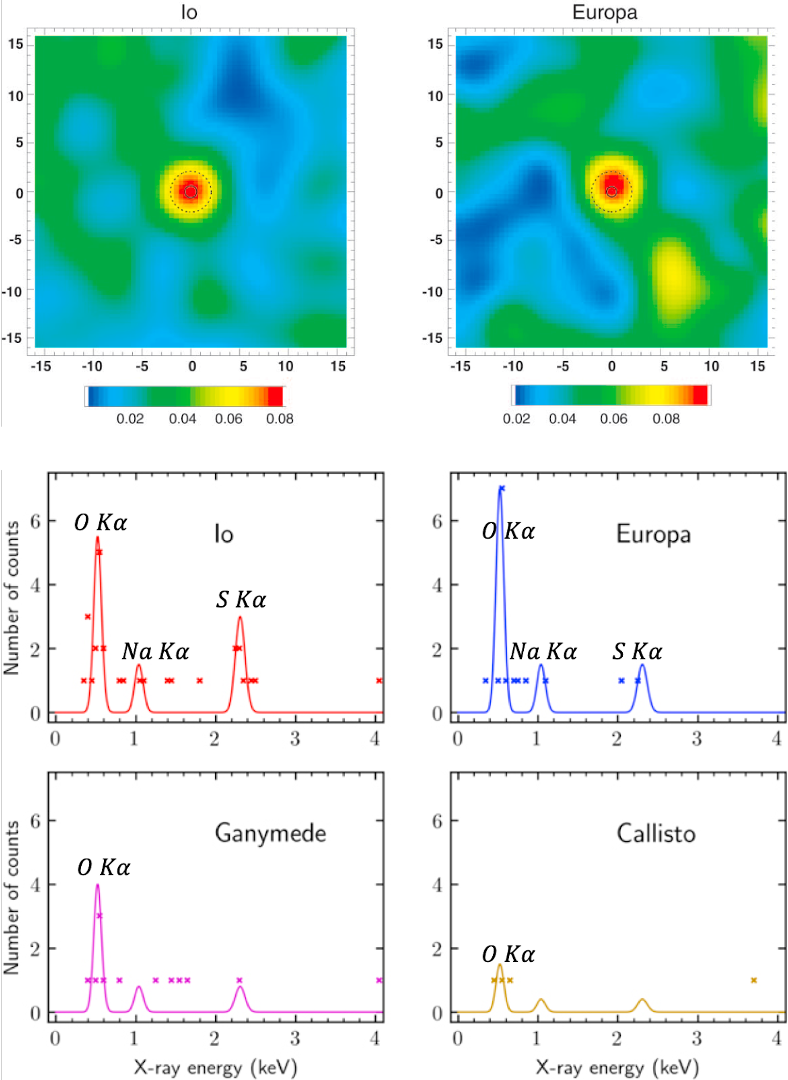}
			\caption{\textbf{\textit{Chandra ACIS} X-ray Observations of Jupiter’s Icy Moons Io, Europa, Ganymede and Callisto}: Top 2 panels: \textit{Chandra ACIS} images of Io and Europa (250eV < E < 2000eV), smoothed by a 2D Gaussian. The axes are in arcsec (100’ = 2995 km), and the scale bar is in units of smoothed counts per image pixel. The solid circle shows the size of the satellite. Lower 4 panels: X-ray spectra of each of the Galilean moons (data points), using 50 eV energy bins, overlaid with Gaussians at the positions of the O K-$\alpha$ (0.525 keV), Na K-$\alpha$ (1.041 keV), and S K-$\alpha$(2.308 keV) lines. Figures are from Elsner et al. (2002)\cite{Elsner2002} and Nulsen et al. (2020)\cite{Nulsen2020}.
}
	\label{IoEuropa}
\end{figure}

While the Ice Giants will be the focus of \textit{NASA’s} flagship planetary science mission through the 2040s, the icy moons of the outer planets are to be the primary focus of \textit{ESA} and \textit{NASA} through the 2030s,  at the start of \textit{LEM’s} lifetime, and before \textit{ATHENA} becomes operational. 

The Galilean satellites play a key role in the PS decadal and ESA Voyage 2050 strategy through both their role as a tracer of Solar System formation and through the questions of their potential habitability. Consequently, \textit{ESA} and \textit{NASA} have both committed flagship missions to the Jovian moons in the form of \textit{JUICE} (an \textit{ESA} L-Class mission) and \textit{Europa Clipper} (a \textit{NASA} Planetary Flagship), which are each due to arrive in the early 2030s (Fig. \ref{OtherSpacecraft}). For the icy moons, understanding the elemental and chemical abundances and how they interact with Jupiter’s extreme radiation environment will be critical to understanding the formation and astrobiological potential of these worlds. The moons offer cornerstones into the circumplanetary disks in which Jupiter formed\cite{PSDec}.

The extended fields of view of \textit{Chandra} and \textit{XMM-Newton} (and \textit{LEM}) mean that the orbits of the Galilean satellites are within the field for most Jupiter observations. The discovery of the X-ray emissions from Io and Europa occurred with the very first \textit{Chandra} observations of Jupiter\cite{Elsner2002}, while a tentative detection of emission from Ganymede and hints of emission from Callisto required combining several \textit{Chandra ACIS} observations\cite{Nulsen2020}. Figure \ref{IoEuropa} shows X-ray images and low-signal spectra from the moons, achieved at the limits of current instrumentation. Here, challenges with signal and background again arise that make it difficult to do characterisation studies of the elemental composition with the current signal. Although the signal does demonstrate proof of concept by showing peaks at the oxygen fluorescence lines expected for a water ice moon like Europa\cite{Nulsen2020}. \textit{LEM} will provide order of magnitude enhancements in signal over background.
 
Beyond Jupiter, the moons of the other outer planets are yet to be observed in the X-ray waveband, although Saturn’s moon Enceladus does reside in radiation belts that lead to energetic particle impacts with the surface of the moon \cite{Roussos2021}.

\begin{figure}
	\centering
		\includegraphics[width=0.48\textwidth, trim={0cm 0cm 0cm 0cm},clip]{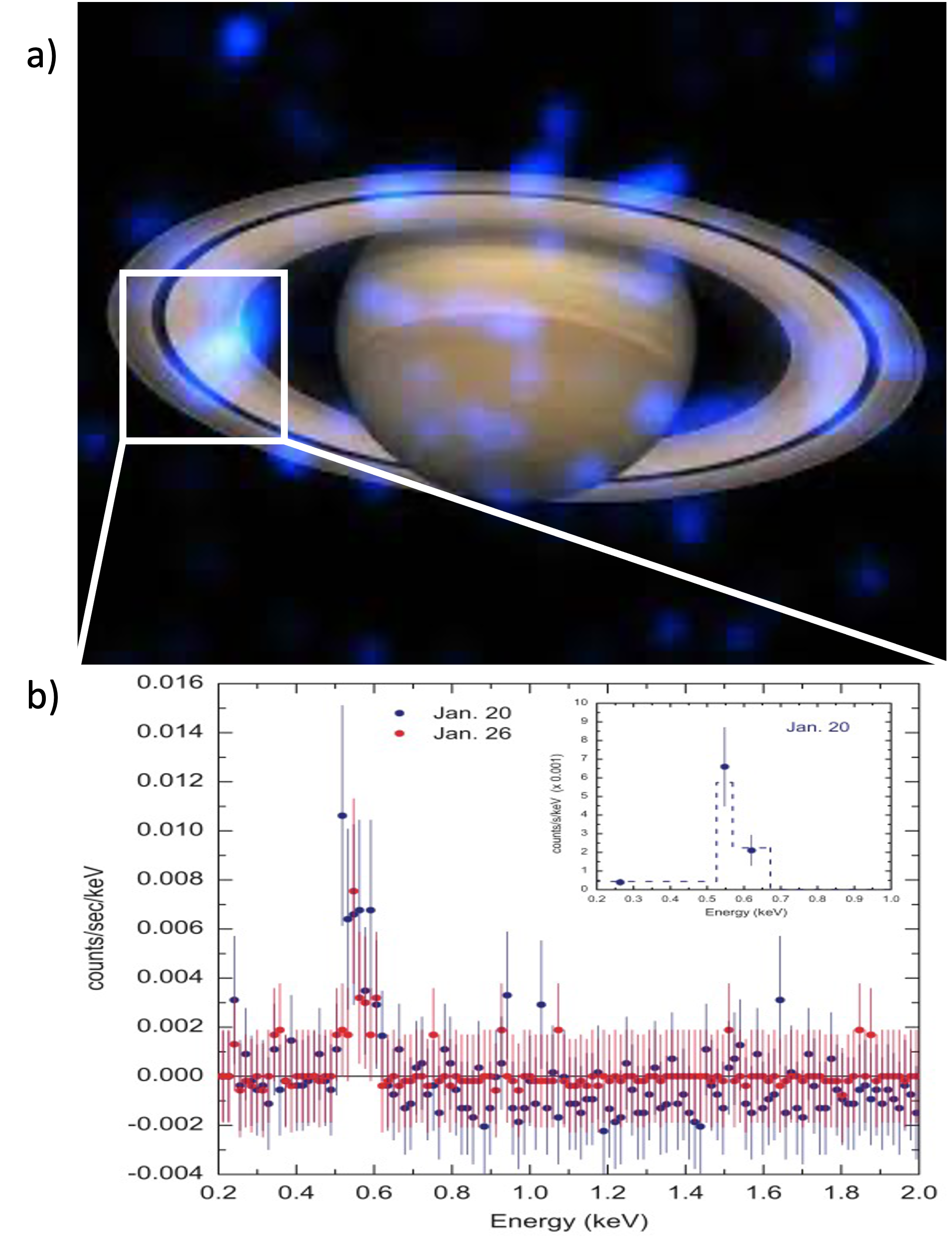}
			\caption{\textbf{Remote X-ray Identification of the Elemental Composition of Saturn's Rings:} Top panel: X-ray Image of Saturn (blue) overlaid on an optical image. Lower Panel: Background-subtracted X-ray spectrum from Saturn's rings as observed by Chandra ACIS-S3, showing a prominent O k-$\alpha$ line, with the inset showing a gaussian model fit to this line (Bhardwaj et al. 2005)\cite{Bhardwaj2005Ring}. Image Credit: NASA/MSFC/CXC/A.Bhardwaj et al.; Optical: NASA/ESA/STScI/AURA.}
	\label{SaturnsRings}
\end{figure}

As well as a key plasma source for the Saturnian system, Enceladus also contributes to the creation of Saturn’s most iconic attribute: its rings. Planetary rings produce fluorescence emissions\cite{Bhardwaj2005Ring}, which can provide a useful probe of the rings composition and formation (Fig. \ref{SaturnsRings}). These emissions may be caused by solar photons or through direct interaction of the planetary radiation belts with the rings. Q2.3 in the PS Decadal highlights the importance of understanding the formation processes that lead to rings and moons: \textit{``Indeed, so common are satellites (and to an extent, rings) about the larger bodies of the outer Solar System, that the question might be not so much why they have rings and moons, but why the terrestrial planets are deficient in satellites and lack rings entirely.''}\cite{PSDec}. Here, studies of the elemental composition and formation of planetary rings and moons is critical to understanding their formation. With sufficient signal to noise, provided by a \textit{LEM}-like instrument, X-rays can play an important role in remotely addressing these questions.

\subsection{What \textit{LEM} will do for Identifying the Building Blocks of the Solar System}

\textit{LEM} will provide unprecedented elemental composition insights across the Solar System, building inventories of the key elements highlighted in the PS Decadal\cite{PSDec}. Table 1 shows a non-exhaustive list of elements that \textit{LEM} will be able to identify across comets, Venus, Mars, the Moon, and the moons, rings and atmospheres of Jupiter, Saturn and Uranus. Many of these measurements will be new insights and enable new understanding of Solar System planet, moon and ring formation that directly supports PS Decadal questions and exoplanetary science (PS Decadal Q12.7), where unified theories of planetary formation drive the current development in the field.

\begin{table}\centering

\begin{tabular}{ccc}\hline
Element & Line Energy (eV) & Shell \\
\hline
C & 277 & K-$\alpha$ \\
N & 392 & K-$\alpha$ \\
O & 525 & K-$\alpha$ \\
Ne & 849 & K-$\alpha$ \\
Na & 1041 & K-$\alpha$ \\
Mg & 1254 & K-$\alpha$ \\
Al & 1487 & K-$\alpha$ \\
Si & 1740 & K-$\alpha$ \\
P & 2014 & K-$\alpha$ \\
P & 48-189 & L \\
S & 53-164 & L \\
Cl & 50-270 & L \\
Ar & 220-326 & L \\
K & 259-381 & L \\
Ca & 302-444 & L \\
Fe & 705-850 & L \\
Xe & 600-1000 & M \\ 
\hline
\label{Table:ElementalFluorescence}
\end{tabular}
\caption{A list of example elemental fluorescence lines that are valuable for planetary formation and habitability and that \textit{LEM} will have sufficient spectral resolution and energy coverage to identify. We note that detection of P and S will depend on the extent of \textit{LEM}'s soft and hard energy response (see section 1).}

\end{table}

\subsubsection{Elemental Composition of the Surfaces of Rocky Bodies}
Fig. \ref{Chandrayaan} demonstrates the state-of-the-art for orbiting spacecraft measurements of lunar elemental abundances. While these observations provide excellent spatial resolution (of order 10-100 km\cite{Pillai2021}), they are limited to relatively thin tracks across the Lunar surface that are typical of all orbiting X-ray measurements of the moon. Given that these instruments are part of an extensive instrument suite, they need to be lighter than a dedicated X-ray observatory and therefore have relatively low sensitivity with effective areas of $\sim$4-100 cm$^2$, and CCD-like spectral resolution of order $\sim$100 eV \cite{Pillai2021}.

Where orbiters typically measure the lunar composition within a small track, \textit{LEM’s} field of view and spatial resolution provides global maps with $\sim$70 km resolution. This will enable \textit{LEM} to probe $\sim$100 lunar craters to explore their different elemental abundances to distinguish bombardment and Solar System formation theories (e.g. PS-Q2.4) and to support local knowledge for \textit{Artemis} and the \textit{Lunar Gateway}. The effective area and spectral resolution provides clear identification of the Mg, Al and Si lines that have historically been blended\cite{Swinyard2009} alongside enhanced sensitivity for previously inaccessible key trace elements defined in the PS Decadal\cite{PSDec} and outlined in Table 1. For example, through  measurements of the Argon L-shell fluorescence lines (220-326 eV - Table 1), \textit{LEM} can further distinguish variation in the lunar atmosphere from the surface.

In the coming years, the \textit{ESA-JAXA BepiColombo} spacecraft will settle into orbit at Mercury utilising the \textit{MIXS} lobster eye X-ray instrument to provide high resolution surface maps of the elemental abundances\cite{Bunce2020}. These offer comparative global maps of elemental composition between Mercury and the Moon, a key component for different Solar System formation theories.

Comets also offer detailed information on the early-stage chemical and dust compositions of the Solar System and the transport of such chemicals during the Solar System's evolution. PS-Q3.5 strategic research highlights the importance of telescopic measurements of comet composition. While coherent scattering of solar X-rays by comet dust and ice particles have been found to contribute significantly to the total emission intensity at energies greater than 1 keV\cite{Snios2014,Snios2018}, fluorescence lines have rarely been usable to provide a tracer of the comet’s composition, because current X-ray telescopes have insufficient spectral resolution. LEM’s spectral resolution will enable these needed studies of fluorescence lines,  distinguishing them from the dominant SWCX lines of cometary emission (see e.g. Fig. \ref{BrynFluorescenceCX} and section 3), providing detailed elemental catalogues of comets from various regions across the Solar System for the first time.

\subsubsection{Elemental and Molecular Composition of Planetary Atmospheres, Rings and Moons}
PS-Q3.5 states that \textit{``Measuring Noble gas concentrations in the Venusian atmosphere would provide an important constraint on degassing from the interior''}. Fig. \ref{AtreyuAtmos} further demonstrates the lack of elemental measurements for Saturn and Uranus as well. For the atmospheres of Venus, Mars, Jupiter, Saturn and Uranus, LEM’s spectral resolution will distinguish carbon (0.28 keV) K-$\alpha$ lines from possible 0.29 keV carbon dioxide s$\rightarrow\pi^*$ lines, potentially enabling the observatory to be used to identify elemental and molecular inventories. Table 2 shows a small sample of spectral lines that might be discernible by \textit{LEM}. Although, we caution that the ability to unambiguously identify these lines will be dependent on line broadening from the source and it may (not) be possible to distinguish e.g. CH$_4$ from CO$_2$. At a minimum this will provide elemental budgets for different planetary bodies.

For Uranus, a $\sim$500 ks \textit{LEM} observation will identify fluorescence and scattered solar emissions from the planet as well as potentially characterising auroral emissions (see section 3). 

\begin{table}\centering

\begin{tabular}{cccc} \hline
Molecule & Element &  Energy (eV) & Transition \\ 
\hline

CO$_2$ & C & 291 \cite{Mclaren1987} & 1s$\rightarrow\pi^*$ \\
CO$_2$ & O & 535 \cite{Hitchcock1987} & 1s$\rightarrow\pi^*$ \\
CO & O & 534 \cite{Sohdi1984} & 1s$\rightarrow\pi^*$ \\
CO & C & 287\cite{Sohdi1984} & 1s$\rightarrow\pi^*$\\
O$_2$ & O & 531\cite{Nordgren1997} & 1s$\rightarrow\pi^*$ \\
CH$_4$ & C & 288\cite{Hitchcock1987} & 1s$\rightarrow$3p \\
CH$_4$ & C-H & 301 \cite{Hitchcock1987} & 1s$\rightarrow\sigma^*$ \\
C$_2$H$_4$ & C=C & 285\cite{Mclaren1987} & 1s$\rightarrow\pi^*$ \\
NH$_3$ & N & 402 \cite{Schirmer1993} & 1s$\rightarrow$3p \\
H$_2$O & O & 535\cite{Schirmer1993} & 1s$\rightarrow$3p\\ 
\hline
\label{Table:MolecularFluorescence}
\end{tabular}
\caption{A sample of molecular X-ray fluorescence lines that exist within the \textit{LEM} energy range}
\end{table}

The combined Uranian ring and planetary system will be a point source for \textit{LEM’s} 15” resolution, so that careful consideration will be needed to disentangle the signature from the rings (hinted at in the \textit{Chandra ACIS} observations\cite{Dunn2021}) from that of the planet. However, Saturn’s rings will occupy pixels either side of the planet so that \textit{LEM’s} observations will go orders of magnitude deeper than \textit{Chandra ACIS} observations, reducing the background and enabling a richer elemental inventory for the rings. Through comparisons with solar activity, it may be possible to distinguish the source process for this emission, distinguishing direct impact of radiation belt particles from solar photon driven fluorescence.

For the Galilean satellites, \textit{LEM} measurements will break model degeneracies to support IR mineralogy by the \textit{JUICE} and \textit{Europa Clipper} spacecraft (Fig. 1), distinguishing between astrobiologically important salts with similar IR spectral profiles. For example,  NaCl and KCl salts have similar IR spectral profiles and can be difficult to distinguish with IR spectra\cite{Brown2013}. Measurements of the relative Na and K abundances by \textit{LEM} will help to break these degeneracies and in turn provide critical insights into the salts that make up Europa’s oceans. LEM’s spectral resolution, low background and heightened sensitivity are key for this, providing order of magnitude enhancements in signal over the observations shown in Figure \ref{IoEuropa} with a similar observing time (a few 100 ks) and helping to identify trace elemental species shown in Table 1.

\textit{LEM} is opportunely timed for these studies, perfectly overlapping the operations of \textit{JUICE} and \textit{Europa Clipper} (Fig. \ref{OtherSpacecraft}). This offers unique synergies between in-situ measurements and global remote monitoring  by \textit{LEM}. For example, X-ray fluorescence detections combined with in situ spacecraft particle flux measurements (e.g. from \textit{JUICE} and \textit{Europa Clipper}) will enable quantification of particles impacts with the surface, validating surface evolution and radiolysis models that are key to understanding the formation, evolution and subsurface oceans of the moons. Q6.5 in the PS Decadal highlights the importance of characterising the impact of particles on the surfaces of moons and \textit{LEM} will remotely enable observations of these impacts.  \textit{ATHENA} would launch after \textit{JUICE} and \textit{Europa Clipper} finish and therefore miss these synergistic opportunities. 

\subsubsection{Solar Composition with LEM}

\textit{LEM} will be able to explore solar composition in two ways: 1. Through indirect observations of the disk-integrated solar spectrum and time-series via Thomson scattering from planetary atmospheres and 2. through unprecedented access to the solar ion abundances via SWCX. We discuss each in turn.

Firstly, while unable to directly observe the Sun, Fig. \ref{JupiterSolar} shows that the solar spectrum and its variability (e.g. flares\cite{GBR2007Disk,Dunn2022Jupiter} and solar cycle \cite{McEntee2022,Wibisono2023}) becomes imprinted on the equatorial emissions from Jupiter and Saturn and constitutes $\sim$90\% of this emission\cite{Maurellis2000}, so \textit{LEM} will enable indirect observations of the Sun. \textit{LEM} observations of these emissions will enable temporal and spectral properties of the disk-integrated solar emission to be explored with never-before-seen spectral resolution. It will then be possible to study the broadening and Doppler shifting of solar spectral lines during intervals of complex solar activity, such as the release of a coronal mass ejection and strong flaring, providing a valuable benchmark for extending to stellar activity and exoplanetary systems. These measurements will occur in the critical 0.2-2 keV range that is essential for comparison with X-ray studies of other stars, but is poorly covered for the Sun by instruments over the last few decades (Fig. \ref{Mason}).

\begin{figure*}[h]
	\centering
		\includegraphics[width=0.7\textwidth, trim={0cm 0cm 0cm 0cm},clip]{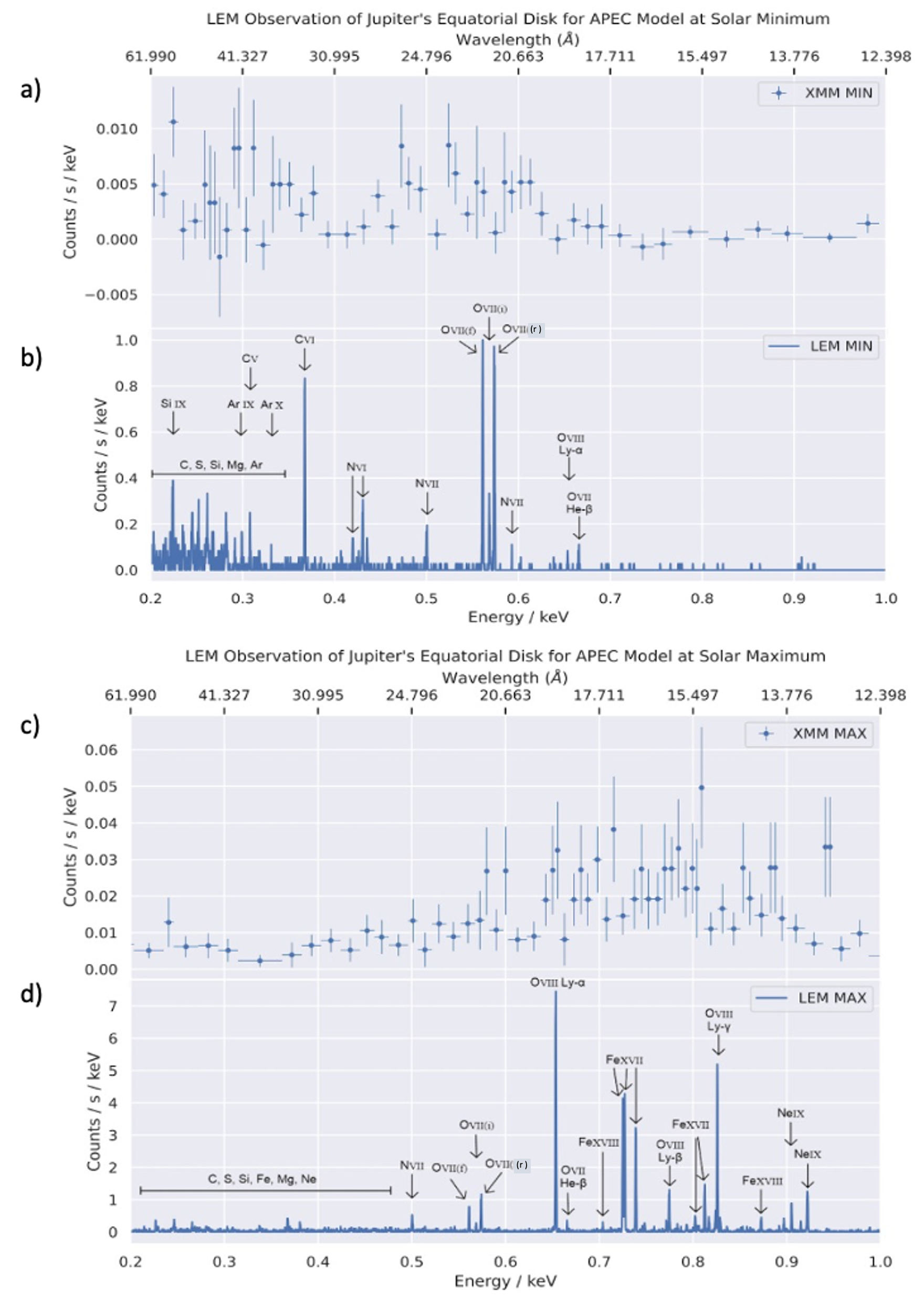}
			\caption{\textbf{Indirect \textit{LEM} Observations of the Solar Spectrum at Solar Minimum and Maximum:} \textit{XMM-Newton EPIC-pn} observations of the Solar X-ray emission Thomson Scattered from hydrogen in Jupiter’s Equatorial atmosphere during Solar Minimum (8 April 2020) and Maximum (15 April 2014) (panels a and c). These observations are best-fit with APEC\cite{Smith2001} models representing the solar corona which are then shown as would be observed by \textit{LEM} (panels b and d). Note: that these models do not include atmospheric fluorescence lines, which are expected to account for ~10\% of the observed emission \cite{Maurellis2000}, but have never previously been distinguished from the scattered solar emission due to limited spectral resolution and sensitivity.  \textit{LEM}  will also reveal the extent and presence of these fluorescence lines for the first time. The solar lines will also be broadened and shifted by solar events, that in turn help to identify the nature of the energetic processes causing them, so that the resulting spectra will be rich with time-variable solar diagnostics. Figures from \cite{Parmar2023}.
}
	\label{LEMAPEC}
\end{figure*}

Figure \ref{LEMAPEC} shows the Jovian equatorial spectrum as observed by \textit{XMM-Newton EPIC-pn} and  \textit{LEM}  at solar minimum and solar maximum. At these times the \textit{XMM} spectra are best fit with solar corona models that replicate the elastically-scattered solar spectrum\cite{Dunn2020a}.   \textit{LEM} will pinpoint and characterise solar emission lines with unprecedented resolution and measure their variability with time. This study, in turn, will provide invaluable ground-truths for observations of stellar populations and the activity cycles of exoplanet-hosting stars.  

\begin{figure*}[h]
	\centering
		\includegraphics[width=0.9\textwidth, trim={0cm 0cm 0cm 0cm},clip]{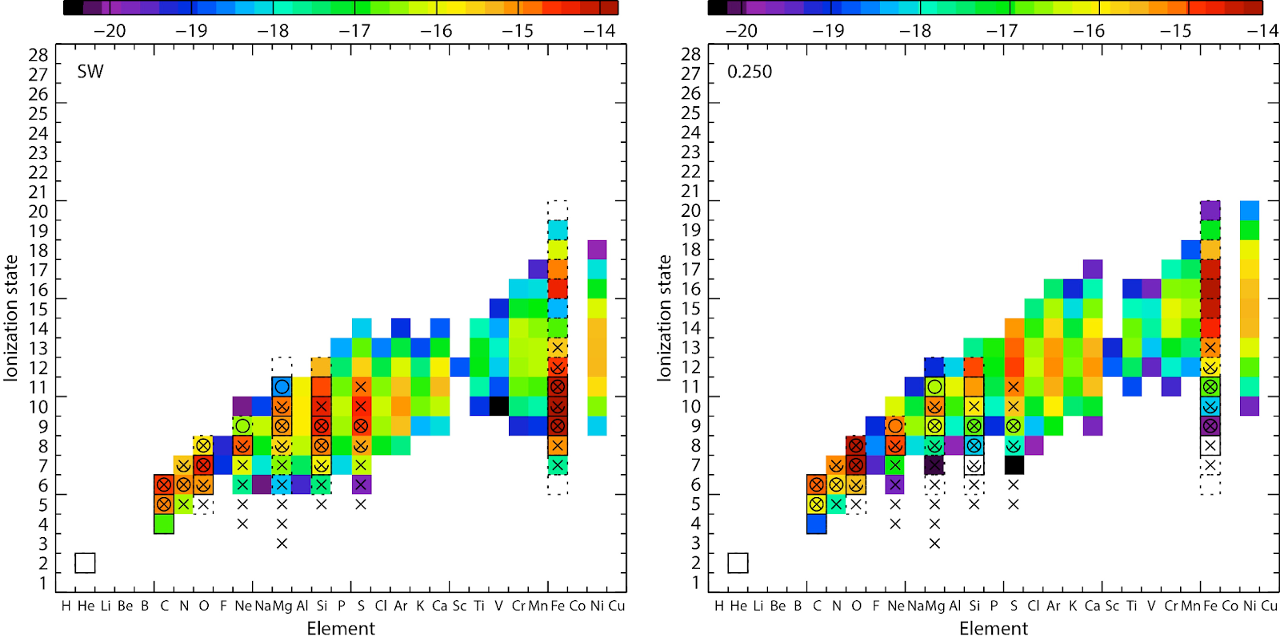}
			\caption{\textbf{\textit{LEM} observations of the solar wind ion abundances in comparison with the species observed in-situ:} Colour bars show the relative X-ray emission strengths of different charge states of different species in a slow solar wind. The authors applied a linear fit to the freeze-in temperature as a function of atomic number\cite{Gloeckler2007}, used the freeze-in temperatures of a species to approximate the relative charge state populations, and then used the X-ray emissivities of those charge states within the \textit{LEM} energy band in order to approximate the relative X-ray brightness of each species/charge state. These values were extracted from the APEC\cite{Smith2001} collisional ionization equilibrium model. The results should be viewed as merely illustrative. The boxes indicate ions measured by \textit{ACE} while the "×" indicate ions measured by \textit{Ulysses}\cite{Schwadron2000}. Dashed boxes are species that are measured, but not well measured by ACE. The full circles indicate ions directly measurable by \textit{LEM} as described in the text (Section 6.2.1) while the half circles are ions that should be measurable by \textit{LEM} by spectral fitting. Note that we have not included most odd ions in our \textit{LEM} simulation because they have not been measured in the solar wind by in situ experiments. Right: The relative X-ray emission strengths of different charge states of different species in an astrophysical plasma with = 0.25 keV. These were extracted from the APEC collisional ionization equilibrium model. The color code is the emissivity in the log of photons. Figure and caption from Kuntz et al. (2023)\cite{Kuntz2023}.
}
	\label{LEMSWIons}
\end{figure*}

Secondly, the process of Solar Wind Charge Exchange, in which a highly charged ion takes an electron from a neutral to produce a characteristic line, is a common source of X-rays throughout the heliosphere (discussed in detail in the next section).  \textit{LEM} 's spectral resolution will provide unprecedented access to the ion species in the solar wind. Kuntz et al. (2023)\cite{Kuntz2023} compare the current measurements of solar wind ion abundances from in-situ spacecraft with the species that \textit{LEM} will detect (Fig. \ref{LEMSWIons}) to show that  \textit{LEM}  will uniquely be able to measure and constrain solar wind ion abundances for a variety of species that have not been previously measured. 

Here, we have discussed using SWCX to probe the solar wind composition; the next section details how this same emission process can be used to study the impact of space weather on a variety of planetary bodies to better understand the nature of the relationship between stars and planets.

\section{Worlds and Suns in Context: How Does the Nature of the Interaction with Stars vary for Different Planetary Bodies? }

Key for both the National Academies’ PS and Heliophysics Decadals is an understanding of what fundamental processes govern the interaction between a planetary body and its surrounding space environment (e.g. Heliophysics Decadal goals 2 and 4; PS Decadal questions: 6 and 8).  \textit{LEM} will provide a unique step-change in this understanding. 

Within our own Solar System, this interaction varies dramatically from an interaction governed by a cycle of reconnection between the solar wind and magnetosphere at Earth\cite{Dungey1961}; to a direct solar wind-atmosphere interaction for Venus\cite{Futaana2017}, Mars\cite{Jakosky2001}, comets and asteroids\cite{Biermann1967}; to complex solar wind interactions between the tilted, offset magnetic fields of Ice Giants\cite{Masters2018,Fletcher2020}; to the extreme cases of internally-governed, rapidly-rotating systems with internal plasma sources such as the Jovian system\cite{Cowley2008,McComas2007}. All of these solar wind interactions happen in the context of the heliosphere as our local astrosphere, and provide essential accessible analogues for an array of astrophysical magnetospheres and astrospheres across the Universe that cannot be visited in-situ or spatially resolved (e.g.,  brown dwarfs, exoplanets and magnetars).  For all of these bodies and structures, charge exchange (CX) provides a large cross section, which offers a sensitive measurement of the interface between hot plasma and neutrals\cite{Cravens1997,Dennerl2010}. CX is the dominant source of X-rays for Solar System planets. \textit{LEM} will provide a paradigm-shift in our capacity to utilise CX to understand the processes that govern the interaction between the solar wind and planetary bodies. For X-ray astronomy generally, this may be the key to solving the 50-year mystery of the X-ray Sky background. 

\subsection{What we know}

\subsubsection{Direct Solar Wind Interactions with the Atmospheres of Venus and Mars}

Atmospheric loss is a critical component of habitability for Earth-like exoplanets and key in the PS decadal. Q6.5 in the PS decadal asks: \textit{``What processes govern atmospheric loss to space?''} and goes on to question \textit{``How Do Atmospheric Dynamics, Such As Martian Dust Storms, Affect the Escape of Gases from Solid Planets and Satellites?''} (Q6.5b). Here, the partnership of global X-ray observations of SWCX with in-situ measurements of planetary conditions is transformative.
 
Within our own Solar System, Venus and Mars provide archetypal rocky planets for which the direct interaction of the solar wind with the planet has substantially altered the conditions and habitability. SWCX provides a particularly sensitive technique that enables direct imaging of global atmospheric stripping for Mars and Venus \cite{Dennerl2002Mars,Dennerl2002Venus}. This offers a valuable global and contextual complement to in-situ single-point measurements by \textit{Maven}\cite{Jakosky2015} and possibly \textit{M-MATISSE} \cite{Sanchez-Cano2022}, and in the near future the fleet of Venus missions: \textit{Veritas, EnVision and DAVINCI+} \cite{Smrekar2022,Widemann2020,Garvin2020}.

\begin{figure}
	\centering
		\includegraphics[width=0.4\textwidth, trim={0cm 0cm 0cm 0cm},clip]{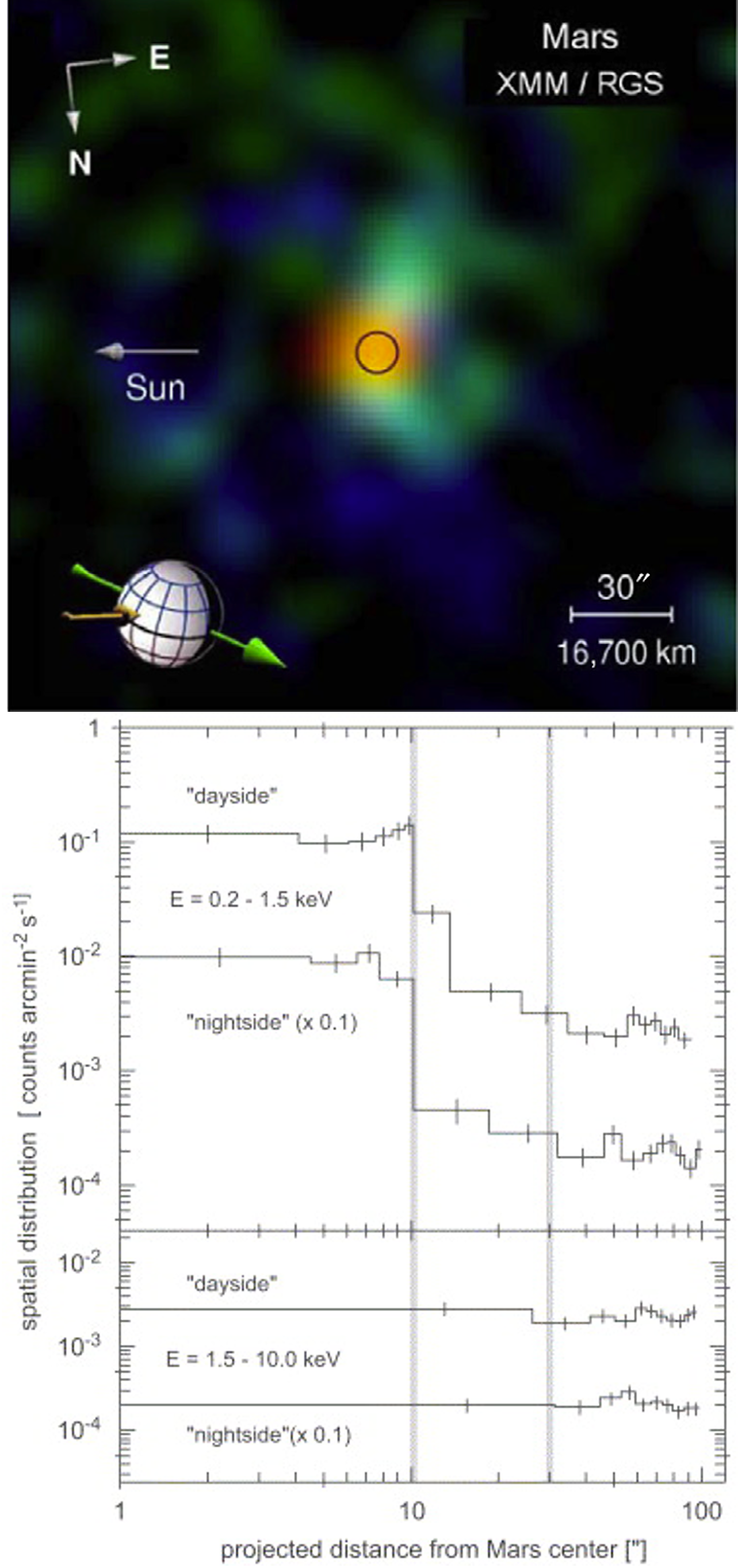}
			\caption{\textbf{Direct X-ray Observations of Atmospheric Loss from Mars}: Upper: XMM-Newton observation of Mars: O$^{6+}$, O$^{7+}$ CX lines in blue, C$^{4+}$, C$^{5+}$ CX lines in green, and fluorescent lines in yellow. Lower: Spatial distribution of the photons around Mars in the soft (E=0.2–1.5 keV) and hard (E=1.5–10.0 keV) energy range, in terms of surface brightness along radial rings around Mars, separately for the “dayside” (offset along projected solar direction >0) and the “nightside” (offset <0); note, however, that the phase angle was only 18.2°. For better clarity the nightside histograms were shifted by one decade downward. The bin size was adaptively determined so that each bin contains at least 28 counts. The thick vertical lines enclose the region between one and three Mars radii. Figures and caption contents from Dennerl et al. (2002,2006)\cite{Dennerl2002Mars,Dennerl2006}.}
	\label{MarsAtmosphereLoss}
\end{figure}

The upper panel of Fig. \ref{MarsAtmosphereLoss} shows observations by XMM-Newton of the halo of X-ray emission that exists around Mars to distances of a few planetary radii, in which solar wind ions charge exchange with the martian upper atmosphere leading to atmospheric loss. This image was acquired during a period of high solar activity but was found to be consistent with previous observations (lower panel of Fig. \ref{MarsAtmosphereLoss}) that also showed an extended halo of emission beyond the planetary disk.

For Venus, where atmospheric loss of water has evolved the planet's climate into a place of temperature extremes, SWCX at Venus has been detected, but is more challenging to explore with current instrumentation, due to the more compact atmosphere and the limitations on spectral resolution and sensitivity of current instrumentation \cite{Dennerl2008}.

\subsubsection{Direct Solar Wind Interactions with Solid Bodies, Dwarf Planets and Comets}

Comets act as remote monitors of solar wind conditions, chronicle the history of the Solar System and serve as natural laboratories for exploring CX mechanisms, offering unique conditions that are impossible to create in a laboratory setting (e.g., low density, temperature environments with negligible magnetic field influence). Moreover, the diverse chemical compositions found among comets, along with their varied mass loss rates, distances to the Sun, and heliocentric latitudes, present an expansive canvas for studying a broad array of interaction environments and solar wind conditions.

\begin{figure}
	\centering
		\includegraphics[width=0.4\textwidth, trim={0cm 0cm 0cm 0cm},clip]{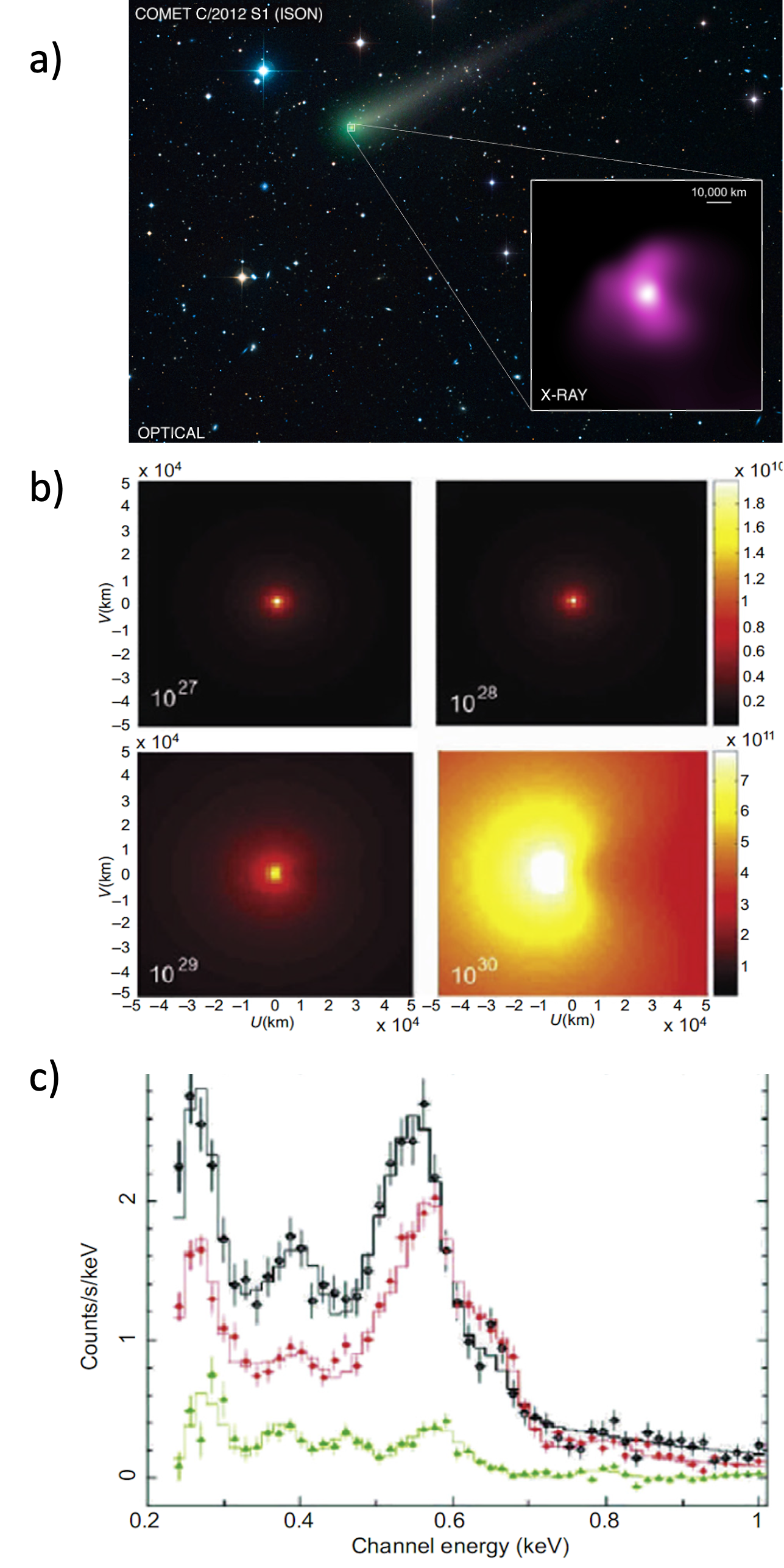}
			\caption{\textbf{Cometary X-ray Emissions:} (a) Optical and X-ray image of comet C/2012 S1 (ISON) (Snios et al., 2016\cite{Snios2016}, NASA/CXC). (b) Morphology as a function of comet gas production rate (in molecules $s^{-1}$ in the lower right of each panel), showing increasing halo emission with gas production rate \cite{Bodewits2004, Lisse2005}. c) \textit{Chandra ACIS} medium resolution CCD X-ray spectra of comets: C/LINEAR 1999 S4 (black),  Comet McNaught–Hartley (red), 2P/Encke (green) multiplied by a factor of 2, with 1$\sigma$ error bars and the best-fit emission line+thermal bremsstrahlung model, convolved with the \textit{ACIS-S} instrument response as a histogram. Best-fit model lines at 284, 380, 466, 552, 590, 648, 796, and 985 are close to those predicted for charge exchange between solar wind C$^{5+}$, C$^{6+}$, C$^{6+}$/N$^{6+}$, O$^{7+}$, O$^{7+}$, O$^{8+}$, O$^{8+}$, and Ne$^{9+}$ ions and neutral gases in the comet's coma\cite{Lisse2001,Krasnopolsky2004,Bhardwaj2007}. Figures from Bhardwaj et al. (2007)}
	\label{CometIson}
\end{figure}

Figure \ref{CometIson} shows an optical and X-ray image of Comet C/2012 S1 (ISON). For comets, SWCX from the coma dominates the X-ray emissions (Fig. \ref{CometIson}c)\cite{Lisse2001,Lisse2005,Krasnopolsky2004,Bodewits2007,Dennerl1997}. The X-ray signatures act as a diagnostic of the gas production rate of comets (e.g. see Fig. \ref{CometIson}b), the solar wind ion population that is incident at the comet (Fig. \ref{CometIson}c), the solar wind freeze-in temperature, the rotation rate of the comet, and potentially the prevalence of dust, which may attenuate the X-ray emissions \cite{Bodewits2004,Lisse1999,Lisse2001,Lisse2005,Dennerl1997, Snios2016}. Also imprinted within cometary SWCX X-ray signatures is information about the nature of the neutrals at the comet; different electron-donor molecular compositions will preferentially populate different energy levels within the X-ray-producing ion. However, to date, X-ray spectral observations have had insufficient resolution to identify differing neutral compositions through the different strengths of SWCX spectral lines. This requires instrumentation with sufficient spectral resolution and sensitivity, such as \textit{LEM}.

\begin{figure}
	\centering
		\includegraphics[width=0.4\textwidth, trim={0cm 0cm 0cm 0cm},clip]{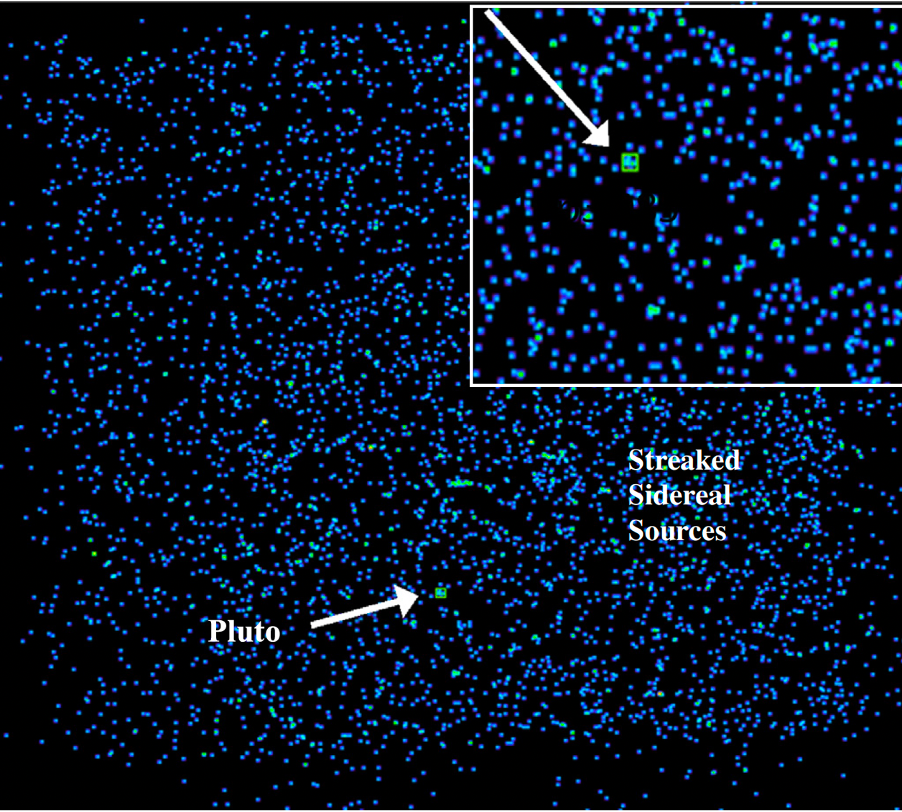}
			\caption{\textbf{X-ray Emissions from the Kuiper Belt: Pluto} Combined Chandra ACIS-S observations of Pluto from 2014-2015 in the Plutocentric frame. The 0.31–0.6 keV events from all 4 epochs (174 ksec total on-target time) have been coadded. X-ray sources that are stationary in sky-coordinates become trails of photons in the Pluto frame. Pluto’s location is shown by the arrow. The inset in the top right corner zooms-in to Pluto’s location in the image. Figures from Lisse et al. (2017)\cite{Lisse2017}.
}
	\label{Pluto}
\end{figure}

Another valuable probe for Solar System evolution and to explore the dominant processes in the heliosphere at regions beyond the reach of most in-situ spacecraft are the Kuiper Belt Objects (KBOs). At 33 au,  detections of X-ray emissions from Pluto have been made by  \textit{Chandra}, which showed an anomalously bright signature that was most likely to be due to CX, but may also be, in part, fluorescence (Fig. \ref{Pluto}) \cite{Lisse2017}. The authors suggest that at Pluto’s distance from the Sun, it will retain outgassed material for much longer than comets are able to at a few au. Consequently, this build up of extended outgassed neutrals, which may include contributions from Charon, provides a much larger neutral population with which the solar wind can directly interact. However, there remains an orders of magnitude discrepancy between modeling predictions and the observed X-ray emissions. Solving this mystery of Pluto's `puzzling' X-ray emissions, and identifying the extent to which KBOs carry extended clouds of outgassed material, is likely to require X-ray observatories with increased sensitivity and reduced background, such as \textit{LEM} will provide.

A little closer to home, the PS Decadal asks the question: \textit{``Q6.5e How Is the Escape of Volatiles from the Moon .. Driven by Photon, Charged Particle, and Micrometeorite Influx?''} Solar wind ion and electron impacts with the moon are expected to be an additional mechanism for X-ray emission and are identified for the far more distant Galilean Satellites (see section 2). Solar wind particle impacts have also been identified at the surface of Mercury \cite{Lindsay2016}. This emission may be identifiable on the Lunar dark side (with fluorescence from solar photons dominating the dayside emission). However, Wargelin et al. (2004)\cite{Wargelin2004} showed that the X-ray emission coincident with the lunar dark side in the Chandra observations was dominated by the SWCX emissions from the terrestrial exosphere between the spacecraft and the moon. To distinguish fluorescence from solar wind particle impacts from the SWCX emissions will require improvements in spectral resolution, like those provided by \textit{LEM}.

\subsubsection{Magnetosheath Solar Wind Interactions with Exospheres - The Influence of Space Weather on Magnetospheres}

The Heliophysics Decadal key science goal 2 is to \textit{``Determine the dynamics and coupling of Earth’s magnetosphere, ionosphere, and atmosphere and their response to solar and terrestrial inputs.''} This also features significantly in the ESA Voyage 2050 document and the long term plan for the agency. 

At the Earth, the magnetosphere provides a protective barrier that largely prevents atmospheric loss. Along the boundary of the magnetosphere, solar wind builds up into a dense region that traces the shape and structure of the global system:  the magnetosheath. This structure provides a valuable analogue for astrophysical shocks, a natural laboratory for fundamental plasma physics and is critical to understanding the impact of space weather on Earth and human technology. 

Within the magnetosheath solar wind ions interact with neutrals from the exosphere and produce X-rays through SWCX. These emissions have been observed as a source of astrophysical X-ray background by the fleet of X-ray observatories to date \cite{Carter2008,Carter2010,Carter2011,Kuntz2015}. The enhanced densities of solar wind ions in the magnetosheath lead this X-ray emission to stand out against the background, enabling the emission to be used to image the terrestrial magnetosheath and identify variation in the global dynamics of the terrestrial magnetosphere with time\cite{sibeck2018}.

This unique capability has led NASA, ESA and the Chinese Academy of Sciences to launch dedicated spacecraft, \textit{LEXI} (launch 2024) and \textit{SMILE}\cite{GBR2018} (launch 2025), to directly image the magnetosheath and study its variation with space weather, season and exospheric processes.  These spacecraft will use wide-field soft X-ray instruments to image the magnetosheath of the Earth. LEXI uses an MCP detector with almost no energy resolution and SMILE uses CCD detectors with low spectral resolution. Both have relatively low effective areas of order $\sim$10 cm$^2$, but with large fields of view that offer a global perspective on the nature of the interaction between Earth and the Solar Wind. These instruments are unlikely to provide the sensitivity and/or spatial resolution with which to study small-scale structure and processes within the magnetosheath.

\subsubsection{Aurorae and CX within the Magnetosphere}

\begin{figure}
	\centering
		\includegraphics[width=0.5\textwidth, trim={0cm 0cm 0cm 0cm},clip]{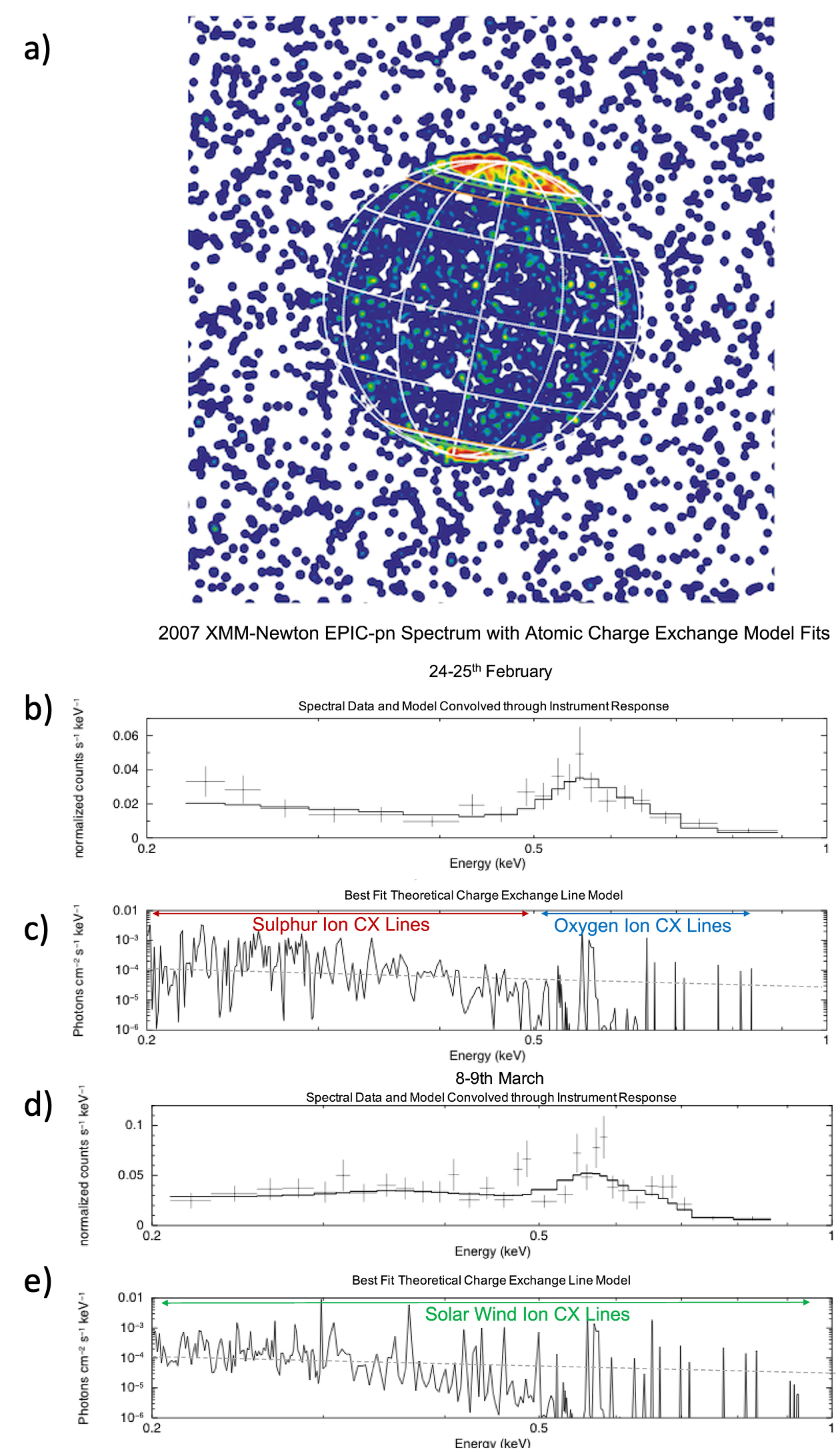}
			\caption{\textbf{Jupiter's X-ray Emissions - Hints of Iogenic and Solar Wind Ion Precipitation:} a) Taken in December 2000, this is the first Chandra HRC X-ray image of Jupiter. X-ray photons have been smeared by double the 0.4” FWHM PSF of Chandra’s HRC instrument. Jupiter is overlaid with a 30° latitude-longitude graticule (white).  b - e) XMM-Newton EPIC-pn Jovian Northern Aurora spectral data from an observation on 24-25 Feb (b) and 8–9 March 2007 (d). (b) and (d) respectively, are the best-fit atomic charge exchange spectral models convolved through the instrument response (black line) and plotted with the XMM-Newton EPIC-pn observation (crosses). (c) and (e) respectively, show the best-fit theoretical atomic charge exchange models of iogenic ions (sulfur + oxygen) or a solar wind ion population\cite{VonSteiger2000}. Figures from Gladstone et al. (2002)\cite{Gladstone2002} and Dunn et al. (2020b)\cite{Dunn2020b}.}
	\label{Jupiter}
\end{figure}

The Jovian system produces a treasure trove of bright, dynamic X-ray emissions from the moons, Io Torus, atmosphere, radiation belts and from its spectacular aurorae, - the most powerful in the Solar System (for a review of all Jovian X-ray emissions observable by  \textit{LEM}  see \cite{Dunn2022Jupiter}). The rapid-rotation, strong magnetic field and mass-loading in Jupiter’s magnetosphere mean the system presents perhaps the most extreme counterpoint to Earth with which to conduct comparative planetology. Where the magnetospheric dynamics of Earth are largely governed by the solar wind, for decades the extent to which the solar wind controls or enters the Jovian system has been debated, carrying broad implications for the importance of internal vs external drivers for magnetospheres across the Universe\cite{McComas2007,Cowley2008}.

\begin{figure*}
	\centering
		\includegraphics[width=0.9\textwidth, trim={0cm 0cm 0cm 0cm},clip]{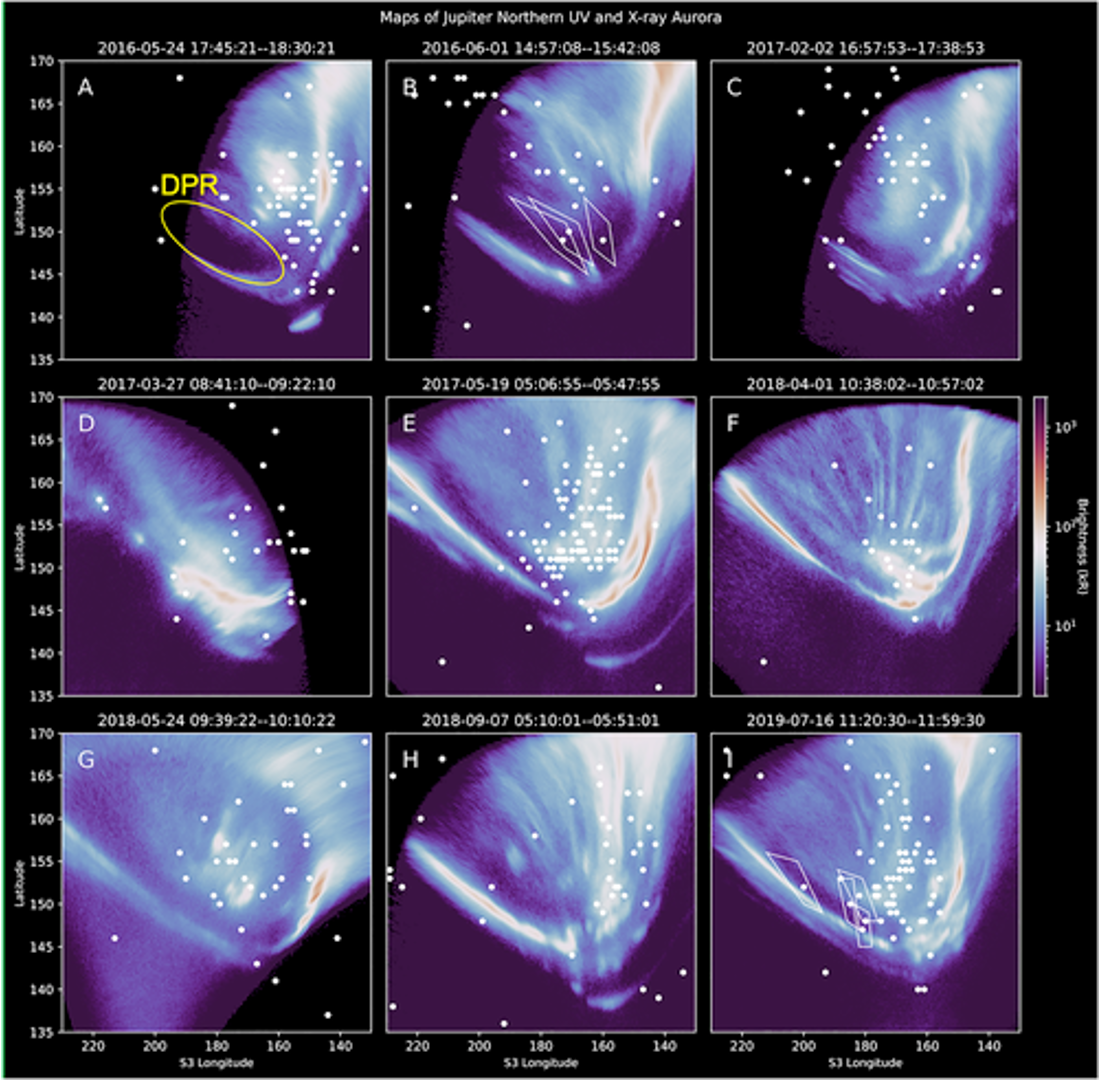}
			\caption{\textbf{Variability in Jupiter’s X-ray and UV aurorae:} Overlaid simultaneous UV (blue-white-red color map) and X-ray photon (white dots) longitude-latitude maps of Jupiter's North Pole, from the Hubble Space Telescope (HST) and Chandra X-ray Observatory High Resolution Camera (CXO-HRC). Dates and times of the observations (UT) are at the top of each panel. Only UV and X-ray emission produced during these times is shown. Panel (a) shows a yellow circle indicating the location of the dark polar region (DPR). Panels (b and i) show white quadrilaterals around example X-ray photons. The vertices of a given quadrilateral are the extremes of the photon's possible projected longitude-latitude location based on the projected 1-sigma (0.4'') point spread function of CXO. Figure and caption contents from Dunn et al. (2022)\cite{Dunn2022XUV}.
}
	\label{JupitersAurora}
\end{figure*}

The PS decadal spotlights the importance of understanding these emissions:\textit{``The excited aurorae in the waveband spanning radio to X-ray spectrum are both a window into the workings of these magnetospheres and one of the main loss processes of the energy of the confined plasma''} (Q7.4b).

Fig. \ref{Jupiter}a shows the first Chandra X-ray image of Jupiter from 1999\cite{Gladstone2002}, with the polar aurorae clearly standing out against the equatorial emissions from Thomson scattered solar emission and fluorescence (see section 2). Fig. \ref{Jupiter}b-e shows that Jupiter’s X-ray aurora can vary from favouring populations of precipitating Iogenic plasma to populations of solar wind ions. Often a spectrum will only marginally favour Iogenic CX over SWCX or vice versa, so that it is challenging to pin-point whether solar wind is precipitating at all and under what conditions. At the heart of such observations are fundamental questions about the nature of magnetospheres and the extent of their openness to the solar wind - is Jupiter entirely closed from the solar wind or open to it and under what conditions does it switch between these regimes? Current X-ray instrumentation is at the cusp of being able to answer this, but is held back by either spectral resolution (for \textit{XMM-Newton EPIC-pn} or \textit{Chandra ACIS}) or effective area (\textit{XMM-Newton RGS}). 

Fig. \ref{JupitersAurora} shows a series of 9 different 20-40-minute joint \textit{Hubble Space Telescope} UV and \textit{Chandra HRC} X-ray observations of Jupiter’s aurora. These highlight how variable the aurora can be, with different components of the auroral emission changing on $\sim$minute, $\sim$hour and $\sim$Jovian rotation (9.9 hr) timescales \cite{Gladstone2002,Elsner2005,GBR2004,GBR2007Aurora,GBR2008,Dunn2016,Weigt2020,Weigt2021,Wibisono2020,Wibisono2021}. Fig. \ref{JupitersAurora}H shows an example of auroral behaviour during a solar wind compression, while Fig. \ref{JupitersAurora}E and Fig. \ref{JupitersAurora}I are under more rarefied solar wind conditions \cite{Dunn2022XUV} - when solar wind ion precipitation appears to provide better fits to the resulting X-ray auroral spectra\cite{Dunn2020b}. However, with current resolution and sensitivity it is not possible to significantly differentiate between the two cases and therefore to understand the changing relationships of Jupiter-like planets to the solar wind. While other wavebands, such as the UV, provide higher photon counts from the planet (e.g. Fig. \ref{JupitersAurora}), they do not have the capacity to spectrally resolve the precipitating ion population. Consequently, only the X-ray waveband can remotely diagnose solar wind precipitation into the atmosphere and \textit{LEM}-like effective area and energy resolution is required to unambiguously identify these SWCX signatures.

\begin{figure}
	\centering
		\includegraphics[width=0.48\textwidth, trim={0cm 0cm 0cm 0cm},clip]{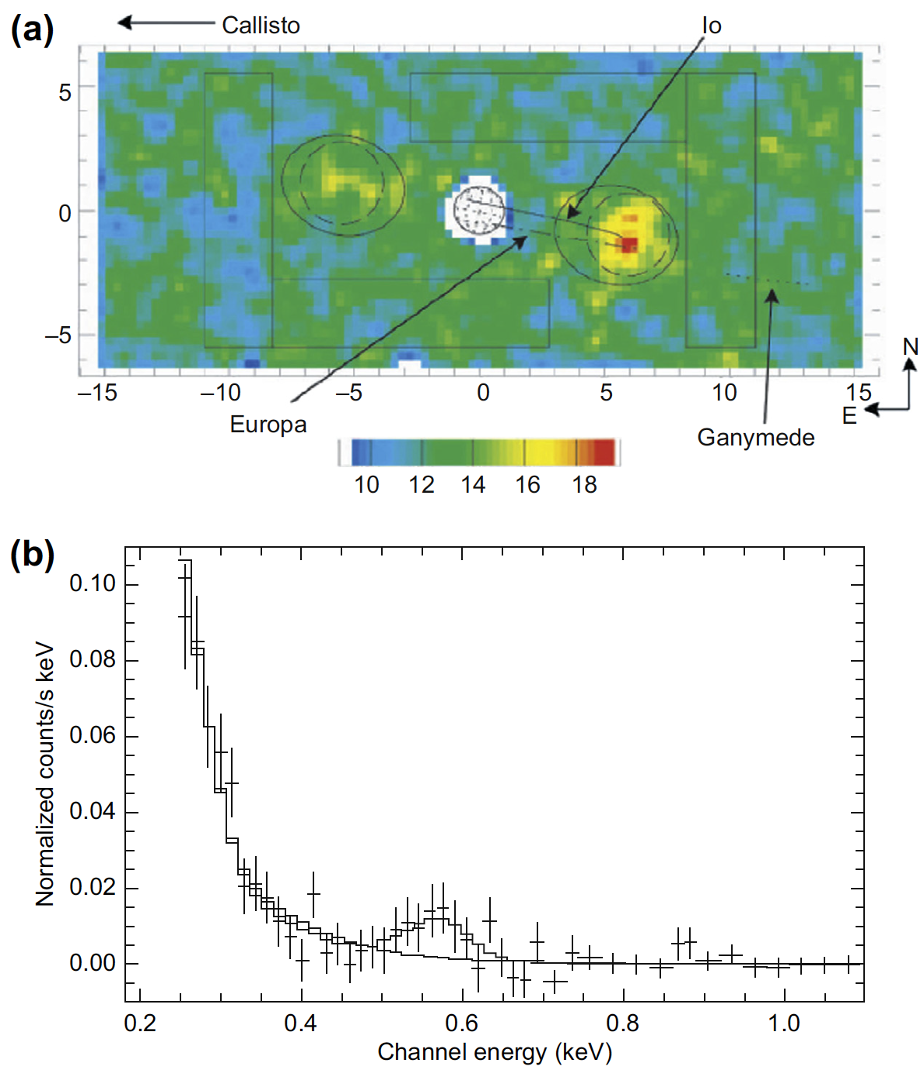}
			\caption{\textbf{X-ray observations of the Io Torus:} Upper : Gaussian -smoothed Chandra HRC-I image of the Io Plasma Torus (2000 December 18). Axes are in units of R$_J$ , and the scale bar is in smoothed counts per image pixel. The paths traced by the moons are shown on the figure for Io (solid line), Europa (dashed line), and Ganymede (dotted line). Lower panel: Background-subtracted \textit{Chandra ACIS} IPT spectrum, with a power-law model fit and also a power-law plus Gaussian model fit. Figures and caption contents from Elsner et al. (2002)\cite{Elsner2002}.}
	\label{IPT}
\end{figure}

As well as the interaction of the solar wind with planetary neutrals, within planetary systems magnetospheric plasma often interacts with neutrals providing further analogues. Encircling Jupiter at Io's orbit is the Io plasma torus and neutral torus. These are produced by volcanic material from the moon, approximately half of which becomes ionised. The very first \textit{Chandra} observations of Jupiter also provided the first X-ray observations of X-ray emission from the Io torus (Fig. \ref{IPT}\cite{Elsner2002}). Within the Solar System this combination of a neutral cloud and hot magnetospheric plasma provides a valuable analogue for plasma-neutral interactions across astrophysics. The source of these emissions is unclear due to current instrument energy resolution, sensitivity, and background, but is thought to be from CX \cite{Elsner2002,Dunn2022Jupiter}. To conclusively identify the cause of the emissions will require heightened spectral resolution and instrument grasp.

\subsubsection{Ice Giant Aurorae}
For the Ice Giants, the tilted, offset magnetic fields tumble through the Solar System providing the most complex magnetospheric interactions with the solar wind. The Uranian aurora is suggested to be produced by the magnetospheric cusp\cite{Lamy2017} and therefore may show signatures of entry by the solar wind into the system. Perhaps the only way to test this remotely, and its seasonal dependence (the Uranus Orbiter and Probe will arrive during a different Uranian season in the mid-2040s), is through X-ray observations. As the cusp rotates into and out of view from Earth, one may expect a periodic SWCX signature to occur in phase with auroral visibility. Identifying this signature against the X-ray background will require a high spectral resolution and sensitivity instrument akin to \textit{LEM}.

\subsubsection{Mapping the Heliosphere, Local Hot Bubble and Measuring Interstellar Neutral Winds}

Beyond the Solar System planets, the heliosphere and local Hot Bubble can be mapped through CX emissions. Key Science Goal 4 in the Heliophysics Decadal is to \textit{``Discover and characterize fundamental processes that occur both within the heliosphere and throughout the universe.”}
The heliosphere is the Sun's own astrosphere, a bubble-like region blown into space and filled with solar wind (SW) plasma. Initially emitted at supersonic speeds (300-800 km/s), the solar wind slows down abruptly at the termination shock ($\sim$80 AU) due to the pressure of the interstellar (IS) medium. The outer heliospheric limit is the heliopause, at first order defined by the pressure equilibrium between the solar wind and IS plasmas, and situated approximately at $\sim$130 au from the Sun \cite{2015ApJS..220...32I}.
 
In the case of heliospheric SWCX, the neutral targets are IS atoms flowing through the heliospheric interface and into the Solar System \cite{2005Sci...307.1447L}. This IS wind, composed mainly by hydrogen (H) and $\sim$15\% helium (He), appears to flow at $\sim$ 25 km/s from the direction (255$^{\circ}$, 5$^{\circ}$) in ecliptic coordinates, due to the relative motion of the Sun with respect to the local IS medium. In Galactic coordinates this translates to (3.2$^{\circ}$, 15.5$^{\circ}$) coincidentally near the Galactic center.

\begin{figure}
	\centering
		\includegraphics[width=0.48\textwidth, trim={0cm 0cm 0cm 0cm},clip]{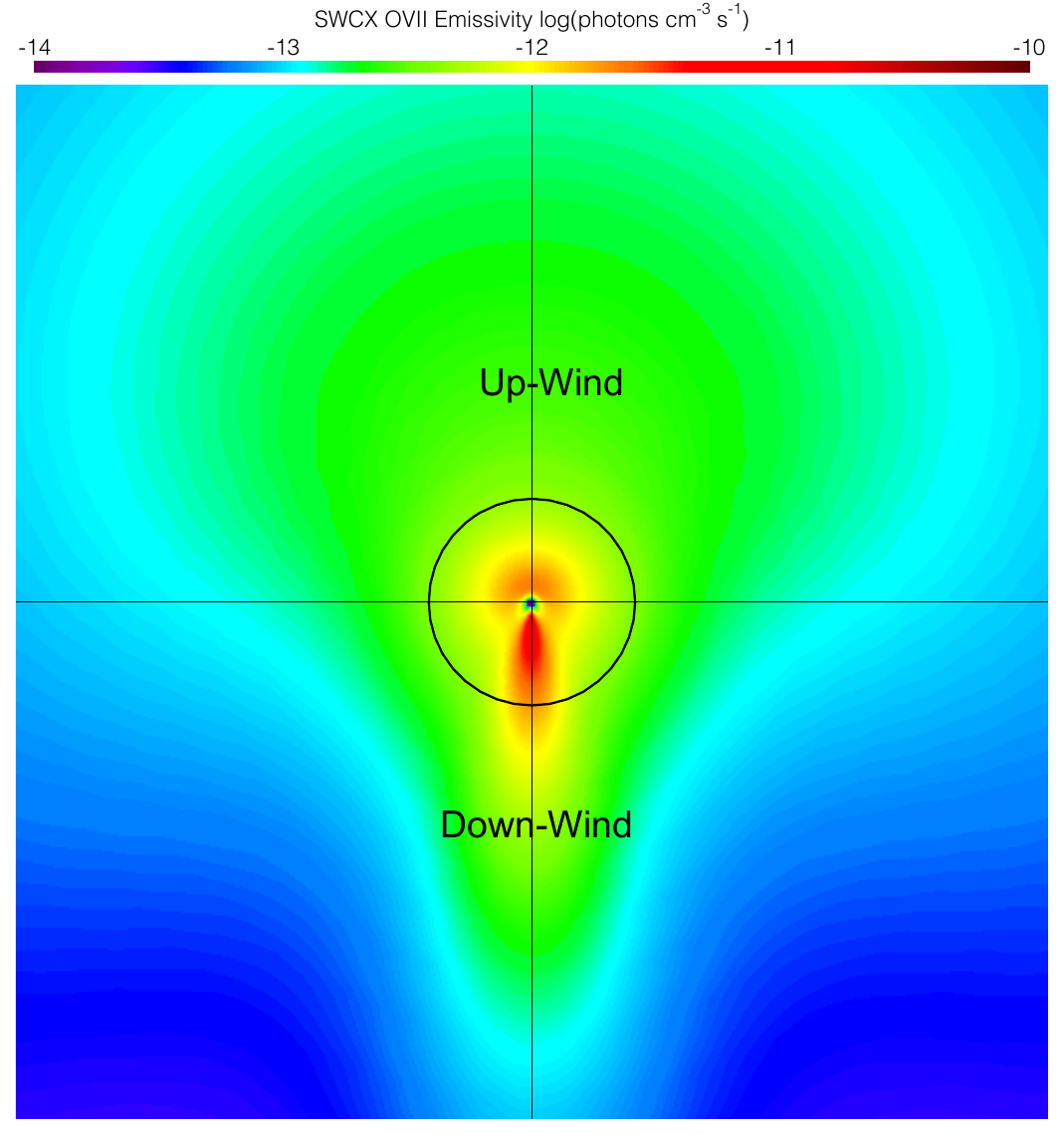}
			\caption{\textbf{SWCX X-ray Emission from the Heliosphere:} Ecliptic plane cut of the total O~VII SWCX volume emissivity in log units of photons/cm$^3$/s for steady-state slow solar wind in solar maximum conditions. The Sun is in the center and the black circle represents roughly the Earth/\textit{LEM} position throughout the year. The Up-Wind (Down-Wind) direction corresponds to early June (December) on the orbit, and points roughly near the Galactic Center (anti-center).}
	\label{fig:SWCX_emissivity}
\end{figure}

Until now, studies of the heliospheric SWCX have been based on the characteristic spatial and temporal variations that differentiate it from the invariable cosmic background (Local Bubble and Galactic halo).

The spatial variations of heliospheric SWCX are mainly due to the distribution of interstellar neutrals in the solar system. To first order, this neutral atom distribution in interplanetary space (and hence the SWCX volume emissivity) is axi-symmetric around the IS flow vector \citep{2012AN....333..341K}, as shown in Figure \ref{fig:SWCX_emissivity}. Hydrogen SWCX production is the strongest towards the incoming flow direction (upwind, or towards the Galactic center $\sim$ 2-3 AU from the Sun, represented by the green area in Figure \ref{fig:SWCX_emissivity}), as it is strongly ionised by CX with solar protons near the Sun. SWCX with He atoms dominates the inner Solar System (<1 AU; orange/red area in Figure \ref{fig:SWCX_emissivity}) and especially a cone-like structure in the downwind direction where He atoms are focused due to the Sun's gravity\footnote{H atoms cannot be focused like He, because they are subjected to solar radiation pressure.}.The Earth crosses this He cone region every year in December. 

A dedicated campaign with the Diffuse X-rays from the Local galaxy (DXL) sounding rocket experiment, enabled measurements of the SWCX excess emission in the He cone \cite{Galeazzi2014}. DXL was launched when the Earth was crossing the denser part of the He cone, and scanned the central region of the cone, near the Galactic anti-center (Figure \ref{fig:SWCX_emissivity}). The X-ray flux observed with DXL was compared to data of the same sky region obtained with ROSAT, in a different observation geometry \footnote{The ROSAT All-sky Survey scanned directions roughly perpedicular to the Sun-Earth direction, therfore essentially missing the bulk of the He cone emission when observing the galactic anti-center.}. The comparison allowed an estimate of the SWCX contribution of $\sim$40\%\ to the soft X-ray background in the 1/4 keV band in the Galactic plane \cite{Galeazzi2014}. 

The temporal variations of the heliospheric SWCX are mainly due to the solar activity and its effects on the solar wind flux and heavy ion composition. A large archive of CCD-resolution data with XMM-Newton spanning almost one solar cycle showed a clear correlation of the diffuse oxygen band (0.5-0.7 keV) flux with the solar sunspot number that was clearly linked to the heliospheric SWCX emission dependence on solar activity \cite{Qu2022}. 

Despite the rich information provided by current observatories, their spectral resolution, sensitivity and grasp struggle to distinguish the distinct spectral characteristics of the heliospheric SWCX from the cosmic X-ray background. The \textit{LEM}  All Sky Survey will change this.

\begin{figure*}
	\centering
		\includegraphics[width=0.98\textwidth, trim={0cm 0cm 0cm 0cm},clip]{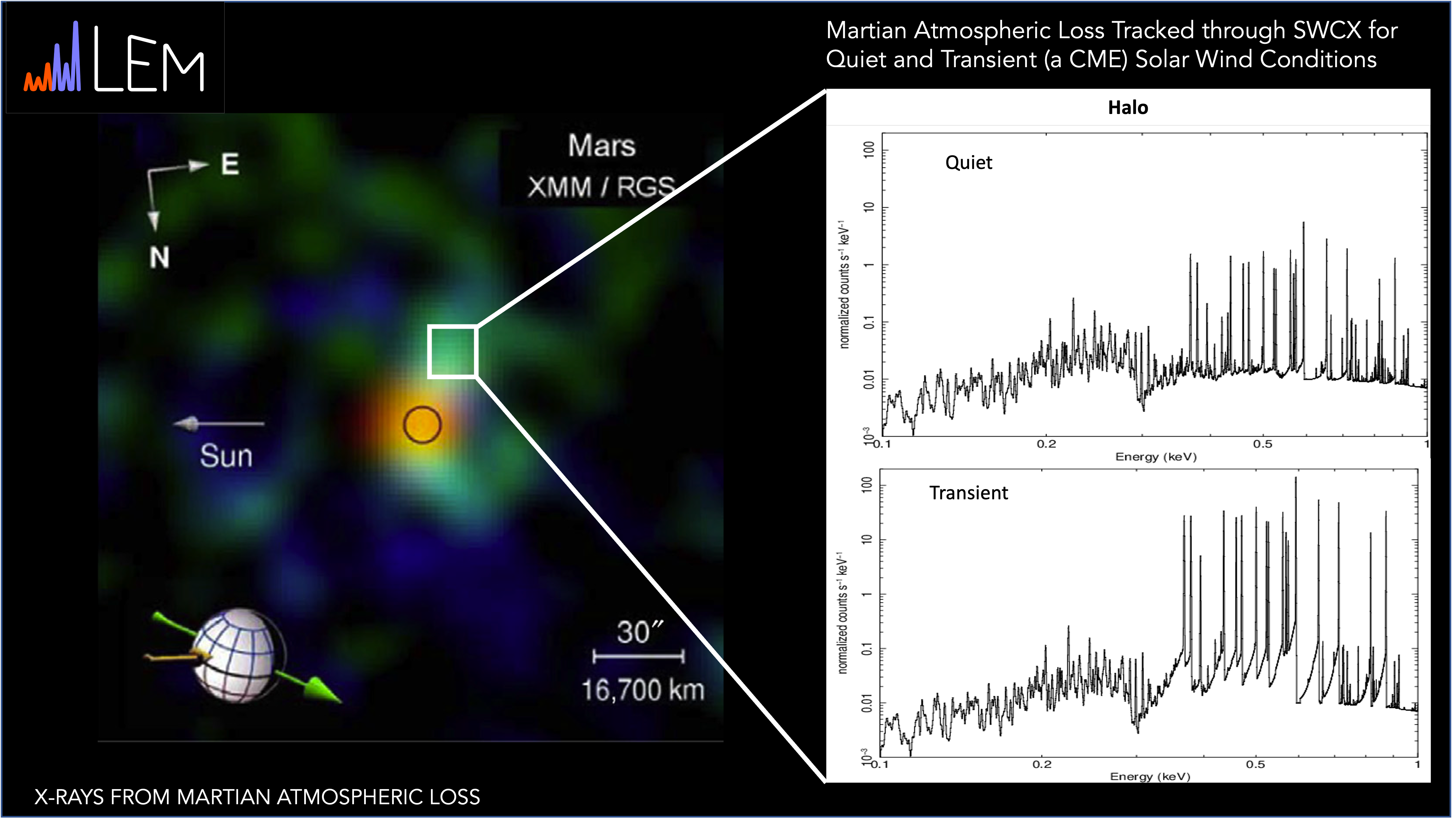}
			\caption{\textbf{Using \textit{LEM} to Connect Loss of the Martian Atmosphere to Solar Wind Conditions}. The white box illustrates a \textit{LEM} pixel on the previously shown XMM-Newton observation of SWCX from the Martian Halo\cite{Dennerl2006}. The spectra on the right show predicted \textit{LEM} observations of SWCX signatures from the Martian halo for quiet and transient (e.g. a Coronal Mass Ejection) solar wind conditions\cite{CarterMars}.}
	\label{LEMMars}
\end{figure*}

\subsection{What \textit{LEM} will tell us about Solar Wind Interactions across the Heliosphere}

\textit{LEM} is perfectly suited to the study of CX emissions with an ideal energy range, spectral resolution, grasp, and sensitivity; it will prove transformative for our understanding of any physical system that produces these emissions. It opens a range of new CX capabilities that have been largely inaccessible for previous instrumentation including measurement of new species, doppler shifts, and thermal broadening of spectral lines. It will also open the door to identification of molecular composition of the neutrals through CX, where different donor neutral atoms will lead the exchanged electron to populate different energy levels in the ion. This produces different line ratios for different molecules.

\subsubsection{Direct Solar Wind Interactions with the Atmospheres of Venus and Mars}

For Mars, through LEM’s high effective area and spectral resolution, it can fully utilise the metastable states (e.g. the forbidden lines) within the ratio of the OVII multiplet to track atmospheric loss at different heights above Mars \cite{CarterMars}. Variation in SWCX line ratios may enable constraints to be placed on the loss of different neutrals at different atmospheric heights. 

Furthermore, through broadening and doppler shifts of the spectral lines, detectable from Mars for the first time by \textit{LEM}, it may be possible to characterise how atmospheric stripping varies globally with different solar wind conditions (e.g. Coronal Mass Ejections (CME) vs quiescent slow solar wind - Fig \ref{LEMMars}). \textit{LEM}’s few day response time is ideal for tracking specific space weather events (e.g. CMEs) to Mars to directly test their impact.  Fig. \ref{LEMMars} showcases the spatial scales over which \textit{LEM} will be able to explore the interaction and shows modeled \textit{LEM} spectra from the Martian Halo \cite{CarterMars}. These highlight how varying solar wind conditions will change the observed  \textit{LEM}  spectrum from the halo of atmospheric loss around Mars.

For Venus, the resolution of \textit{XMM-Newton EPIC-pn} and \textit{Chandra ACIS} or the sensitivity of \textit{XMM-Newton RGS} make it much more challenging to study the SWCX emissions\cite{Dennerl2008}. \textit{LEM}’s solar avoidance angle will permit observations of Venus during limited windows each year, while its enhancements in spectral resolution and effective area will permit step-changes in studies of Venus’s atmospheric loss.

\textbf{Partnering LEM observations with simultaneous in-situ measurements by spacecraft at Mars or Venus during the 2030s (see Fig. \ref{OtherSpacecraft}) will break the degeneracies in the SWCX models, quantifying the loss of the Martian and Venusian atmosphere through SWCX}.

\subsubsection{Direct Solar Wind Interactions with Solid Bodies, Dwarf Planets and Comets}

\begin{figure*}
	\centering
		\includegraphics[width=0.98\textwidth, trim={0cm 0cm 0cm 0cm},clip]{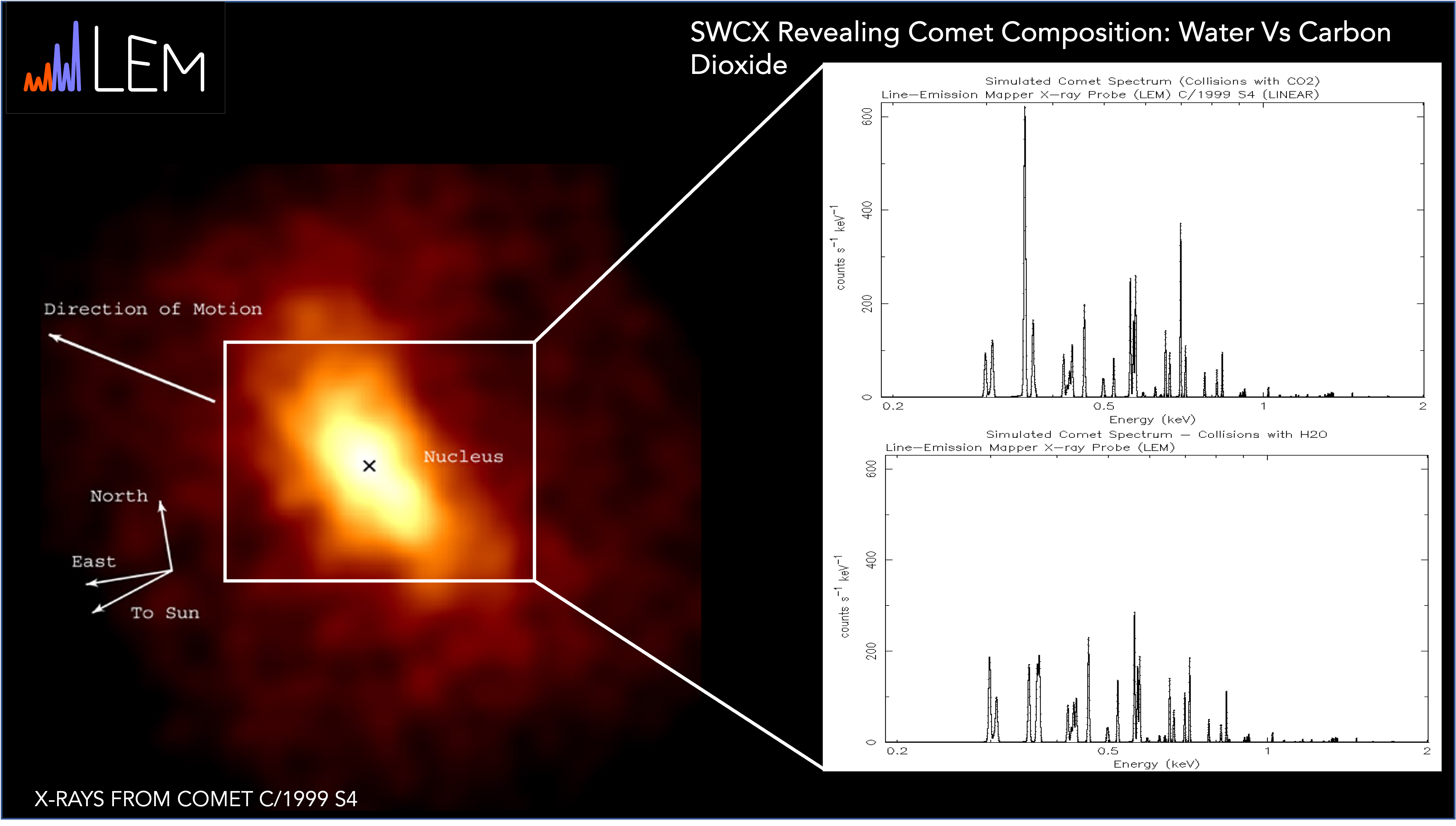}
			\caption{\textbf{Using  \textit{LEM}  Observations of SWCX to Identify  Comet Composition and Solar Wind Conditions:} Left: X-ray Image of Comet C/1999 S4 by \textit{Chandra ACIS}. The white box shows the central array of $\sim$1 eV resolution  \textit{LEM}  pixels, while the remainder of the  \textit{LEM}  field of view has dimensions of almost twice the \textit{Chandra ACIS} observation. On the right, the two spectra show two different models of the cometary X-ray emission convolved with the  \textit{LEM}  response,  highlighting  \textit{LEM} ’s ability to not only identify the SWCX lines, but to distinguish the neutral coma molecular composition (CO$_2$ - upper, H$_2$O - lower) through the different line ratios - enabling heliophysics studies in tandem with investigations into planetary composition and Solar System formation. Image Credit: NASA/CXC/Lisse/Deskins}
	\label{LEMComet}
\end{figure*}

\textit{LEM} is ideally suited to study comets. With the necessary large grasp and spectral resolution, \textit{LEM} will enable entirely new scientific investigations for comets. One particular strength of \textit{LEM} for cometary science is that \textit{LEM} provides high spectral resolution by non-dispersive means, as opposed to gratings where the substantial extent of the cometary X-ray emission degrades the spectral resolution.

The archetypal \textit{Chandra} cometary X-ray spectrum from C/1999 S4\cite{Lisse2001}(Fig. \ref{CometIson} and Fig. \ref{LEMComet}) can be directly contrasted with the high resolution \textit{LEM} spectra in Fig. \ref{LEMComet}. \textit{LEM’s} capabilities mean that it will be possible not only to measure the gas production rate for the comet, but also to constrain the composition of the gas being produced. Fig. \ref{LEMComet} contrasts the \textit{LEM} spectrum for a Hydrogen neutral population against a CO neutral population, showing the clear variation in line strengths that  \textit{LEM}  will, for the first time, measure. By measuring the SWCX signatures from a variety of comets,  \textit{LEM}  will also identify ion composition of the solar wind in new regions across the heliosphere. These observations can also be used to probe the solar ion abundances and velocities at different latitudes from the ecliptic and across the heliosphere.

\begin{figure*}
	\centering
		\includegraphics[width=0.98\textwidth, trim={0cm 0cm 0cm 0cm},clip]{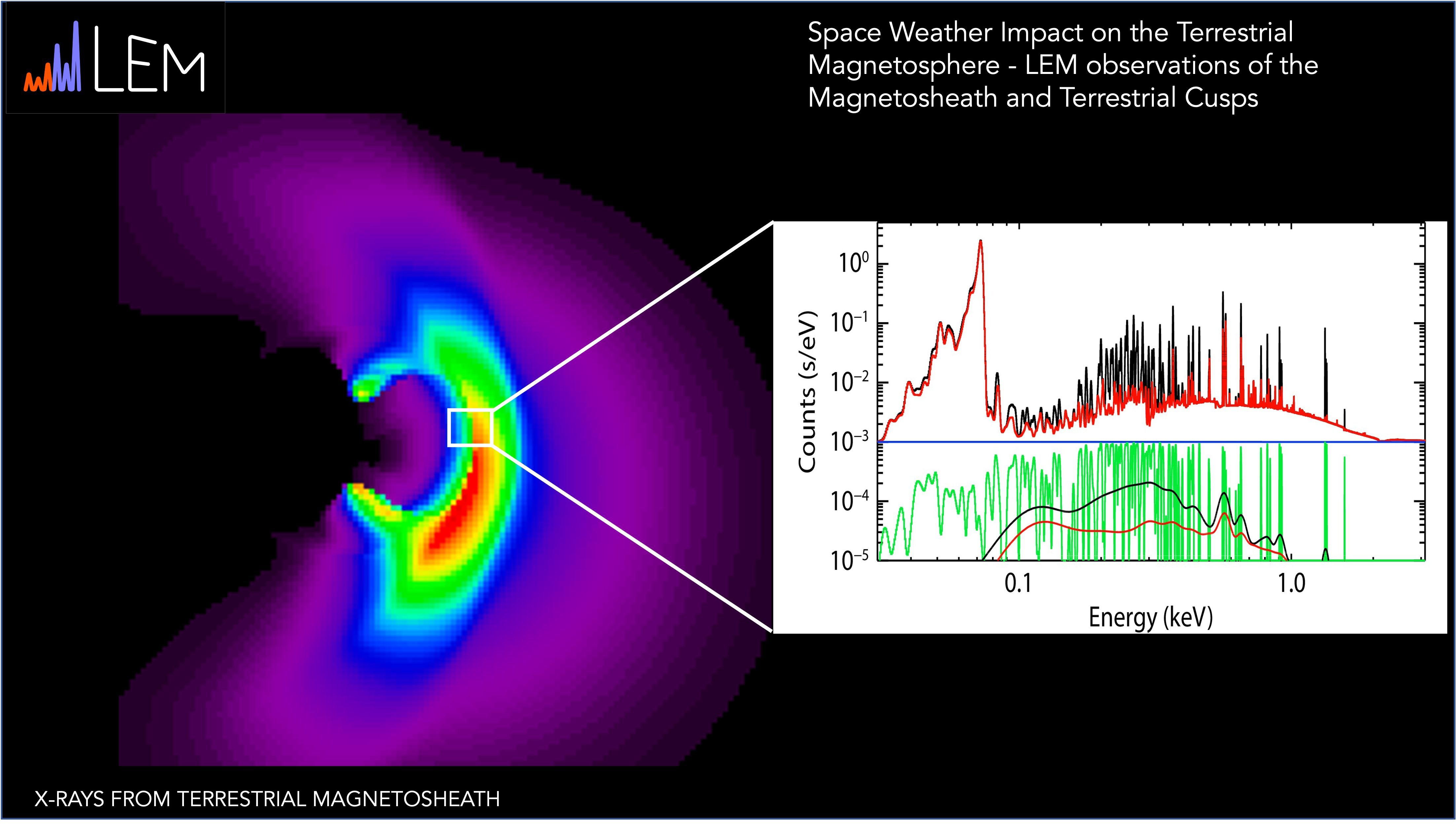}
			\caption{\textbf{Using \textit{LEM} to Explore how Magnetospheric Dynamics and the Terrestrial Magnetosheath are Governed by Space Weather:}  Left: The relative SWCX emission as seen from the vantage point of \textit{LEM} at its greatest elongation. The white box shows the size of the \textit{LEM} FOV. This simulation was for a median solar wind flux and for a northern hemisphere winter. Details of the simulation are provided in Kuntz et al. (2023) \cite{Kuntz2023}. 	Right: A comparison of the total spectra observed by \textit{LEM} (upper black curve) including the magnetosheath for a median solar wind and the background emission (upper red curve) and anticipated instrumental background (blue curve). The spectrum has been convolved with the \textit{LEM} response and line spread functions. The required \textit{LEM} bandpass is 0.2 - 2.0 keV while the expected \textit{LEM} bandpass is 0.05 - 2.5 keV. The green curve is the ratio of the magnetospheric component to the total spectrum, shifted downwards by a factor of 1000. The SMILE SXI spectrum (lower black and red curves) has been constructed for the same spectrum and emission region but does not include a instrumental background. The resolution of the SMILE spectrum is typical of that achieved by CCD-based X-ray instruments. Figures from Kuntz et al. (2023)}
	\label{LEMMag}
\end{figure*}

For the moon, Q6.5 in the PS decadal highlights the importance of probing how charged particle impacts at the Moon affect the surface. Where studies by Chandra were unable to distinguish neutral fluorescence from SWCX lines, \textit{LEM} will be able to do this,  probing the Lunar dark side that has historically been largely unavailable. Electron and proton impacts lead to thick target bremsstrahlung (TTB) and particle induced X-ray emissions (PIXE). For \textit{LEM} the distinguishing spectral features (continuum for TTB  vs purely line emission for PIXE) will be resolvable. Observations of these emissions on the lunar darkside will offer remote global measurements of charged particle impacts with the surface and can be directly compared with solar wind conditions to inform development of predictive models.

In the regions of the heliosphere beyond the known planets, with the reduced background and heightened spectral resolution, \textit{LEM} is likely to enable further detections of Pluto and may usher in an era where the extended atmospheres and degassing of Kuiper Belt objects can be explored through X-ray spectra.

\subsubsection{Magnetosheath Solar Wind Interactions with Exospheres}

Critical to the Heliophysics decadal are questions of coupling between the terrestrial Magnetosphere and Solar Wind.  Here, \textit{LEM’s} L1 orbit, high effective area, and spectral resolution, offer detailed SWCX observations of small-scale processes and solar wind conditions within the magnetosheath and the cusps. This significantly complements the global (but low spatial and spectral resolution) magnetosheath and cusp X-ray images provided by dedicated missions such as \textit{SMILE}\cite{GBR2018} and \textit{LEXI}\cite{Walsh2020}. Fig. \ref{LEMMag} shows both the size of the \textit{LEM} field of view on a terrestrial magnetosphere X-ray emission simulation and the predicted \textit{LEM} spectrum from the magnetosheath. Fig. \ref{LEMMag} also shows the step-change in spectral capability of \textit{LEM} in comparison with \textit{SMILE}, although \textit{LEM} does not permit the wide-field global imaging of \textit{SMILE}.

\begin{figure*}
	\centering
		\includegraphics[width=0.98\textwidth, trim={0cm 0cm 0cm 0cm},clip]{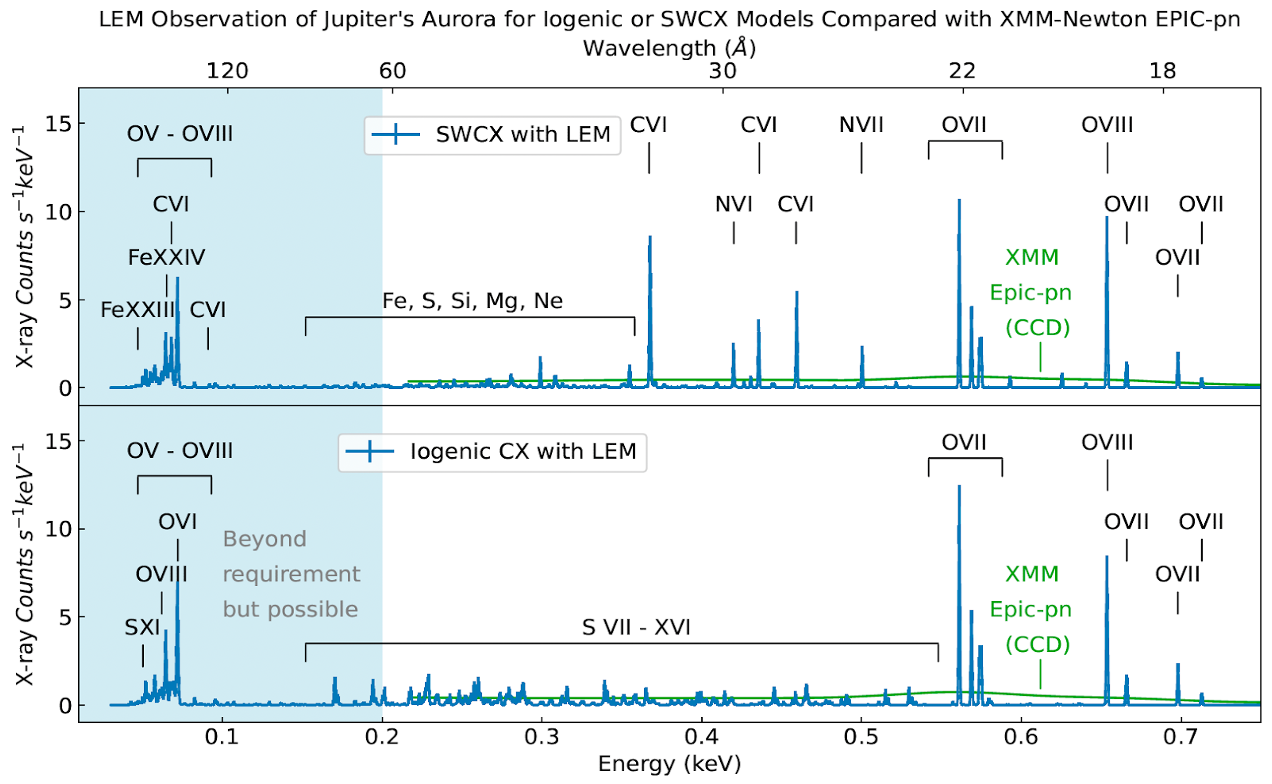}
			\caption{\textbf{Using \textit{LEM}  to Identify the Sources of Planetary Aurorae}. The panels show simulated \textit{LEM} spectra (blue) from two different models that each are able to achieve excellent fits to the Jovian auroral spectrum, as observed by \textit{XMM-Newton EPIC-pn} (green line). The models each identify a fundamentally different source population. The upper model is produced by CX from solar wind ions precipitating into the Jovian Atmosphere while the lower model is produced by CX from Iogenic plasma precipitation. In most instances they are  statistically indistiguishable for a given \textit{Chandra ACIS} or \textit{XMM-Newton EPIC-pn} observation of Jupiter.  \textit{LEM}  will unambiguously distinguish the precipitating ion species in  Jupiter's aurorae and, through comparison with simultaneous in-situ data from \textit{JUICE} and/or \textit{Europa Clipper}, will identify how the X-ray auroral source population varies with magnetospheric and solar wind conditions. Shaded in blue is an energy range that the  \textit{LEM}  microcalorimeter has access to, but that is not an instrument requirement for the mission. Here,  \textit{LEM}  may provide access to spectral lines and species that have not previously been accessible.}
	\label{JupiterAuroraSpec}
\end{figure*}

For understanding the dynamics that govern the terrestrial magnetosphere, \textit{LEM} will provide unique small scale structure observations ($\pm$0.22 R$_E$) across 3.7 × 3.7 R$_E$ of the magnetosheath at high cadence ($\sim$3 minute timescales). This complements \textit{SMILE's} and \textit{LEXI’s} global imaging of the magnetosheath, providing detailed studies of magnetosheath structure at smaller scales. 

The scales of \textit{LEM}'s resolution are close to the scales at which \textit{NASA's Magnetospheric MultiScale (MMS) mission} measures turbulence-driven magnetic reconnection \cite{Stawarz2019,Stawarz2021}, so that it's possible that \textit{LEM} measurements of Doppler shifts may probe outflows from small-scale reconnection events across the magnetosheath, although we note that electron acceleration normally dominates over ion acceleration on small scales.  

For studies of this nature, \textit{LEM} will fly simultaneous with \textit{NASA's Helioswarm}. \textit{Helioswarm} will seek to explore multiple scales of turbulence and fundamental plasma processes throughout the solar wind and magnetosphere. There will be a rich array of multi-scale magnetospheric observations of reconnection, waves and turbulence possible through the combination of \textit{Helioswarm} and \textit{LEM}. Synchronised magnetosheath observations by both would provide a rich scientific legacy.

\begin{figure*}
	\centering
		\includegraphics[width=0.98\textwidth, trim={0cm 0cm 0cm 0cm},clip]{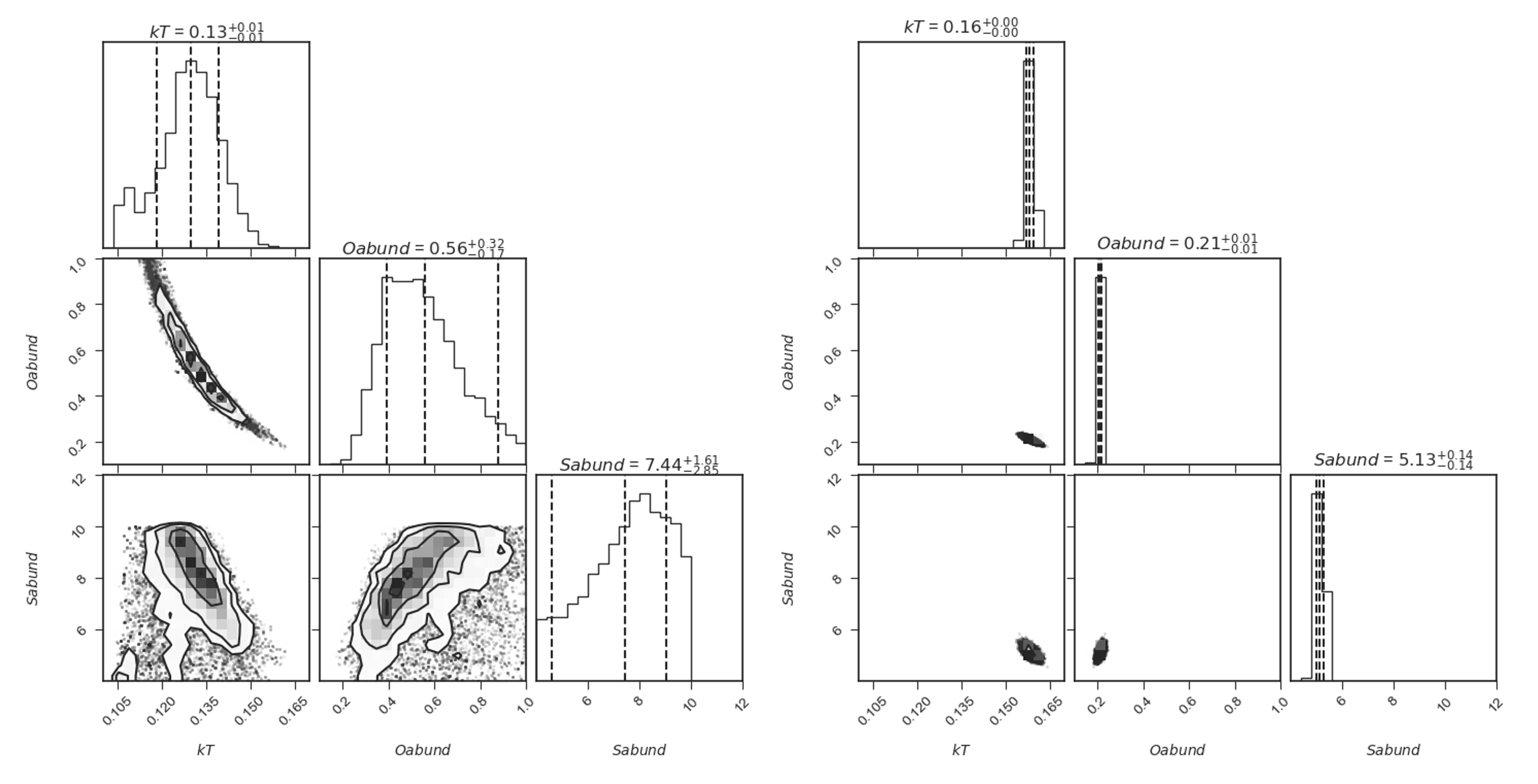}
			\caption{\textbf{LEM's Step-Change Improvement over Existing Instrumentation for CX Spectra:} Markov Chain Monte Carlo posterior distributions for simulated \textit{XMM-Newton EPIC-pn} and \textit{LEM} observations of atomic charge exchange in Jupiter’s aurora. The left corner plot shows \textit{XMM-Newton EPIC-pn} MCMC posterior distributions, while the right shows these distributions for  \textit{LEM}, demonstrating how tightly the parameters can be constrained through the high spectral resolution measurements. Here, the initial model parameters for the simulations were an oxygen abundance of 0.2 times the solar abundance, a sulfur abundance of 5 times the solar abundance, and a temperature of 0.16 keV. With \textit{XMM-Newton EPIC-pn}, we are unable to retrieve the initial parameters for the model, where \textit{LEM} will enable this with error bars at the second significant figure. Models are from the AtomDB ACX Package. Figures are from \cite{Parmar2023}.}
	\label{MCMC}
\end{figure*}

\subsubsection{Auroral Emissions and CX within Magnetospheres}

\textit{LEM} will be critical to address if and when the solar wind precipitates into the Jovian atmosphere, and possibly to identify the mysteriously absent Jovian magnetospheric cusp. By doing so, it will help to address PS Decadal questions on the nature of the solar wind relationship with giant, fast-rotating planets and the sources of their auroral emissions (e.g. PS-Q7.4b). 

For most observations, Jupiter’s X-ray auroral spectrum is dominated by lines from precipitating sulphur and oxygen ions  - Iogenic material that fills the vast Jovian magnetosphere but originates in Io's volcanoes. However, for some observations, there are hints of SWCX, suggestive of entry of the solar wind into the planet’s atmosphere. The spectral resolution and effective area of \textit{XMM-Newton} and \textit{Chandra} have been insufficient to distinguish SWCX from Iogenic CX components. Fig. \ref{JupiterAuroraSpec} shows that LEM’s resolution will, for the first time, clearly distinguish the different ion species precipitating into the Jovian atmosphere. 

Fig. \ref{MCMC} quantifies \textit{LEM} 's improvement in accuracy and precision over current instrument resolution. Currently, the energy resolution of instruments do not enable retrieval of the initial atomic charge exchange model parameters, where \textit{LEM} will retrieve the correct parameters to a precision of the second significant figure.

We are currently in a golden era for gas giant research, with vast $\sim$decade-long orbital datasets from \textit{Cassini} at Saturn\cite{Dougherty2009}, and \textit{Juno} at Jupiter \cite{Bolton2017}. Such in-situ spacecraft have rewritten the textbooks - e.g. on Jupiter's internal magnetic field\cite{Moore2018}, with direct consequences for revising our observing strategies of auroral, disk and moon-related X-ray emissions\cite{Dunn2022Jupiter}. Breakthroughs have been hampered by lack of contiguous high spectral and spatial resolution, but \textit{LEM} will usher in this new era of discovery. It's launch will coincide with the arrival of the new generation of in-situ missions to the Jovian system: \textit{ESA's JUICE} and \textit{NASA's Europa Clipper} spacecraft.

In partnership with ESA's \textit{JUICE} and \textit{NASA's Europa Clipper} spacecraft, it will be possible to compare magnetospheric and solar wind conditions and processes with the resulting auroral emissions, identifying drivers of the observed X-ray aurora (as shown in e.g. \cite{YaoDunn2021}). In turn, such studies will offer essential insights into the processes driving X-ray emissions for more distant rapidly-rotating magnetospheres.

\textit{LEM} offers another step-change in capability: through measurements of the broadening and shifts of the ion CX lines (Fig. \ref{ParmarBroadeningAndShifts}) it will be possible to measure the plasma conditions at the point of collision with the atmosphere. The Juno spacecraft’s flights over the poles of Jupiter have provided a cornucopia of new insights into the physical processes governing magnetosphere-ionosphere coupling and the transfer of energy to the atmosphere, including MV potential drops \cite{Clark2020} and the nature of the processes and particle distributions across the currents in the main oval \cite{Sulaiman2022}. However, from $\sim$1 R$_J$ above the atmosphere, Juno is unable to directly sample the plasma at the point of collision with the atmosphere and thus to measure directly the fundamental acceleration processes between the spacecraft and atmosphere.

X-ray observations provide remote sensing of the plasma at the point of collision with the atmosphere, sampling the magnetospheric plasma as it leaves the magnetospheric regime and enters the atmosphere. High spectral resolution provides measurements of the thermal and collisional velocities (and therefore energies) of the ions as they undergo atmospheric collisions. Using \textit{XMM-Newton’s RGS} instrument, Branduardi-Raymont et al. (2007)\cite{GBR2007Aurora} modeled the combined spectral lines from Jupiter’s auroral and equatorial emission during solar maximum in 2003. The work found that the line broadening was consistent with ion velocities of $\sim$5000 km/s. The low effective area of \textit{RGS}, challenges of a moving-target observed through a grating, and time variability of Jupiter have meant that it is difficult to reproduce such analyses for observations beyond that to-date brightest observation seen (Nov 2003) \cite{Wibisono2023thesis}.

Fig. \ref{ParmarBroadeningAndShifts} shows predicted \textit{LEM} observations of thermally broadened S and O CX model spectra fitted to the Northern aurora. Beyond thermal velocities of $\sim$1000 km/s, \textit{LEM’s} capacity to resolve the OVII and OVIII lines becomes similar to RGS but with the important order of magnitude enhancement in effective area and the access to the sulphur and carbon lines below 0.3 keV, that appears to have inhibited such investigation for \textit{RGS}. These measurements of the ion energies will be key to identifying the processes governing magnetosphere-ionosphere coupling for giant planets.

For the X-ray emissions from the Io Plasma Torus,  \textit{LEM}'s grasp and spectral resolution will be transformative, removing background and resolving the nature of the emission process. In turn, this will enable the X-rays to be used as a diagnostic tool for the plasma source of the Jovian magnetosphere.

Finally, for the Ice Giants, \textit{LEM}'s high spectral resolution will solve the challenges with the background for Uranus (Fig \ref{Uranus}. Further, \textit{LEM's} spectral resolution will distinguish potential SWCX signatures indicative of Uranian cusp aurorae from the scattered solar emission and fluorescent emissions from the rest of the atmosphere. This offers a potentially transformative insight into the nature of the interaction between Ice Giants and the Solar Wind, and the possibility to identify the cause of Uranian auroral emissions.

\subsubsection{Mapping the Heliosphere and Local Hot Bubble}

Having explored how \textit{LEM} will transform our understanding of the nature of the solar wind interaction with all types of planetary body in the Solar System, we turn to how it will impact our understanding of the Heliosphere as a whole. 

\begin{figure}
    \centering
    \includegraphics[width=\columnwidth]{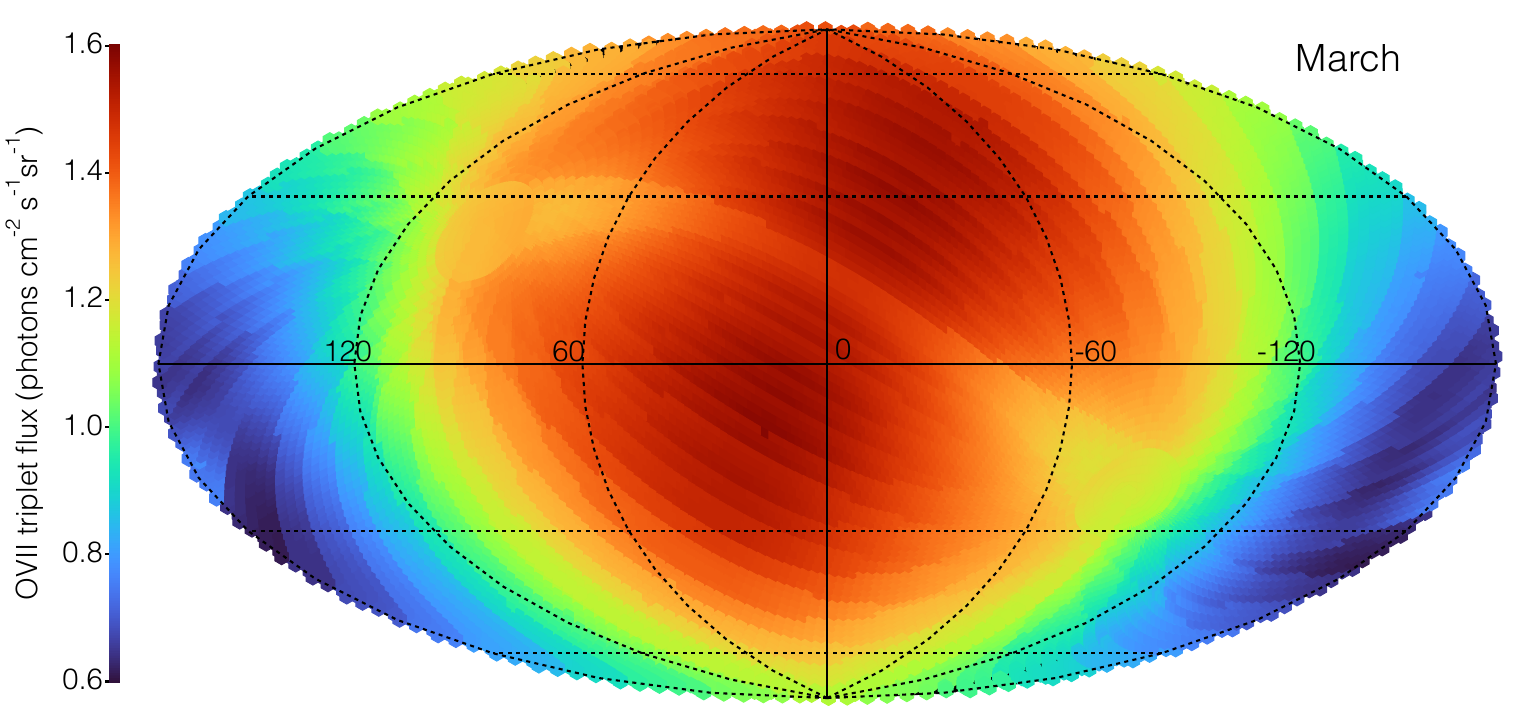}
    \includegraphics[width=\columnwidth]{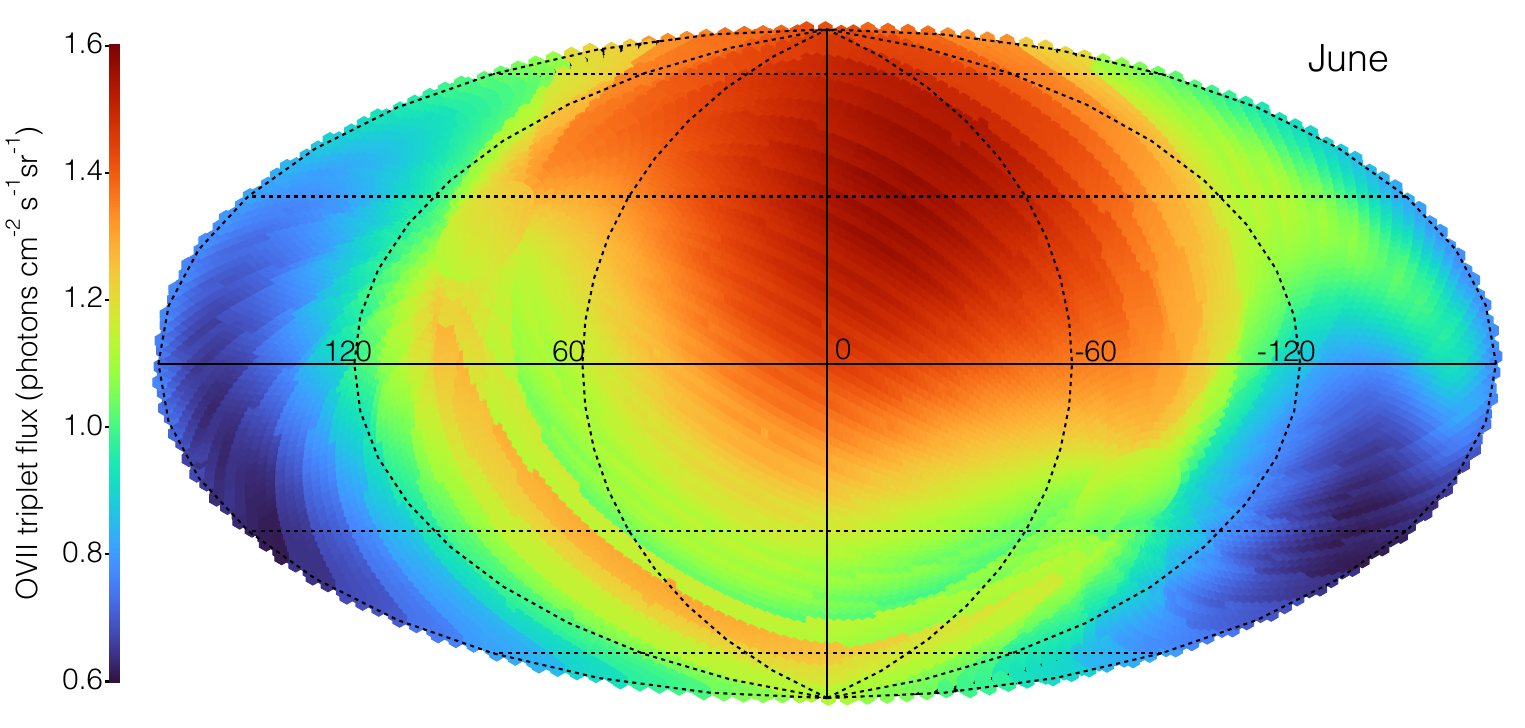}
    \includegraphics[width=\columnwidth]{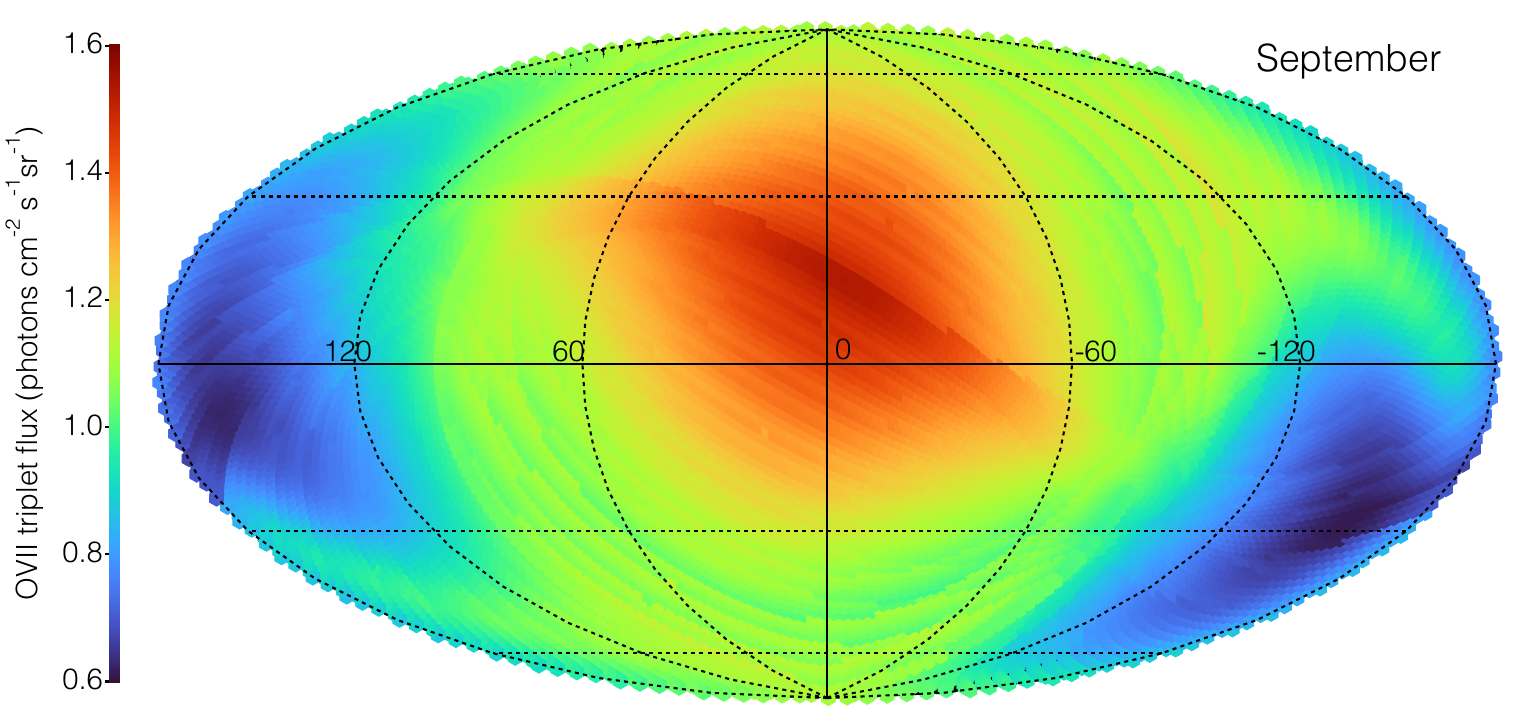}
    \includegraphics[width=\columnwidth]{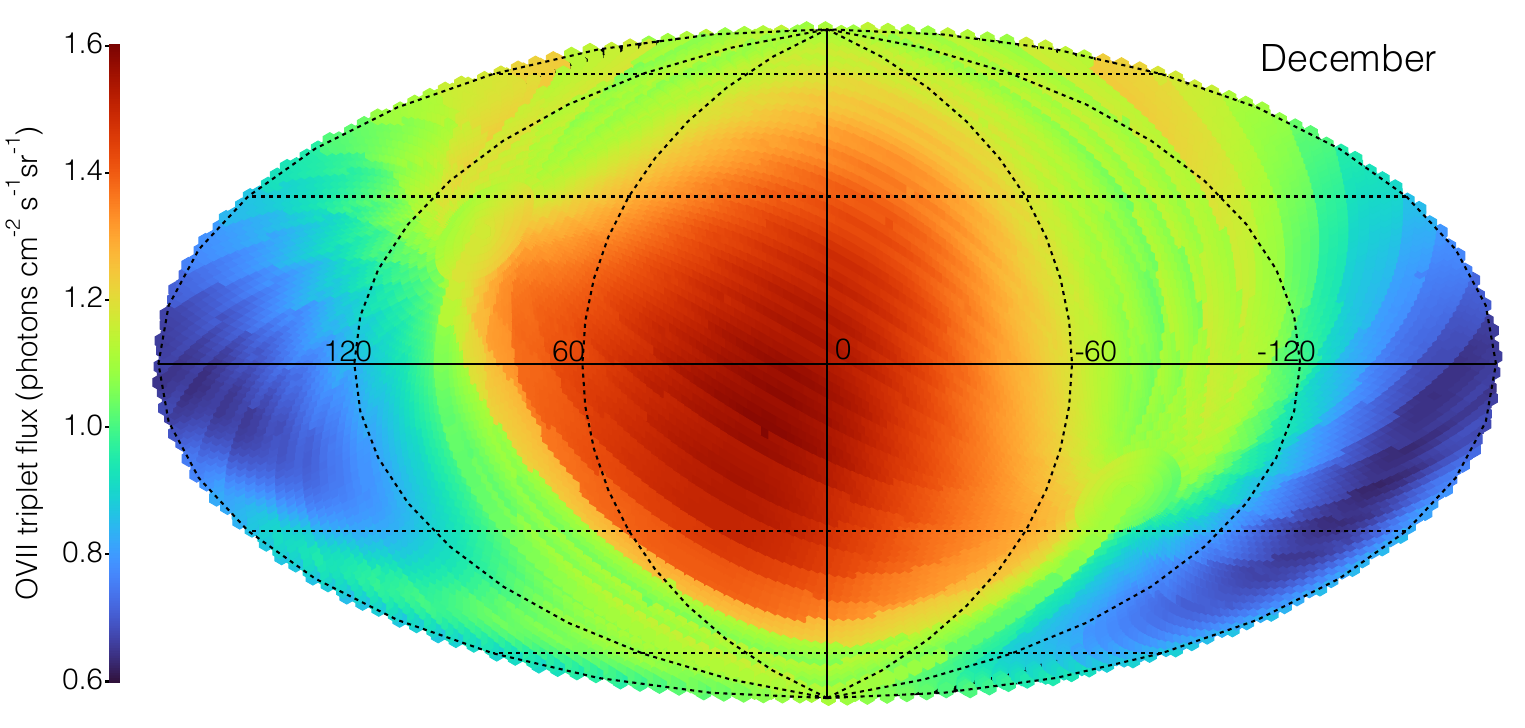}
    \caption{\textbf{Average \textit{LEM} All-Sky Survey OVII SWCX maps} in units of photons/cm$^2$/s/sr for steady-state slow solar wind conditions (solar maximum). The maps are in zero-centered galactic Aitoff coordinates with longitude increasing to the left. From top to bottom, the maps are calculated using a continuous 6-month period on \textit{LEM's} orbit starting in March, June, September and December, and respecting the Solar aspect angle constrains. The discontinuities are due to the change of observer position for each tile scan.}
    \label{fig:swcx_lass}
\end{figure}

For an observatory like \textit{LEM}, orbiting the Sun, an all-sky map of SWCX emission will appear very different depending on the position/time of year, since the map traces different neutral column densities (as shown in Fig. \ref{fig:SWCX_emissivity}) due to parallax effects \citep{2006A&A...460..289K}. In Fig. \ref{fig:swcx_lass} we present average all-sky maps of the SWCX OVII triplet emission, constructed over periods of 6 months (the time it will take \textit{LEM} for a complete scan of the sky) starting at different times of the year, highlighting the heliospheric parallax effects. These maps are calculated for steady-state slow solar wind parameters \citep{Koutroumpa2023}, roughly corresponding to solar maximum activity.

\begin{figure*}
\centering
\includegraphics[angle=0, width=2.0\columnwidth]{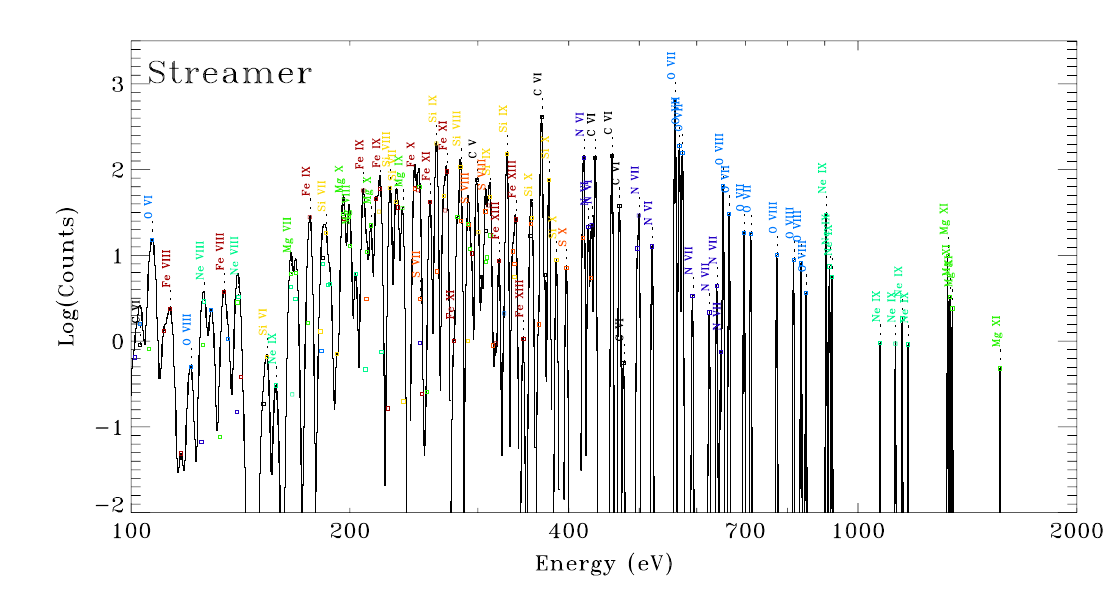}
\includegraphics[angle=0, width=2.0\columnwidth]{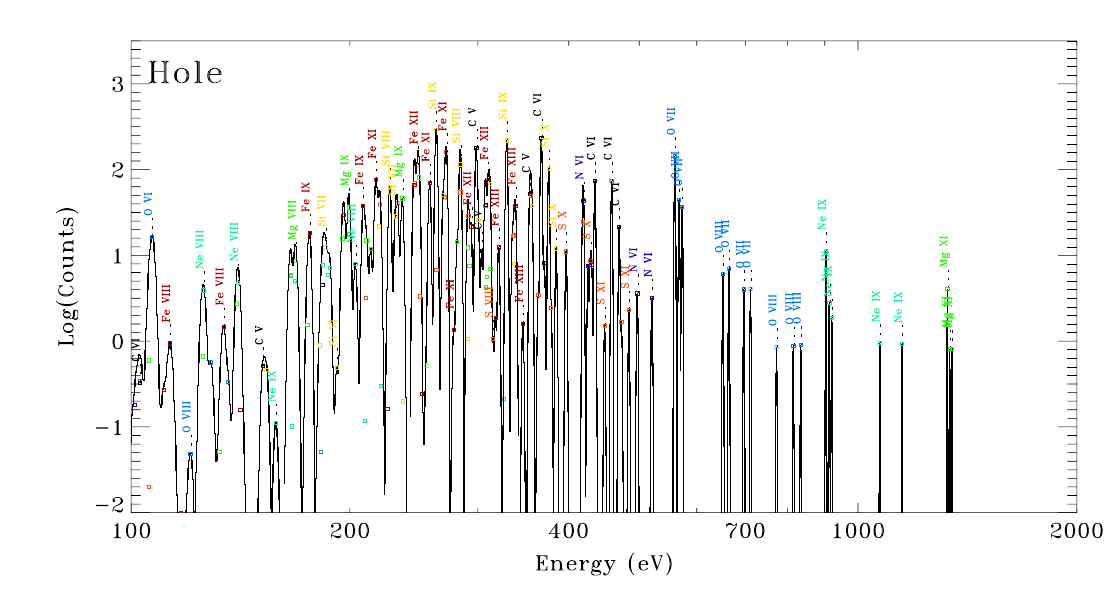}
\caption{\textbf{Theoretical heliospheric SWCX spectrum as observed by \textit{LEM}}: \textbf{Upper}: SWCX spectrum for streamer (slow) type of SW in counts for a 100 s observation of a $10^\circ$ by $10^\circ$ region. \textbf{Lower}: SWCX spectrum for coronal hole (fast) SW type (see text). The \lem\ response and line spread function has been applied. Major lines have been labelled. The boxes indicate the strength of an individual line; if it is not at the observed peak, then the observed line is strongly contaminated with other lines.}
\label{fig:swcx_SWtype}
\end{figure*}

In reality, the SWCX emission will have more complex characteristics, involving temporal variations due to the SW intrinsic variability, in addition to the spatial variations of the neutral distributions. The \textit{LEM} All-Sky Survey, in high resolution spectroscopic mode, will provide innovative diagnostics for the heliospheric SWCX science. A few key features that may be studied involve:

\textit{Resolved SWCX lines below 0.3 keV}: The bulk of the SWCX emission (and the Local Bubble) is produced below 0.3 keV, which current instruments cannot access due to limited sensitivity/grasp at those energies. \textit{LEM's} efficiency and spectral resolution will allow for the first time to identify the key spectral lines and measure their relative strengths (Figure~\ref{fig:swcx_SWtype}). 
   
\textit{Latitudinal distribution of the solar wind (SW)}: the solar wind has a sharp bimodal configuration during solar minimum, with a narrow zone of high density, high temperature, slow ($\sim$ 390 km/s) SW around the ecliptic/solar equator, while at high ecliptic latitudes large coronal holes emit low density, low temperature, fast (> 600 km/s) SW (e.g. Ulysses data in Figure 1 of~ \cite{2008GeoRL..3518103M}). The two SW types have considerably different ion composition that imprints on the SWCX spectral signal (Figure~\ref{fig:swcx_SWtype}). In the slow SW the high-charge-state ions (e.g. C$^{6+}$, N$^{7+}$, O$^{7, 8+}$, Ne$^{9+}$), that produce the higher energy spectral lines C VI, N VII, O VII, O VIII, and Ne IX, have higher relative abundances, while in the fast SW the high-charge-state ions abundances are lower. On the other hand, the lower charge states (e.g. C$^{5+}$, O$^{6+}$ producing C V, O VI) have opposite behaviour, with increased relative abundances in the fast SW. The composition changes between slow and fast SW result in a change of hardness ratio in the SWCX spectra. By comparing lines-of-sight near the ecliptic that are completely embedded in the slow solar wind, and high latitude lines-of-sight that cross mainly coronal hole type solar wind with only a small contribution of equatorial slow SW, we may provide a complementary analysis of the latitudinal distribution of the SW.  The extended mission that will encompass a large period of the solar cycle covering the maximum and minimum activity periods will provide particularly powerful diagnostic data. 

\textit{Localized enhancements/Doppler measurements due to ICMEs (Interplanetary Coronal Mass Ejections)}: ICMEs can have both very high velocities, and significant changes in SW ion composition \cite{Carter2010, 2004JGRA..109.9104R}. Detecting transient SWCX events, through localized enhancements and/or Doppler shift measurements, in repeated All-Sky Survey pointings of the same regions will put constraints on the SW propagation models used by the heliophysics community for space weather predictions at Earth, as well as to plan observations at Mars\cite{Dennerl2006}, Jupiter\cite{Dunn2016}, Saturn\cite{GBR2010} and Uranus\cite{Dunn2021}.

\begin{figure}
\centering
\includegraphics[angle=0, width=0.9\columnwidth]{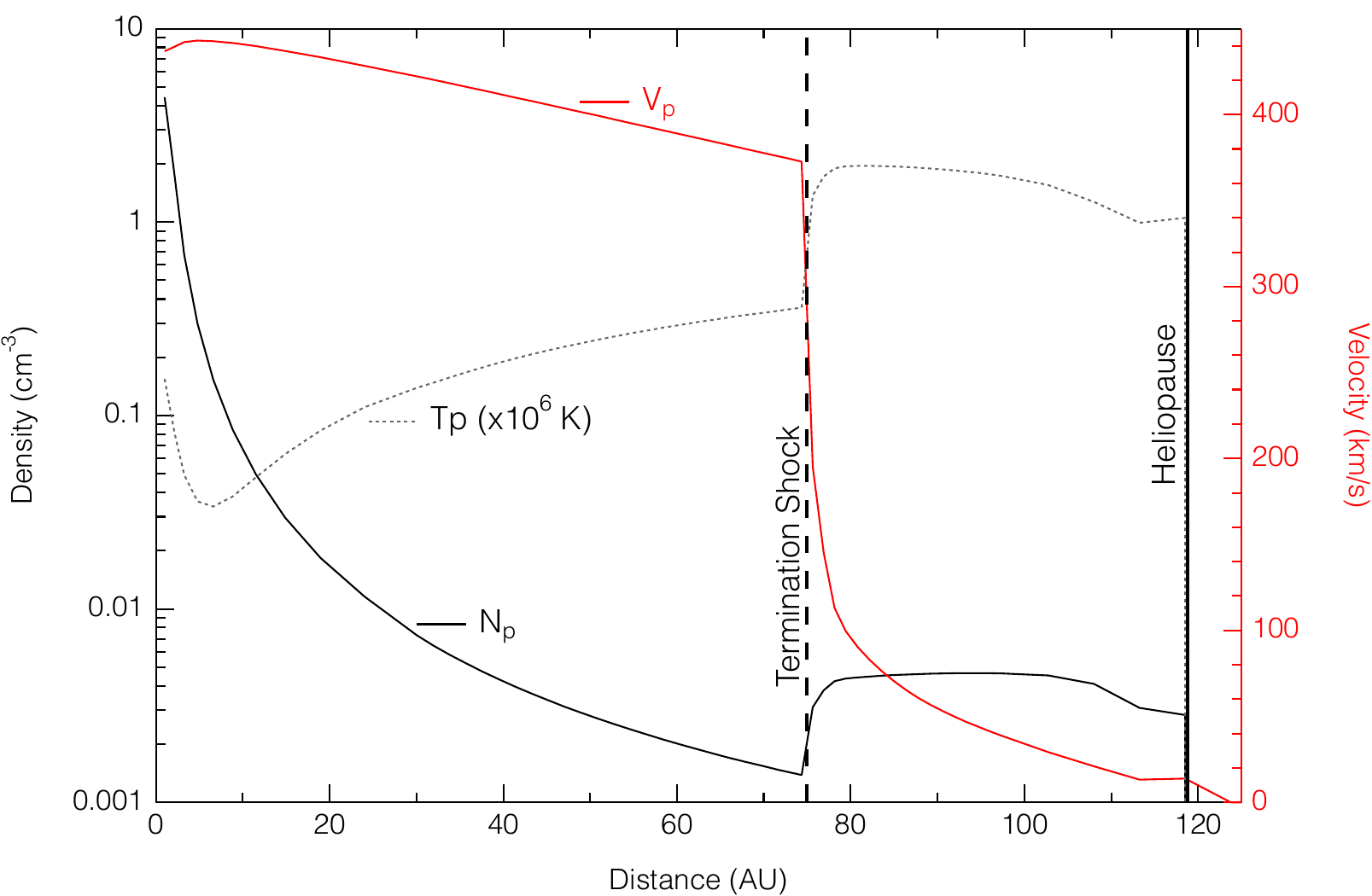}
\includegraphics[angle=0, width=0.9\columnwidth]{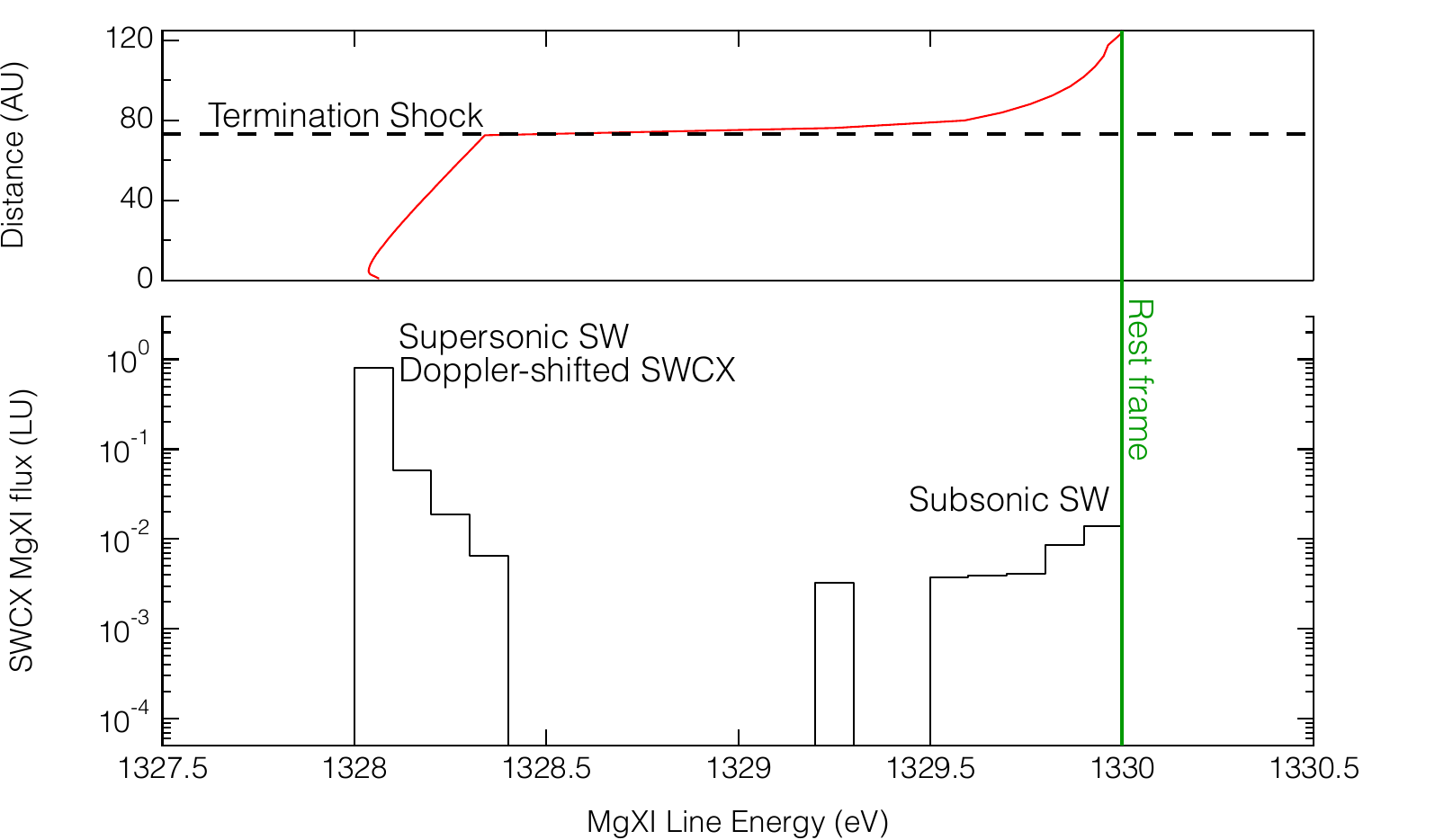}
\caption{\textbf{\textit{LEM}'s Capability to Resolve Doppler Shifts on SWCX Lines to Distinguish the Supersonic and Subsonic Solar Wind Populations Remotely:} \textit{Top}: Solar wind proton density (full black curve), temperature (dotted black curve), and velocity (full red curve; right axis) in the direction of the nose of the heliosphere, based on the model from \cite{2015ApJS..220...32I}. The vertical lines represent the termination shock (dashed) and heliopause (full). \textit{Bottom}: SWCX flux (in photons/cm$^2$/s/sr) of the Mg XI line at 1.33 keV, binned in 0.1 eV bins. The line is separated in two components, one Doppler-shifted, produced by supersonic solar wind ions up to $\sim$ 80 AU (top sub-panel), and one near the rest frame, produced by subsonic solar wind ions after they cross the termination shock. }
\label{fig:swcx_MgXI}
\end{figure}

\begin{figure}
\centering
\includegraphics[angle=0, trim=1cm 12cm 1cm 2.5cm, width=1.0\columnwidth]{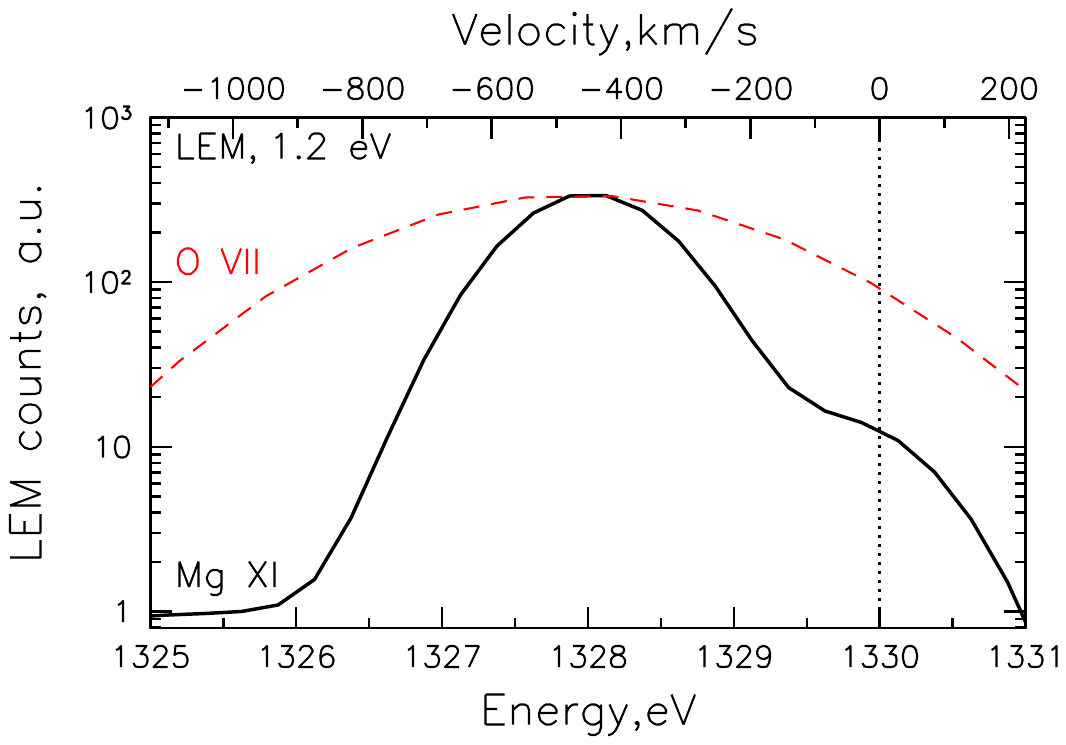}
\caption{\textbf{Exploration of the Importance of the Mg~XI SWCX Line as a Diagnostic of Solar Wind Conditions:} Convolution of the predicted velocity profile (shown in Figure \ref{fig:swcx_MgXI}) for the Solar Wind Charge Exchange (SWCX) Mg~XI line with \lem\, response function for the inner array (FWHM=1.2 eV). For comparison, the red dashed line shows the convolved SWCX velocity profile for O~VII,  demonstrating the unique power of the Mg~XI line for resolving the subsonic Solar wind component.}
\label{fig:swcx_MgXI_lem}
\end{figure}

\textit{Doppler measurements of the inner heliosphere and the outer heliosheath regions}: the heliospheric interface is a complex region of great interest for the physics of the solar wind interaction with the interstellar medium (cf. Voyagers’ studies of the termination shock and heliosheath \cite{2005Sci...309.2017S}, and future plans of an Interstellar Probe mission \cite{2023SSRv..219...18B}). At the termination shock the supersonic SW (400-800 km/s) decelerates abruptly to speeds <50 km/s, while it is heated from a few thousands K to nearly 2 MK (Fig.~\ref{fig:swcx_MgXI} - top). In the meanwhile the interstellar hydrogen population undergoes charge exchange with protons piled up in the heliosheath (region between the termination shock and the heliopause), resulting in a filtration and modification of its distribution (density, velocity, and temperature). The SWCX emission from the heliosheath will be near the rest-frame of the spectral line energies, while the inner heliosphere SWCX emission will be red-shifted by $\sim$ 2eV, especially for the higher energy lines (e.g. Mg~XI, Fig. \ref{fig:swcx_MgXI} - bottom). Measurements with \textit{LEM} (Fig. \ref{fig:swcx_MgXI_lem}) near the nose of the heliosphere $(l,b) \sim (3.2, 15.5)$, where the velocity change and thus the Doppler shift is the largest, with accumulated (but not necessarily continuous) exposures during the All-Sky Survey, will enable detection of the heliosheath SWCX emission and put constraints on the heliospheric interaction models.

\section{Concluding Remarks}

NASA's X-ray observatories have a rich legacy of transformative planetary studies and synergistic partnerships with in-situ spacecraft.  In  \textit{LEM}, the planetary science and heliophysics community have a new  instrument that appears bespoke to study planetary bodies and the wider heliosphere. The science that \textit{LEM} enables is diverse, spanning key fundamental strategic questions from the decadal studies \cite{PSDec,HelioDec}, while providing step-changes in our understanding of almost every planetary system within the Solar System.

Within this white paper we have provided an overview of some of the scientific studies that  \textit{LEM}  will enable, from its capacity to resolve key fluorescence lines from elements and molecules that offer valuable insights for Solar System formation and habitability, to the global understanding and context it can uniquely offer for the nature of the solar wind interaction with planetary bodies. It is clear that \textit{LEM} is a multi-faceted planetary and heliophysics mission that will provide an extensive scientific return for many fields and communities.



\section{References}
\small


\vspace{-6mm}
\parindent=0cm
\baselineskip=12pt

\begin{thebibliography}{100}
\bibitem{PSDec}
National Academies of Sciences, Engineering, and Medicine. (2022). Origins, worlds, and life: a decadal strategy for planetary science and astrobiology 2023-2032.

\bibitem{HelioDec}
Baker, D. N., Charo, A., \& Zurbuchen, T. (2013). Science for a technological society: The 2013–2022 decadal survey in solar and space physics. Space Weather, 11(2), 50-51.

\bibitem{Kraft22}
Kraft, R., Markevitch, M., Kilbourne, C., et al.\ (2022), The Line Emission Mapper White Paper, arXiv:2211.09827.

\bibitem{Adler1977}
Adler, I., J.I. Trombka, \ (1977), Orbital chemistry—lunar surface analysis from the X-ray and gamma ray remote sensing experiments, Phys. Chem. Earth, 10 (1977), p. 17

\bibitem{Adler1973}
Adler, I., J.I. Trombka, P. Lowman, R. Schmadebeck, H. Blodget, E. Eller, L. Yin, R. Lamothe, G. Osswakd, J. Gerard, P. Gorenstein, P. Bjorkhom, H. Gursky, B. Harris, J. Arnold, A. Metzer, R. Reedy, \ (1973),
Apollo 15 and 16 results of the integrated geochemical experiment The Moon, 7 (1973), p. 487

\bibitem{Andre1977}
Andre, C.G., M.J. Bielefeld, E. Eliason, L.A. Soderblom, I. Adler, J.A. Philpotts \ (1977), Lunar surface chemistry: a new imaging technique, Science, 197 (1977), p. 986

\bibitem{Atreya2022}
Atreya, S. K., Crida, A., Guillot, T., Li, C., Lunine, J. I., Madhusudhan, N., ... \& Wong, M. H. (2022). The origin and evolution of Saturn: A post-Cassini perspective. arXiv preprint arXiv:2205.06914.

\bibitem{Bertucci2008}
Bertucci, C., Achilleos, N., Dougherty, M. K., Modolo, R., Coates, A. J., Szego, K., ... \& Young, D. T. (2008). The magnetic memory of Titan's ionized atmosphere. Science, 321(5895), 1475-1478.


\bibitem{Bhardwaj2005Ring}
Bhardwaj, A., Elsner, R. F., Waite Jr, J. H., Gladstone, G. R., Cravens, T. E., \& Ford, P. G. (2005). The discovery of oxygen K$\alpha$ X-ray emission from the rings of Saturn. The Astrophysical Journal, 627(1), L73.

\bibitem{Bhardwaj2005SolarJupiter}
Bhardwaj, A., Branduardi‐Raymont, G., Elsner, R. F., Gladstone, G. R., Ramsay, G., Rodriguez, P., ... \& Cravens, T. E. (2005). Solar control on Jupiter's equatorial X‐ray emissions: 26–29 November 2003 XMM‐Newton observation. Geophysical Research Letters, 32(3).

\bibitem{Bhardwaj2005SolarSaturn}
Bhardwaj, A., Elsner, R. F., Waite Jr, J. H., Gladstone, G. R., Cravens, T. E., \& Ford, P. G. (2005). Chandra observation of an X-ray flare at Saturn: evidence of direct solar control on Saturn’s disk X-ray emissions. The Astrophysical Journal, 624(2), L121.

\bibitem{Bhardwaj2006}
Bhardwaj, A., Elsner, R. F., Gladstone, G. R., Waite Jr, J. H., Branduardi‐Raymont, G., Cravens, T. E., \& Ford, P. G. (2006). Low‐to middle‐latitude X‐ray emission from Jupiter. Journal of Geophysical Research: Space Physics, 111(A11).

\bibitem{Bhardwaj2007}
Bhardwaj, A., Elsner, R. F., Gladstone, G. R., Cravens, T. E., Lisse, C. M., Dennerl, K., ... \& Kharchenko, V. (2007). X-rays from solar system objects. Planetary and Space Science, 55(9), 1135-1189.

\bibitem{Bierhaus2018}
Bierhaus, E. B., Clark, B. C., Harris, J. W., Payne, K. S., Dubisher, R. D., Wurts, D. W., ... \& OSIRIS-REx Team. (2018). The OSIRIS-REx spacecraft and the touch-and-go sample acquisition mechanism (TAGSAM). Space Science Reviews, 214, 1-46.

\bibitem{Biermann1967}
Biermann, L., Brosowski, B., \& Schmidt, H. U. (1967). The interaction of the solar wind with a comet. Solar Physics, 1, 254-284.

\bibitem{Bodewits2004}
Bodewits, D., Juhász, Z., Hoekstra, R., \& Tielens, A. G. G. M. (2004). Catching some Sun: Probing the solar wind with cometary X-ray and far-ultraviolet emission. The Astrophysical Journal, 606(1), L81.

\bibitem{Bodewits2007}
Bodewits, D., Christian, D. J., Torney, M., Dryer, M., Lisse, C. M., Dennerl, K., ... \& Hoekstra, R. (2007). Spectral analysis of the Chandra comet survey. Astronomy \& Astrophysics, 469(3), 1183-1195.

\bibitem{Bolton2017}
Bolton, S. J., Adriani, A., Adumitroaie, V., Allison, M., Anderson, J., Atreya, S., ... \& Wilson, R. (2017). Jupiter’s interior and deep atmosphere: The initial pole-to-pole passes with the Juno spacecraft. Science, 356(6340), 821-825.

\bibitem[Brandt et al.(2023)]{2023SSRv..219...18B} Brandt, P.~C., Provornikova, E., Bale, S.~D., et al.\ 2023, \ssr, 219, 18. \url{https://doi.org/10.1007/s11214-022-00943-x}


\bibitem{GBR2004}
Branduardi-Raymont, G., Elsner, R. F., Gladstone, G. R., Ramsay, G., Rodriguez, P., Soria, R., \& Waite, J. H. (2004). First observation of Jupiter by XMM-Newton. Astronomy \& Astrophysics, 424(1), 331-337.

\bibitem{GBR2007Aurora}
Branduardi-Raymont, G., Bhardwaj, A., Elsner, R. F., Gladstone, G. R., Ramsay, G., Rodriguez, P., ... \& Cravens, T. E. (2007). A study of Jupiter's aurorae with XMM-Newton. Astronomy \& Astrophysics, 463(2), 761-774.

\bibitem{GBR2007Disk}
Branduardi-Raymont, G., Bhardwaj, A., Elsner, R. F., Gladstone, G. R., Ramsay, G., Rodriguez, P., ... \& Cravens, T. E. (2007). Latest results on Jovian disk X-rays from XMM-Newton. Planetary and Space Science, 55(9), 1126-1134.

\bibitem{GBR2008}
Branduardi‐Raymont, G., Elsner, R. F., Galand, M., Grodent, D., Cravens, T. E., Ford, P., ... \& Waite Jr, J. H. (2008). Spectral morphology of the X‐ray emission from Jupiter's aurorae. Journal of Geophysical Research: Space Physics, 113(A2).

\bibitem{GBR2010}
Branduardi-Raymont, G., Bhardwaj, A., Elsner, R. F., \& Rodriguez, P. (2010). X-rays from Saturn: a study with XMM-Newton and Chandra over the years 2002–05. Astronomy \& Astrophysics, 510, A73.

\bibitem{GBR2018}
Branduardi-Raymont, C. Wang, C. P. Escoubet et al., SMILE Definition Study Report, European Space Agency, ESA/SCI, 1, 2018; https://doi.org/10.5270/esa.smile.definition\_study\_report-2018-12

\bibitem{GBR2022}
Branduardi‐Raymont, G. (2022). X‐ray views of our solar system. Astronomische Nachrichten, 343(4), e210101.

\bibitem{Brown2013}
Brown, M. E., \& Hand, K. P. (2013). Salts and radiation products on the surface of Europa. The Astronomical Journal, 145(4), 110.

\bibitem{Bunce2008}
Bunce, E. J., Arridge, C. S., Clarke, J. T., Coates, A. J., Cowley, S. W. H., Dougherty, M. K., ... \& Talboys, D. L. (2008). Origin of Saturn's aurora: Simultaneous observations by Cassini and the Hubble Space Telescope. Journal of Geophysical Research: Space Physics, 113(A9).

\bibitem{Bunce2020}
Bunce, E. J., Martindale, A., Lindsay, S., Muinonen, K., Rothery, D. A., Pearson, J., ... \& Yeoman, T. (2020). The BepiColombo mercury imaging X-ray spectrometer: Science goals, instrument performance and operations. Space Science Reviews, 216, 1-38.

\bibitem{CarterMars}
J. A. Carter, K. Dennerl, W. Dunn, D. Bodewits, A. Bogdan, G. Branduardi-Raymont, Y. Ezoe, C. Feldman, A. Foster, R. Gladstone, C. Jackman, D. Koutroumpa, R. Kraft, K. Kuntz, C. Lisse, S. McEntee, J.-U. Ness, F. S. Porter, S. Wolk, S. F. Sembay, (in prep), The exosphere of Mars can tracked by a high-spectral resolution telescope, such as the Line Emission Mapper

\bibitem{Carter2008}
Carter, J. A., \& Sembay, S. (2008). Identifying XMM-Newton observations affected by solar wind charge exchange. Part I. Astronomy \& Astrophysics, 489(2), 837-848.

\bibitem{Carter2010}
Carter, J. A., Sembay, S., \& Read, A. M. (2010). A high charge state coronal mass ejection seen through solar wind charge exchange emission as detected by XMM–Newton. Monthly Notices of the Royal Astronomical Society, 402(2), 867-878.

\bibitem{Carter2011}
Carter, J. A., Sembay, S., \& Read, A. M. (2011). Identifying XMM-Newton observations affected by solar wind charge exchange–Part II. Astronomy \& Astrophysics, 527, A115.

\bibitem{Clarke1998}
Clarke, J. T., Ballester, G., Trauger, J., Ajello, J., Pryor, W., Tobiska, K., ... \& Gérard, J. C. (1998). Hubble Space Telescope imaging of Jupiter's UV aurora during the Galileo orbiter mission. Journal of Geophysical Research: Planets, 103(E9), 20217-20236.

\bibitem{Clark2020}
Clark, G., Mauk, B. H., Kollmann, P., Paranicas, C., Bagenal, F., Allen, R. C., ... \& Westlake, J. H. (2020). Heavy ion charge states in Jupiter's polar magnetosphere inferred from auroral megavolt electric potentials. Journal of Geophysical Research: Space Physics, 125(9), e2020JA028052.

\bibitem{Cravens1997}
Cravens, T. E. (1997). Comet Hyakutake X‐ray source: Charge transfer of solar wind heavy ions. Geophysical Research Letters, 24(1), 105-108.

\bibitem{Cravens2005}
Cravens, T. E., Howell, E., Waite Jr, J. H., \& Gladstone, G. R. (1995). Auroral oxygen precipitation at Jupiter. Journal of Geophysical Research: Space Physics, 100(A9), 17153-17161.


\bibitem{Cravens2001}
Cravens, T. E., \& Maurellis, A. N. (2001). X‐ray emission from scattering and fluorescence of solar X‐rays at Venus and Mars. Geophysical research letters, 28(15), 3043-3046.

\bibitem{Cravens2003}
Cravens, T. E., Waite, J. H., Gombosi, T. I., Lugaz, N., Gladstone, G. R., Mauk, B. H., \& MacDowall, R. J. (2003). Implications of Jovian X‐ray emission for magnetosphere‐ionosphere coupling. Journal of Geophysical Research: Space Physics, 108(A12).

\bibitem{Cravens2006}
Cravens, T. E., Clark, J., Bhardwaj, A., Elsner, R., Waite Jr, J. H., Maurellis, A. N., ... \& Branduardi‐Raymont, G. (2006). X‐ray emission from the outer planets: Albedo for scattering and fluorescence of solar X rays. Journal of Geophysical Research: Space Physics, 111(A7).


\bibitem{Cowley2008}
Cowley, S. W. H., Badman, S. V., Imber, S. M., \& Milan, S. E. (2008). Comment on" Jupiter: A fundamentally different magnetospheric interaction with the solar wind" by DJ McComas and F. Bagenal. Geophysical Research Letters, 35(10).

\bibitem{Demidova2007}
S.I. Demidova, M.A. Nazarov, C.A. Lorenz, G. Kurat, F. Brandsttter, Th. Ntaflos (2007),  Chemical composition of lunar meteorites and the lunar crust, Petrology, 15 (2007), pp. 386-407

\bibitem{Dennerl1997}
Dennerl, K., Englhauser, J., \& Trumper, J. (1997). X-ray emissions from comets detected in the Rontgen X-ray satellite all-sky survey. Science, 277(5332), 1625-1630.

\bibitem{Dennerl2002Mars}
Dennerl, K. (2002). Discovery of X–rays from Mars with Chandra. Astronomy \& Astrophysics, 394(3), 1119-1128.

\bibitem{Dennerl2002Venus}
Dennerl, K., Burwitz, V., Englhauser, J., Lisse, C., \& Wolk, S. (2002). Discovery of X–rays from Venus with Chandra. Astronomy \& Astrophysics, 386(1), 319-330.

\bibitem{Dennerl2006}
Dennerl, K., Lisse, C. M., Bhardwaj, A., Burwitz, V., Englhauser, J., Gunell, H., ... \& Rodríguez-Pascual, P. M. (2006). First observation of Mars with XMM-Newton-High resolution X-ray spectroscopy with RGS. Astronomy \& Astrophysics, 451(2), 709-722.

\bibitem{Dennerl2008}
Dennerl, K. (2008). X-rays from Venus observed with Chandra. Planetary and Space Science, 56(10), 1414-1423.

\bibitem{Dennerl2010}
Dennerl, K. (2010). Charge transfer reactions. Space Science Reviews, 157, 57-91.

\bibitem{Dougherty2009}
Dougherty, M., \& Esposito, L. (2009). Saturn from Cassini-huygens.

\bibitem{Dungey1961}
Dungey, J. W. (1961). Interplanetary magnetic field and the auroral zones. Physical Review Letters, 6(2), 47.

\bibitem{Dunn2016}
Dunn, W. R., Branduardi‐Raymont, G., Elsner, R. F., Vogt, M. F., Lamy, L., Ford, P. G., ... \& Jasinski, J. M. (2016). The impact of an ICME on the Jovian X‐ray aurora. Journal of Geophysical Research: Space Physics, 121(3), 2274-2307.

\bibitem{Dunn2020a}
Dunn, W. R., Branduardi‐Raymont, G., Carter‐Cortez, V., Campbell, A., Elsner, R., Ness, J. U., ... \& Achilleos, N. (2020). Jupiter's X‐ray emission during the 2007 solar minimum. Journal of Geophysical Research: Space Physics, 125(6), e2019JA027219.

\bibitem{Dunn2020b}
Dunn, W. R., Gray, R., Wibisono, A. D., Lamy, L., Louis, C., Badman, S. V., ... \& Kraft, R. (2020). Comparisons between Jupiter's X‐ray, UV and radio emissions and in‐situ solar wind measurements during 2007. Journal of Geophysical Research: Space Physics, 125(6), e2019JA027222.

\bibitem{Dunn2021}
Dunn, W. R., Ness, J. U., Lamy, L., Tremblay, G. R., Branduardi‐Raymont, G., Snios, B., ... \& Wibisono, A. D. (2021). A low signal detection of x‐rays from uranus. Journal of Geophysical Research: Space Physics, 126(4), e2020JA028739.

\bibitem{Dunn2022Jupiter}
Dunn, W.R. (2022). X-ray Emissions from the Jovian System. In: Bambi, C., Santangelo, A. (eds) Handbook of X-ray and Gamma-ray Astrophysics. Springer, Singapore. https://doi.org/10.1007/978-981-16-4544-0\_73-1

\bibitem{Dunn2022Ice}
Dunn, W. R. (2022). X-ray Emissions from the Ice Giants and Kuiper Belt. In Handbook of X-ray and Gamma-ray Astrophysics (pp. 1-23). Singapore: Springer Nature Singapore.

\bibitem{Dunn2022XUV}
Dunn, W. R., Weigt, D. M., Grodent, D., Yao, Z. H., May, D., Feigelman, K., ... \& Ray, L. C. (2022). Jupiter's X‐Ray and UV Dark Polar Region. Geophysical Research Letters, 49(11), e2021GL097390.

\bibitem{Dunn2023}
Dunn, W., Berland, G., Roussos, E., Clark, G., Kollmann, P., Turner, D., ... \& Kraft, R. P. (2023). Exploring Fundamental Particle Acceleration and Loss Processes in Heliophysics through an Orbiting X-ray Instrument in the Jovian System. arXiv preprint arXiv:2303.02161.

\bibitem{Elsner2002}
Elsner, R. F., Gladstone, G. R., Waite, J. H., Crary, F. J., Howell, R. R., Johnson, R. E., ... \& Weisskopf, M. C. (2002). Discovery of soft X-ray emission from Io, Europa, and the Io plasma torus. The Astrophysical Journal, 572(2), 1077.

\bibitem{Elsner2005}
Elsner, R. F., Lugaz, N., Waite Jr, J. H., Cravens, T. E., Gladstone, G. R., Ford, P., ... \& Majeed, T. (2005). Simultaneous Chandra X ray, Hubble Space Telescope ultraviolet, and Ulysses radio observations of Jupiter's aurora. Journal of Geophysical Research: Space Physics, 110(A1).

\bibitem{Fletcher2020}
Fletcher, L. N., Helled, R., Roussos, E., Jones, G., Charnoz, S., André, N., ... \& Turrini, D. (2020). Ice Giant Systems: The scientific potential of orbital missions to Uranus and Neptune. Planetary and Space Science, 191, 105030.

\bibitem{Futaana2017}
Futaana, Y., Stenberg Wieser, G., Barabash, S., \& Luhmann, J. G. (2017). Solar wind interaction and impact on the Venus atmosphere. Space Science Reviews, 212, 1453-1509.

\bibitem{Galeazzi2014} Galeazzi, M., Chiao, M., Collier, M.~R., et al.\ 2014, \nat, 512, 171. doi:10.1038/nature13525

\bibitem{Garvin2020}
Garvin, J., Getty, S., Arney, G., Johnson, N., Kohler, E., Schwer, K., ... \& Sekerak, M. (2020, March). DAVINCI+: Deep Atmosphere Venus Investigation of Noble gasses, Chemistry, and Imaging Plus. In Venus Exploration Analysis Group (VEXAG).

\bibitem{Gladstone1998}
Gladstone, G. R., Waite Jr, J. H., \& Lewis, W. S. (1998). Secular and local time dependence of Jovian X ray emissions. Journal of Geophysical Research: Planets, 103(E9), 20083-20088.


\bibitem{Gladstone2002}
Gladstone, G. R., Waite Jr, J. H., Grodent, D., Lewis, W. S., Crary, F. J., Elsner, R. F., ... \& Cravens, T. E. (2002). A pulsating auroral X-ray hot spot on Jupiter. Nature, 415(6875), 1000-1003.

\bibitem{Gloeckler2007}
Gloeckler, G., \& Geiss, J. (2007). The composition of the solar wind in polar coronal holes. The Composition of Matter, 139-152.

\bibitem{Goswami2009}
Goswami, J. N., \& Annadurai, M. (2009). Chandrayaan-1: India's first planetary science mission to the moon. Current science, 486-491.

\bibitem{Grande2009}
Grande, M., Maddison, B. J., Howe, C. J., Kellett, B. J., Sreekumar, P., Huovelin, J., ... \& Wieczorek, M. (2009). The C1XS X-ray spectrometer on Chandrayaan-1. Planetary and Space Science, 57(7), 717-724.

\bibitem{Haskin1991}
L. Haskin, P. Warren (1991), Lunar Source Book: A User’s Guide to the Moon, Cambridge Univ. Press, Cambridge, England (1991), pp. 357–474


\bibitem{Hitchcock1987}
Hitchcock, A. P., \& Ishii, I. (1987). Carbon K-shell excitation spectra of linear and branched alkanes. Journal of Electron Spectroscopy and Related Phenomena, 42(1), 11-26.

\bibitem[Izmodenov \& Alexashov(2015)]{2015ApJS..220...32I} Izmodenov, V.~V. \& Alexashov, D.~B.\ 2015, \apjs, 220, 32. \url{https://doi.org/10.1088/0067-0049/220/2/32}


\bibitem{Jakosky2001}
Jakosky, B. M., \& Phillips, R. J. (2001). Mars' volatile and climate history. nature, 412(6843), 237-244.

\bibitem{Jakosky2015}
Jakosky, B. M., Lin, R. P., Grebowsky, J. M., Luhmann, J. G., Mitchell, D. F., Beutelschies, G., ... \& Zurek, R. (2015). The Mars atmosphere and volatile evolution (MAVEN) mission. Space Science Reviews, 195, 3-48.

\bibitem{Jia2018}
Jia, X., Kivelson, M. G., Khurana, K. K., \& Kurth, W. S. (2018). Evidence of a plume on Europa from Galileo magnetic and plasma wave signatures. Nature Astronomy, 2(6), 459-464.

\bibitem{Kato2010}
Kato, M., Sasaki, S., Takizawa, Y., \& Kaguya Project Team. (2010). The Kaguya mission overview. Space science reviews, 154, 3-19.

\bibitem{Kollmann2021}
Kollmann, P., Clark, G., Paranicas, C., Mauk, B., Roussos, E., Nénon, Q., ... \& Rymer, A. (2021). Jupiter's ion radiation belts inward of Europa's orbit. Journal of Geophysical Research: Space Physics, 126(4), e2020JA028925.

\bibitem{Koutroumpa2010}
Koutroumpa, D., Smith, R. K., Edgar, R. J., Kuntz, K. D., Plucinsky, P. P., \& Snowden, S. L. (2010). XMM-Newton observations of MBM 12: more constraints on the solar wind charge exchange and local bubble emissions. The Astrophysical Journal, 726(2), 91.


\bibitem[Koutroumpa(2012)]{2012AN....333..341K} Koutroumpa, D.\ 2012, Astronomische Nachrichten, 333, 341. \url{https://doi.org/10.1002/asna.201211666}

\bibitem[Koutroumpa et al.(2006)]{2006A&A...460..289K} Koutroumpa, D., Lallement, R., Kharchenko, V., et al.\ 2006, \aap, 460, 289. \url{https://doi.org/10.1051/0004-6361:20065250}

\bibitem[Koutroumpa(2023)]{Koutroumpa2023} Koutroumpa, D.\ 2023, Earth Planet. Phys., 8(1), 1–14. \url{https://doi.org/10.26464/epp2023056}

\bibitem{Krasnopolsky2004}
Krasnopolsky, V. A., Greenwood, J. B., \& Stancil, P. C. (2004). X-ray and extreme ultraviolet emissions from comets. Space science reviews, 113, 271-373.

\bibitem{Kuntz2023}
Küntz, K. D., Koutroumpa, D., Dunn, W. R., Foster, A., Porter, F. S., Sibeck, D. G., \& Walsh, B. (2023). The magnetosheath at high spectral resolution. Earth and Planetary Physics.

\bibitem{Kuntz2015}
{Kuntz}, K.~D., {Collado-Vega}, Y.~M., {Collier}, M.~R., {Connor}, H.~K., {Cravens}, T.~E., {Koutroumpa}, D., {Porter}, F.~S., {Robertson}, I.~P., {Sibeck}, D.~G., {Snowden}, S.~L., {Thomas}, N.~E., {Walsh}, B.~M., (2015). "The Solar Wind Charge-exchange Production Factor for Hydrogen." The Astrophysical Journal 808, no. 2, 143.

\bibitem{Kuntz1997}
Kuntz, K. D., S. L. Snowden, and F. Verter (1997). "X-ray Shadows by High-latitude Molecular Clouds. I. Cartography." The Astrophysical Journal 484, no. 1, 245.



\bibitem[Lallement et al.(2005)]{2005Sci...307.1447L} Lallement, R., Qu{\'e}merais, E., Bertaux, J.~L., et al.\ 2005, Science, 307, 1447. \url{https://doi.org/10.1126/science.1107953}

\bibitem{Lamy2017}
Lamy, L., Prangé, R., Hansen, K. C., Tao, C., Cowley, S. W. H., Stallard, T. S., ... \& Pogorelov, N. (2017). The aurorae of Uranus past equinox. Journal of Geophysical Research: Space Physics, 122(4), 3997-4008.


\bibitem{Lebreton2005}
Lebreton, J. P., Witasse, O., Sollazzo, C., Blancquaert, T., Couzin, P., Schipper, A. M., ... \& Pérez-Ayúcar, M. (2005). An overview of the descent and landing of the Huygens probe on Titan. Nature, 438(7069), 758-764.

\bibitem{Linder1998}
Linder, D. R., Coates, A. J., Woodliffe, R. D., Alsop, C., Johnstone, A. D., Grande, M., ... \& Young, D. T. (1998). The Cassini CAPS electron spectrometer. Measurement techniques in space plasmas: Particles, 102, 257-262.

\bibitem{Lindsay2016}
Lindsay, S. T., James, M. K., Bunce, E. J., Imber, S. M., Korth, H., Martindale, A., \& Yeoman, T. K. (2016). MESSENGER X-ray observations of magnetosphere–surface interaction on the nightside of Mercury. Planetary and Space Science, 125, 72-79.

\bibitem{Lisse1996}
Lisse, C. M., Dennerl, K., Englhauser, J., Harden, M., Marshall, F. E., Mumma, M. J., ... \& West, R. G. (1996). Discovery of X-ray and extreme ultraviolet emission from comet C/Hyakutake 1996 B2. Science, 274(5285), 205-209

\bibitem{Lisse1999}
Lisse, C. M., Christian, D., Dennerl, K., Englhauser, J., Trümper, J., Desch, M., ... \& Snowden, S. (1999). X-ray and extreme ultraviolet emission from comet P/Encke 1997. Icarus, 141(2), 316-330.

\bibitem{Lisse2001}
Lisse, C. M., Christian, D. J., Dennerl, K., Meech, K. J., Petre, R., Weaver, H. A., \& Wolk, S. J. (2001). Charge exchange-induced X-ray emission from Comet C/1999 S4 (LINEAR). Science, 292(5520), 1343-1348.

\bibitem{Lisse2005}
Lisse, C. M., Christian, D. J., Dennerl, K., Wolk, S. J., Bodewits, D., Hoekstra, R., ... \& Weaver, H. (2005). Chandra observations of comet 2P/Encke 2003: First detection of a collisionally thin, fast solar wind charge exchange system. The Astrophysical Journal, 635(2), 1329.

\bibitem{Lisse2017}
Lisse, C. M., McNutt Jr, R. L., Wolk, S. J., Bagenal, F., Stern, S. A., Gladstone, G. R., ... \& Ennico, K. A. (2017). The puzzling detection of X-rays from Pluto by Chandra. Icarus, 287, 103-109.

\bibitem{Mason2020}
Mason, J. P., Woods, T. N., Chamberlin, P. C., Jones, A., Kohnert, R., Schwab, B., ... \& Warren, H. (2020). MinXSS-2 CubeSat mission overview: Improvements from the successful MinXSS-1 mission. Advances in Space Research, 66(1), 3-9.

\bibitem{Masters2018}
Masters, A. (2018). A more viscous‐like solar wind interaction with all the giant planets. Geophysical Research Letters, 45(15), 7320-7329.

\bibitem{MaukFox2010}
Mauk, B. H., \& Fox, N. J. (2010). Electron radiation belts of the solar system. Journal of Geophysical Research: Space Physics, 115(A12).

\bibitem{Maurellis2000}
Maurellis, A. N., Cravens, T. E., Gladstone, G. R., Waite, J. H., \& Acton, L. W. (2000). Jovian X‐ray emission from solar X‐ray scattering. Geophysical Research Letters, 27(9), 1339-1342.

\bibitem{McComas2007}
McComas, D. J., \& Bagenal, F. (2007). Jupiter: A fundamentally different magnetospheric interaction with the solar wind. Geophysical Research Letters, 34(20).



\bibitem[McComas et al.(2008)]{2008GeoRL..3518103M} McComas, D.~J., Ebert, R.~W., Elliott, H.~A., et al.\ 2008, \grl, 35, L18103. \url{https://doi.org/10.1029/2008GL034896}


\bibitem{McEntee2022}
McEntee, S. C., Jackman, C. M., Weigt, D. M., Dunn, W. R., Kashyap, V., Kraft, R., ... \& Gallagher, P. T. (2022). Comparing Jupiter’s Equatorial X‐Ray Emissions With Solar X‐Ray Flux Over 19 Years of the Chandra Mission. Journal of Geophysical Research: Space Physics, 127(12), e2022JA030971.

\bibitem{Mclaren1987}
McLaren, R., Clark, S. A. C., Ishii, I., \& Hitchcock, A. P. (1987). Absolute oscillator strengths from K-shell electron-energy-loss spectra of the fluoroethenes and 1, 3-perfluorobutadiene. Physical Review A, 36(4), 1683.

\bibitem{Morris1983}
Morris, R.V., Score, R., Dardano, C., Heiken, G., 1983. Handbook of Lunar Soils. JSC Publ. No. 19069. Planetary Materials Branch Publ. 67. NASA Johnson Space Center, Houston. p. 914.

\bibitem{Moore2018}
Moore, K. M., Yadav, R. K., Kulowski, L., Cao, H., Bloxham, J., Connerney, J. E., ... \& Levin, S. M. (2018). A complex dynamo inferred from the hemispheric dichotomy of Jupiter’s magnetic field. Nature, 561(7721), 76-78.

\bibitem{Narendranath2011}
Narendranath, S., Athiray, P. S., Sreekumar, P., Kellett, B. J., Alha, L., Howe, C. J., ... \& Wieczorek, M. A. (2011). Lunar X-ray fluorescence observations by the Chandrayaan-1 X-ray Spectrometer (C1XS): Results from the nearside southern highlands. Icarus, 214(1), 53-66.

\bibitem{Ness2000}
Ness, J. U., \& Schmitt, J. H. (2000). A search for X-ray emission from Saturn, Uranus and Neptune. arXiv preprint astro-ph/0001131.

\bibitem{Ness2004a}
Ness, J. U., Schmitt, J. H. M. M., Wolk, S. J., Dennerl, K., \& Burwitz, V. (2004). X-ray emission from Saturn. Astronomy \& Astrophysics, 418(1), 337-345.

\bibitem{Ness2004b}
Ness, J. U., Schmitt, J. H. M. M., \& Robrade, J. (2004). Detection of Saturnian X-ray emission with XMM-Newton. Astronomy \& Astrophysics, 414(3), L49-L52.

\bibitem{Nittler2020}
Nittler, L. R., Frank, E. A., Weider, S. Z., Crapster-Pregont, E., Vorburger, A., Starr, R. D., \& Solomon, S. C. (2020). Global major-element maps of Mercury from four years of MESSENGER X-ray Spectrometer observations. Icarus, 345, 113716.


\bibitem{Nordgren1997}
Nordgren, J., Glans, P., Gunnelin, K., Guo, J., Skytt, P., Såthe, C., \& Wassdahl, N. (1997). Resonant soft X-ray fluorescence spectra of molecules. Applied Physics A: Materials Science \& Processing, 65(2).

\bibitem{Nulsen2020}
Nulsen, S., Kraft, R., Germain, G., Dunn, W., Tremblay, G., Beegle, L., ... \& Vance, S. (2020). X-ray emission from Jupiter’s Galilean moons: A tool for determining their surface composition and particle environment. The Astrophysical Journal, 895(2), 79.

\bibitem{Ouyang2010}
Ouyang, Z., Li, C., Zou, Y., Zhang, H., Lü, C., Liu, J., ... \& Gao, G. (2010). Primary scientific results of Chang’E-1 lunar mission. Science China Earth Sciences, 53, 1565-1581.

\bibitem{Parmar2023}
Parmar. V., W. Dunn, A Foster, J. Carter, M. Markevitch, R. Kraft, D. Koutroumpa, A. Bogdan, G. Branduardi-Raymont, K. Kuntz, S. Wolk, F. S. Porter, C. Lisse, K. Dennerl, G. R. Gladstone, C. Jackman, D. Weigt, S. McEntee, D. Bodewits,  C. Feldman, J-U. Ness, R. Cumbee, N. Achilleos, Z. Yao et al., (in review), Fundamental Atomic and Plasma Physics Enabled through Microcalorimeter Resolution of Jupiter’s Aurora with the Line Emission Mapper, Earth and Planetary Physics.

\bibitem{Pillai2021}
Pillai, N. S., Narendranath, S., Vadodariya, K., Tadepalli, S. P., Radhakrishna, V., Tyagi, A., ... \& Vadawale, S. (2021). Chandrayaan-2 Large Area Soft X-ray Spectrometer (CLASS): calibration, in-flight performance and first results. Icarus, 363, 114436.

\bibitem{Plucinsky2018}
Plucinsky, P. P., Bogdan, A., Marshall, H. L., \& Tice, N. W. (2018, July). The complicated evolution of the ACIS contamination layer over the mission life of the Chandra X-ray Observatory. In Space Telescopes and Instrumentation 2018: Ultraviolet to Gamma Ray (Vol. 10699, pp. 1497-1511). SPIE.

\bibitem{Porquet2010}
Porquet, D., Dubau, J., \& Grosso, N. (2010). He-like ions as practical astrophysical plasma diagnostics: From stellar coronae to active galactic nuclei. Space Science Reviews, 157, 103-134.

\bibitem[Qu et al.(2022)]{Qu2022} Qu, Z., Koutroumpa, D., Bregman, J.~N., et al.\ 2022, \apj, 930, 21. doi:10.3847/1538-4357/ac6349

\bibitem{Raab2016}
Raab, W., Branduardi-Raymont, G., Wang, C., Dai, L., Donovan, E., Enno, G., ... \& Zheng, J. (2016, July). SMILE: A joint ESA/CAS mission to investigate the interaction between the solar wind and Earth's magnetosphere. In Space telescopes and instrumentation 2016: Ultraviolet to gamma ray (Vol. 9905, p. 990502). SPIE.

\bibitem{Racca2002}
Racca, G. D., Marini, A., Stagnaro, L., Van Dooren, J., Di Napoli, L., Foing, B. H., ... \& Sjöberg, F. (2002). SMART-1 mission description and development status. Planetary and space science, 50(14-15), 1323-1337.


\bibitem[Richardson \& Cane(2004)]{2004JGRA..109.9104R} Richardson, I.~G. \& Cane, H.~V.\ 2004, Journal of Geophysical Research (Space Physics), 109, A09104. \url{https://doi.org/10.1029/2004JA010598}

\bibitem{Roussos2021}
Roussos, E., \& Kollmann, P. (2021). The radiation belts of Jupiter and Saturn. Magnetospheres in the solar system, 499-514.

\bibitem{Sanchez-Cano2022}
Sanchez-Cano, B., Opgenoorth, H., Leblanc, F., Andrews, D., \& Lester, M. (2022). The M-MATISSE mission: Mars magnetosphere ATmosphere ionosphere and surface SciencE. 44th COSPAR Scientific Assembly. Held 16-24 July, 44, 421.


\bibitem{Schirmer1993}
Schirmer, J., Trofimov, A. B., Randall, K. J., Feldhaus, J., Bradshaw, A. M., Ma, Y., ... \& Sette, F. (1993). K-shell excitation of the water, ammonia, and methane molecules using high-resolution photoabsorption spectroscopy. Physical Review A, 47(2), 1136.

\bibitem{Schwadron2000}
Schwadron, N. A., \& Cravens, T. E. (2000). Implications of solar wind composition for cometary X-rays. The Astrophysical Journal, 544(1), 558.

\bibitem{sibeck2018}
{Sibeck}, David G., {Allen}, R., {Aryan}, H., {Bodewits}, D., {Brandt}, P., {Branduardi-Raymont}, G., {Brown}, G., {Carter}, J.~A., {Collado-Vega}, Y.~M., {Collier}, M.~R., {Connor}, H.~K.,{Cravens}, T.~E., {Ezoe}, Y., {Fok}, M. -C., {Galeazzi}, M., {Gutynska}, O., {Holmstr{\"o}m}, M., {Hsieh}, S. -Y., {Ishikawa}, K., {Koutroumpa}, D., {Kuntz}, K.~D., {Leutenegger}, M., {Miyoshi}, Y., {Porter}, F.~S., {Purucker}, M.~E., {Read}, A.~M., {Raeder}, J., {Robertson}, I.~P., {Samsonov}, A.~A., {Sembay}, S., {Snowden}, S.~L., {Thomas}, N.~E., {von Steiger}, R., {Walsh}, B.~M., {Wing}, S., (2018). "Imaging Plasma Density Structures in the Soft X-Rays Generated by Solar Wind Charge Exchange with Neutrals". Space Science Review 214(4), 79.

\bibitem{Smith2001}
Smith, R. K., Brickhouse, N. S., Liedahl, D. A., \& Raymond, J. C. (2001). Collisional plasma models with APEC/APED: emission-line diagnostics of hydrogen-like and helium-like ions. The Astrophysical Journal, 556(2), L91.

\bibitem{Smith2005}
Smith, R. K., Edgar, R. J., Plucinsky, P. P., Wargelin, B. J., Freeman, P. E., \& Biller, B. A. (2005). Chandra observations of MBM 12 and models of the local bubble. The Astrophysical Journal, 623(1), 225.

\bibitem{Smrekar2022}
Smrekar, S., Hensley, S., Nybakken, R., Wallace, M. S., Perkovic-Martin, D., You, T. H., ... \& Mazarico, E. (2022, March). VERITAS (Venus emissivity, radio science, InSAR, topography, and spectroscopy): a discovery mission. In 2022 IEEE Aerospace Conference (AERO) (pp. 1-20). IEEE.


\bibitem{Sohdi1984}
Sodhi, R. N., \& Brion, C. E. (1984). Reference energies for inner shell electron energy-loss spectroscopy. Journal of electron spectroscopy and related phenomena, 34(4), 363-372.

\bibitem{Snios2016}
Snios, B., Kharchenko, V., Lisse, C. M., Wolk, S. J., Dennerl, K., \& Combi, M. R. (2016). Chandra observations of Comets C/2012 S1 (ISON) and C/2011 L4 (PanSTARRS). The Astrophysical Journal, 818(2), 199.

\bibitem{Snios2014}
Snios, B., Lewkow, N., \& Kharchenko, V. (2014). Cometary emissions induced by scattering and fluorescence of solar X-rays. Astronomy \& Astrophysics, 568, A80.

\bibitem{Snios2018}
Snios, B., Lichtman, J., \& Kharchenko, V. (2018). The presence of dust and ice scattering in X-ray emissions from Comets. The Astrophysical Journal, 852(2), 138.

\bibitem{Stawarz2019}
Stawarz, J. E., Eastwood, J. P., Phan, T. D., Gingell, I. L., Shay, M. A., Burch, J. L., ... \& Franci, L. (2019). Properties of the turbulence associated with electron-only magnetic reconnection in Earth’s magnetosheath. The Astrophysical journal letters, 877(2), L37.

\bibitem{Stawarz2021}
Stawarz, J. E., Eastwood, J. P., Phan, T. D., Gingell, I. L., Pyakurel, P. S., Shay, M. A., ... \& Le Contel, O. (2022). Turbulence-driven magnetic reconnection and the magnetic correlation length: Observations from Magnetospheric Multiscale in Earth's magnetosheath. Physics of Plasmas, 29(1).

\bibitem[Stone et al.(2005)]{2005Sci...309.2017S} Stone, E.~C., Cummings, A.~C., McDonald, F.~B., et al.\ 2005, Science, 309, 2017. \url{https://doi.org/10.1126/science.1117684}

\bibitem{Sulaiman2022}
Sulaiman, A. H., Mauk, B. H., Szalay, J. R., Allegrini, F., Clark, G., Gladstone, G. R., ... \& Bolton, S. J. (2022). Jupiter's low‐altitude auroral zones: Fields, particles, plasma waves, and density depletions. Journal of Geophysical Research: Space Physics, 127(8), e2022JA030334.

\bibitem{Swinyard2009}
Swinyard, B. M., Joy, K. H., Kellett, B. J., Crawford, I. A., Grande, M., Howe, C. J., ... \& SMART, T. (2009). X-ray fluorescence observations of the Moon by SMART-1/D-CIXS and the first detection of Ti K$\alpha$ from the lunar surface. Planetary and Space Science, 57(7), 744-750.

\bibitem{Tinetti2018}
Tinetti, G., Drossart, P., Eccleston, P., Hartogh, P., Heske, A., Leconte, J., ... \& Herrero, E. (2018). A chemical survey of exoplanets with ARIEL. Experimental astronomy, 46, 135-209.

\bibitem{Trombka2000}
Trombka, J. I., Squyres, S. W., Bruckner, J., Boynton, W. V., Reedy, R. C., McCoy, T. J., ... \& Petaev, M. (2000). The elemental composition of asteroid 433 Eros: Results of the NEAR-Shoemaker X-ray spectrometer. science, 289(5487), 2101-2105.

\bibitem{Tsuda2013}
Tsuda, Y., Yoshikawa, M., Abe, M., Minamino, H., \& Nakazawa, S. (2013). System design of the Hayabusa 2—Asteroid sample return mission to 1999 JU3. Acta Astronautica, 91, 356-362.

\bibitem{VonSteiger2000}
Von Steiger, R., Schwadron, N. A., Fisk, L. A., Geiss, J., Gloeckler, G., Hefti, S., ... \& Zurbuchen, T. H. (2000). Composition of quasi‐stationary solar wind flows from Ulysses/Solar Wind Ion Composition Spectrometer. Journal of Geophysical Research: Space Physics, 105(A12), 27217-27238.

\bibitem{Waite1997}
Waite Jr, J. H., Gladstone, G. R., Lewis, W. S., Drossart, P., Cravens, T. E., Maurellis, A. N., ... \& Miller, S. (1997). Equatorial X-ray emissions: Implications for Jupiter's high exospheric temperatures. Science, 276(5309), 104-108.

\bibitem{Waite2004}
Waite, J. H., Lewis, W. S., Kasprzak, W. T., Anicich, V. G., Block, B. P., Cravens, T. E., ... \& Yelle, R. V. (2004). The Cassini ion and neutral mass spectrometer (INMS) investigation. The Cassini-Huygens Mission: Orbiter In Situ Investigations Volume 2, 113-231.

\bibitem{Waite2006}
Waite Jr, J. H., Combi, M. R., Ip, W. H., Cravens, T. E., McNutt Jr, R. L., Kasprzak, W., ... \& Tseng, W. L. (2006). Cassini ion and neutral mass spectrometer: Enceladus plume composition and structure. science, 311(5766), 1419-1422.

\bibitem{Wakeford2020}
Wakeford, H. R., \& Dalba, P. A. (2020). The exoplanet perspective on future ice giant exploration. Philosophical Transactions of the Royal Society A, 378(2187), 20200054.

\bibitem{Walsh2020}
Walsh, B., Collier, M. R., Busk, S., Connor, H. K., Kuntz, K. D., McShane, J., ... \& Thomas, N. (2020, December). The Lunar Environment Heliospheric X-ray Imager (LEXI)-A mission for global magnetospheric imaging. In AGU Fall Meeting Abstracts (Vol. 2020, pp. SM029-01).

\bibitem{Wargelin2004}
Wargelin, B. J., Markevitch, M., Juda, M., Kharchenko, V., Edgar, R., \& Dalgarno, A. (2004). Chandra observations of the “dark” Moon and geocoronal solar wind charge transfer. The Astrophysical Journal, 607(1), 596.

\bibitem{Weigt2020}
Weigt, D. M., Jackman, C. M., Dunn, W. R., Gladstone, G. R., Vogt, M. F., Wibisono, A. D., ... \& Kraft, R. P. (2020). Chandra Observations of Jupiter's X‐ray auroral emission during Juno Apojove 2017. Journal of Geophysical Research: Planets, 125(4), e2019JE006262.

\bibitem{Weigt2021}
Weigt, D. M., Jackman, C. M., Vogt, M. F., Manners, H., Dunn, W. R., Gladstone, G. R., ... \& McEntee, S. C. (2021). Characteristics of Jupiter's X‐Ray Auroral Hot Spot Emissions Using Chandra. Journal of Geophysical Research: Space Physics, 126(9), e2021JA029243.

\bibitem{Wibisono2020}
Wibisono, A. D., Branduardi‐Raymont, G., Dunn, W. R., Coates, A. J., Weigt, D. M., Jackman, C. M., ... \& Fleming, D. (2020). Temporal and spectral studies by XMM‐Newton of Jupiter's X‐ray auroras during a compression event. Journal of Geophysical Research: Space Physics, 125(5), e2019JA027676.

\bibitem{Wibisono2021}
Wibisono, A. D., Branduardi-Raymont, G., Dunn, W. R., Kimura, T., Coates, A. J., Grodent, D., ... \& Haythornthwaite, R. P. (2021). Jupiter’s X-ray aurora during UV dawn storms and injections as observed by XMM–Newton, Hubble, and Hisaki. Monthly Notices of the Royal Astronomical Society, 507(1), 1216-1228.

\bibitem{Wibisono2023}
Wibisono, A. D., Branduardi-Raymont, G., Coates, A. J., Dunn, W. R., \& French, R. J. (2023). Jupiter’s equatorial X-ray emissions over two solar cycles. Monthly Notices of the Royal Astronomical Society, 521(4), 5596-5603.

\bibitem{Wibisono2023thesis}
Wibisono, A. D. (2023). Temporal and Spectral Studies of Jupiter's Auroral and Equatorial X-ray Emissions (Doctoral dissertation, UCL (University College London)).

\bibitem{Widemann2020}
Widemann, T., Ghail, R., Wilson, C. F., \& Titov, D. V. (2020, December). EnVision: Europe's proposed mission to Venus. In Agu fall meeting abstracts (Vol. 2020, pp. P022-02).

\bibitem{Wong2004}
Wong, M. H., Mahaffy, P. R., Atreya, S. K., Niemann, H. B., \& Owen, T. C. (2004). Updated Galileo probe mass spectrometer measurements of carbon, oxygen, nitrogen, and sulfur on Jupiter. Icarus, 171(1), 153-170.

\bibitem{YaoDunn2021}
Yao, Z., Dunn, W. R., Woodfield, E. E., Clark, G., Mauk, B. H., Ebert, R. W., ... \& Bolton, S. J. (2021). Revealing the source of Jupiter’s x-ray auroral flares. Science Advances, 7(28), eabf0851.


\bibitem{Zou2014}
Zou, X., Li, C., Liu, J., Wang, W., Li, H., \& Ping, J. (2014). The preliminary analysis of the 4179 Toutatis snapshots of the Chang’E-2 flyby. Icarus, 229, 348-354.




\end{thebibliography}
\end{document}